\documentclass[11pt]{article}

\makeatletter
\def\maxwidth{ %
  \ifdim\Gin@nat@width>\linewidth
    \linewidth
  \else
    \Gin@nat@width
  \fi
}
\makeatother

\RequirePackage{amsmath,amssymb}
\RequirePackage{amsthm}
\RequirePackage{natbib}
\RequirePackage[colorlinks,allcolors=blue]{hyperref}

\usepackage{fullpage}
\usepackage{setspace}
\usepackage{bbm,graphicx,bm} %subfigure
\usepackage{enumitem}
\usepackage{multirow}
\usepackage[all,import]{xy}
\usepackage{appendix}
\usepackage{comment}
\usepackage{color}
\usepackage{soul}
\usepackage{pdflscape}
\usepackage{subcaption}

\usepackage{booktabs}
\usepackage{longtable}
\usepackage{array}
\usepackage[table]{xcolor}
\usepackage{wrapfig}
\usepackage{float}
\usepackage{colortbl}
\usepackage{pdflscape}
\usepackage{tabu}
\usepackage{threeparttable}
\usepackage{threeparttablex}
\usepackage[normalem]{ulem}
\usepackage{makecell}
\usepackage{afterpage}
\usepackage{nicefrac}

\theoremstyle{definition}
\newtheorem{assumption}{Assumption}

\newtheorem{proposition}{Proposition}

\newcommand\indep{\protect\mathpalette{\protect\independenT}{\perp}}
\def\independenT#1#2{\mathrel{\rlap{$#1#2$}\mkern2mu{#1#2}}}
\newcommand{\bbone}{\ensuremath{\mathbbm{1}}}
\newcommand{\R}{\ensuremath{\mathbb{R}}}

\newcommand{\I}{\ensuremath{\mathcal{I}}}

\newcommand{\calM}{\ensuremath{\mathcal{M}}}

\newcommand{\calP}{\ensuremath{\mathcal{P}}}
\newcommand{\E}{\ensuremath{\mathbb{E}}}
\newcommand{\calE}{\ensuremath{\mathcal{E}}}

\newcommand{\calX}{\ensuremath{\mathcal{X}}}

\newcommand{\calL}{\ensuremath{\mathcal{L}}}

\newcommand{\Var}{\text{Var}}

\newcommand{\logit}{\text{logit}}

\def\super{\textsuperscript}

\def\b1{\boldsymbol{1}}

\definecolor{RED}{RGB}{255,0,0}

\usepackage{soul}
\usepackage[utf8]{inputenc}

\title{Varying impacts of letters of recommendation on college admissions:\\
Approximate balancing weights for subgroup effects in observational studies\thanks{email: \texttt{ebenmichael@berkeley.edu}. We thank Greg Dubrow, Chad Hazlett, Amy Jarich, Jared Murray, Olufeme Ogundole, and James Pustejovsky for helpful conversations and thoughtful comments. We also thank Elsa Augustine, Charles Davis, and Audrey Tiew for excellent research assistance.
This work was supported in part by by the William T. Grant Foundation and by the Institute of Education Sciences, U.S. Department of Education, through Grant R305D200010. The opinions expressed are those of the authors and do not represent views of the Institute or the U.S. Department of Education.}}

\author{Eli Ben-Michael, Avi Feller, and Jesse Rothstein \\[1em] UC Berkeley and Harvard University}
\date{February 2021}
\begin{document}
\singlespacing

\maketitle
\thispagestyle{empty}
\pagenumbering{gobble}

\begin{abstract}
In a pilot program during the 2016-17 admissions cycle, the University of California, Berkeley invited many applicants for freshman admission to submit letters of recommendation. 
We use this pilot as the basis for an observational study of the impact of submitting letters of recommendation on subsequent admission, with the goal of estimating how impacts vary across pre-defined subgroups. 
Understanding this variation is challenging in observational studies, however, because estimated impacts reflect both actual treatment effect variation and differences in covariate balance across groups. 
To address this, we develop balancing weights that directly optimize for ``local balance'' within subgroups while maintaining global covariate balance between treated and control units. 
We then show that this approach has a dual representation as a form of inverse propensity score weighting with a hierarchical propensity score model. 
In the UC Berkeley pilot study, our proposed approach yields excellent local and global balance, unlike more traditional weighting methods, which fail to balance covariates within subgroups.
We find that the impact of letters of recommendation increases with the predicted probability of admission, with mixed evidence of differences for under-represented minority applicants.
\end{abstract}

\clearpage
\pagenumbering{arabic}
\onehalfspacing
% \doublespacing
\section{Introduction and motivation}

In a pilot program during the 2016-17 admissions cycle, the University of California, Berkeley invited many applicants for freshman admission to submit letters of recommendation (LORs) as part of their applications.
UC Berkeley had (and has) a ``holistic review'' admissions process, which attempts to examine the whole applicant, taking account of any contextual factors and obstacles overcome without over-reliance on quantitative measures like SAT scores \citep{hout2005berkeley}.
Unlike other highly selective universities, however, UC Berkeley had never previously asked applicants to submit letters from teachers and guidance counselors. 

The new approach proved controversial within the university. The LORs were intended to help identify students from non-traditional backgrounds who might otherwise be overlooked \citep{UC_Diversity_2017}. But there was also legitimate concern that applicants from disadvantaged backgrounds might not have access to adults who could write strong letters, and that the use of letters would further disadvantage these students \citep{Chalfant_letter_2017}.

In this paper, we use the Berkeley pilot as the basis for an observational study of the impact of submitting a letter of recommendation on subsequent admission. 
Our goal is to assess how impacts vary across pre-defined subgroups, in order to inform the debate over the Berkeley policy and similar debates at other universities.

Assessing treatment effect heterogeneity is challenging in non-randomized settings because variation in estimated impacts reflects both actual treatment effect variation and differences in covariate balance across groups. 
Inverse Propensity Score Weighting (IPW) is one standard approach for addressing this: 
first estimate a propensity score model via logistic regression, including treatment-by-subgroup interaction terms; construct weights based on the estimated model; and then compare IPW estimates across subgroups \citep[see][]{green2014examining, Lee2019}. 
Estimated weights from traditional IPW methods, however, are only guaranteed to have good covariate balancing properties asymptotically.
Balancing weights estimators, by contrast, instead find weights that directly minimize a measure of covariate imbalance, often yielding better finite sample performance \citep{Zubizarreta2015,Athey2018a,Hirshberg2019_amle,benmichael_balancing_review}.
Both balancing weights and traditional IPW face a curse of dimensionality when estimating subgroup effects:
it is difficult to achieve exact balance on all covariates within each subgroup, or, equivalently, balance all covariate-by-subgroup interactions.
 
We therefore develop an approximate balancing weights approach tailored to estimating subgroup treatment effects, with a focus on the UC Berkeley LOR pilot study.
Specifically, we present a convex optimization problem that finds
weights that directly target the level of local imbalance within each subgroup --- ensuring \emph{approximate} local covariate balance --- while guaranteeing \emph{exact} global covariate balance between the treated and control samples.
We show that controlling local imbalance 
controls the estimation error of subgroup-specific effects, allowing us to better isolate treatment effect variation.
We also show that, even when the target estimand is the overall treatment effect, ensuring both exact global balance and approximate local balance reduces the overall estimation error.

Next, we demonstrate that this proposal has a dual representation as inverse propensity weighting with a hierarchical propensity score model, building on recent connections between balancing weights and propensity score estimation \citep{Chattopadhyay2019}. %Zhao2016a, Tan2017, benmichael2019_multisynth
In particular, finding weights that minimize both global and local imbalance corresponds to estimating a propensity score model in which the subgroup-specific parameters are partially pooled toward a global propensity score model.
Any remaining imbalance after weighting may lead to bias. To adjust for this, we also combine the weighting approach with an outcome model, analogous to bias correction for matching \citep{rubin1973bias, Athey2018a}.

After assessing our proposed approach's properties, we use it to estimate the impacts of letters of recommendation during the 2016 UC Berkeley undergraduate admissions cycle.
% \avi{General} 
Based on the Berkeley policy debate, we focus on variation in the effect on admissions rates based on under-represented minority (URM) status and on 
the \emph{a priori} predicted probability of admission, estimated using data from the prior year's admissions cycle.
First, we show that the proposed weights indeed yield excellent local and global balance, while traditional propensity score weighting methods yield poor local balance. 
We then find evidence that the impact of letters increases with the predicted probability of admission.
Applicants who are very unlikely to be admitted see little benefit from letters of recommendation, while applicants on the cusp of acceptance see a larger, positive impact.

The evidence on the differential effects by URM status is more mixed.
Overall, the point estimates for URM and non-URM applicants are close to each other. However, these estimates are noisy and mask 
important variation by \emph{a priori} probability of admission.
For applicants with the highest baseline admission probabilities, we estimate larger impacts for non-URM than URM applicants, though these estimates are sensitive to augmentation with an outcome model.
For all other applicants, we estimate the reverse: larger impacts for URM than non-URM applicants.
Since URM status is correlated with the predicted probability of admission, this leads to a Simpson's Paradox-type pattern for subgroup effects, with a slightly larger point estimate for non-URM applicants pooled across groups \citep{Bickel1975, vanderweele2011interpretation}.

The fact that results hinge on higher-order interactions suggests caution but also highlights the advantages of our design-based approach with pre-specified subgroups \citep{rubin2008objective}. 
Since we separate the design and analysis phases, we can carefully assess covariate balance and overlap in the subgroups of interest --- and can tailor the weights to target these quantities directly.  
By contrast, many black-box machine learning methods estimate the entire conditional average treatment effect function, without prioritizing estimates for pre-specified subgroups \citep{Carvalho2019}.
While these approaches require minimal input from the researcher, they can over-regularize the estimated effects for subgroups of interest, making it difficult to isolate higher-order interactions.
As we discuss in Section \ref{sec:augment}, we argue that researchers should view these weighting and outcome modeling methods as complements rather than substitutes. Even without pre-specified subgroups, there can be substantial gains from combining weighting and outcome modeling for treatment effect variation \citep{nie2017quasi, Kunzel2019}. Our proposed augmented estimator is especially attractive because we can incorporate pre-defined subgroups in the weighting estimator while still leveraging off-the-shelf machine learning methods for outcome modeling.

Finally, we conduct extensive robustness and sensitivity checks, detailed in Appendix \ref{sec:robustness_appendix}. In addition to alternative estimators and sample definitions, we conduct a formal sensitivity analysis for violations of the ignorability assumption, adapting the recent proposal from \citet{soriano2019sensitivity}. We also explore an alternative approach that instead leverages unique features of the UC Berkeley pilot study, which included an additional review without the letters of recommendation from a sample of 10,000 applicants. 
Overall, our conclusions are similar across a range of approaches. Thus, we believe our analysis is a reasonable first look at this question, albeit best understood alongside other approaches that rest on different assumptions \citep[such as those in][]{rothstein_lor2017}.

The paper proceeds as follows. In the next section we introduce the letter of recommendation pilot program at UC Berkeley. 
Section \ref{sec:prelim} introduces the problem setup and notation, and discusses related work. Section \ref{sec:approx_weights} proposes and analyzes the approximate balancing weights approach. 
Section \ref{sec:sims} presents a simulation study.
Section \ref{sec:results} presents empirical results on the effect of letters of recommendation. Section \ref{sec:discussion} concludes with a discussion about possible extensions. 
The appendix includes additional analyses and theoretical discussion.

% Motivating example: 
\subsection{A pilot program for letters of recommendation in college admissions}
\label{sec:lor}

As we discuss above, there is considerable debate over the role of letters of recommendation in college admissions. LORs have the potential to offer insight into aspects of the applicant not captured by the available quantitative information or by the essays that applicants submit \citep{kuncel2014meta}. 
At the same time, letters from applicants from disadvantaged backgrounds or under-resourced high schools may be less informative or prejudicial against the applicant, due, e.g., to poor writing or grammar, or to lower status of the letter writer; see \citet{schmader2007linguistic} as an example.

These issues were central in the UC Berkeley debates over the policy.  
One Academic Senate committee, following an inquiry into the ``intended and unintended consequences of the interim pilot admissions policy, especially for underrepresented minority students (URMs),'' concluded that ``the burden of proof rests on those who want to implement the new letters of recommendation policy, and should include a test of statistical significance demonstrating measurable impact on increasing diversity in undergraduate admissions'' \citep{UC_Diversity_2017}.
The UC system-wide faculty senate ultimately limited the use of LORs following the pilot (though before any results were available), citing concerns that ``LORs conflict with UC principles of access and fairness, because students attending under-resourced schools or from disadvantaged backgrounds will find it more difficult to obtain high-quality letters, and could be disadvantaged by a LOR requirement'' \citep{Chalfant_letter_2017}.

The UC Berkeley LOR pilot study is a unique opportunity to assess these questions; \citet{rothstein_lor2017} discusses implementation details.
For this analysis, we restrict our sample to non-athlete California residents who applied to either the College of Letters and Science or the College of Engineering at UC Berkeley in the 2016 admissions cycle. This leaves 40,541 applicants, 11,143 of whom submitted LORs.
We focus our analysis on the impacts for applicants who both were invited to and subsequently did submit LORs.\footnote{We could use the methods discussed here to explore a range of different quantities. For this target, the net effect of LORs on admission includes differential rates of submission of a letter given invitation. While non-URM applicants submitted letters at a higher rate than URM applicants, the majority of the discrepancy arises from applicants who were unlikely to be admitted \emph{a priori} \citep{rothstein_lor2017}.}

\subsubsection{Selection into treatment}
\label{sec:selection}

\begin{figure}[tb]
  \centering \includegraphics[width=.8\maxwidth]{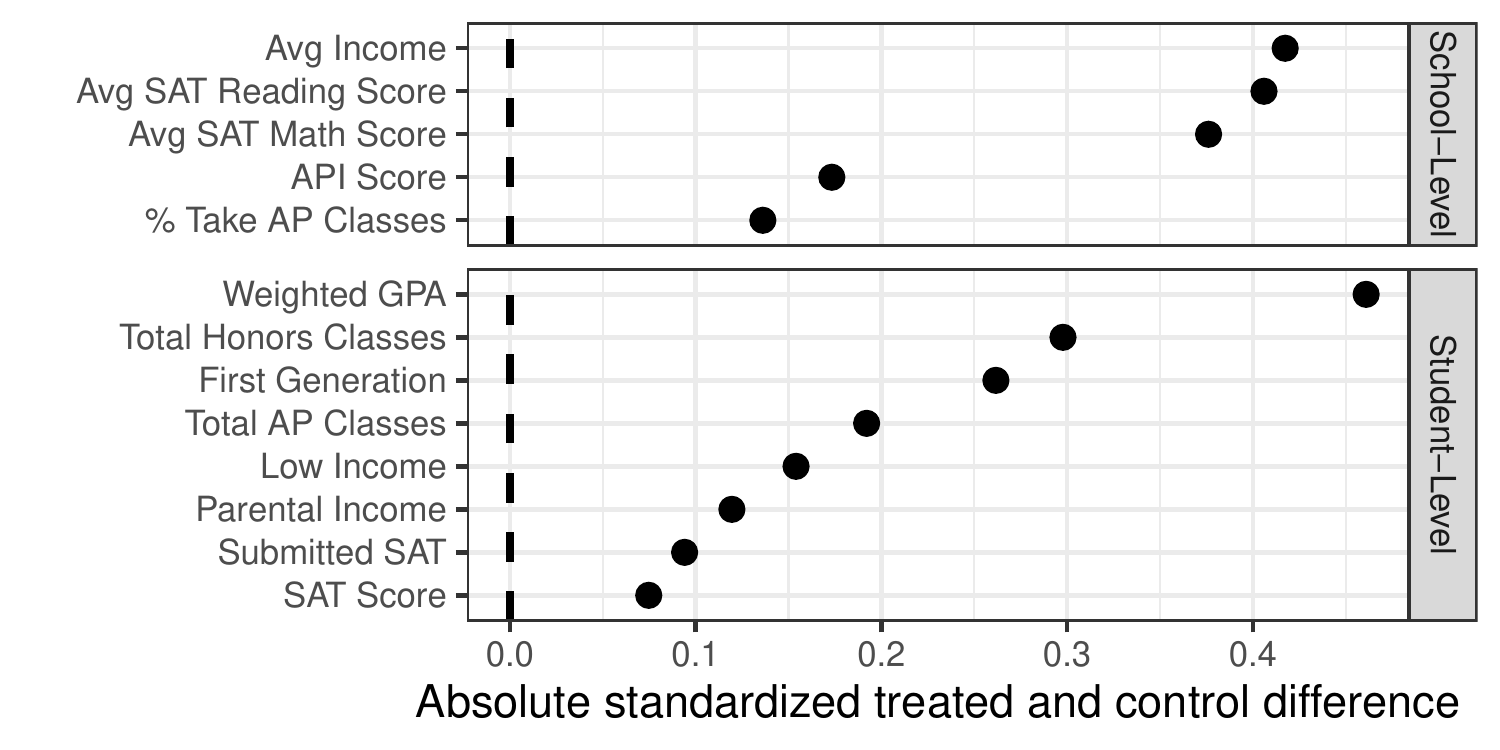}
\caption{Absolute difference in means, standardized by the pooled standard deviation, between applicants submitting and not submitting letters of recommendation for several key covariates. By design, applicants submitting letters of recommendation disproportionately have a ``Possible'' score from the first reader (70\% of treated applicants vs. 4\% of untreated applicants).}
\label{fig:std_diff}
\end{figure}

UC Berkeley uses a two-reader evaluation system. Each reader scores applicants on a three-point scale, as ``No,'' ``Possible,'' or ``Yes.'' Application decisions are based on the combination of these two scores and the major to which a student has applied. In the most selective majors (e.g., mechanical engineering), an applicant typically must receive two ``Yes'' scores to be admitted, while in others a single ``Yes'' is sufficient. In the LOR pilot, applicants were invited to submit letters based in part on the first reader score, and the LORs, if submitted, were made available to the second reader. 

As in any observational study of causal effects, selection into treatment is central. Decisions to submit letters were a two-step process. Any applicant who received a ``Possible'' score from the first reader was invited. In addition, due to concerns that first read scores would not be available in time to be useful, an index of student- and school-level characteristics was generated, and applicants with high levels of the index were invited as well.\footnote{The index was generated from a logistic regression fit to data from the prior year's admissions cycle, predicting whether an applicant received a ``Possible'' score (versus either a ``No'' or a ``Yes''). Applicants with predicted probabilities from this model greater than 50\% were invited to submit LORs. Because we observe all of the explanatory variables used in the index, this selection depends only on observable covariates.
A small share of applicants with low predicted probabilities received first reads after January 12, 2017, the last date that LOR invitations were sent, and were not invited even if they received ``Possible'' scores.}
Of the 40,451 total applicants, 14,596 were invited to submit a letter. Approximately 76\% of those invited to submit letters eventually submitted them, and no applicant submitted a letter who was not invited to.

For this analysis, we assume that submission of LORs is effectively random conditional on the first reader score and on both student- and school-level covariates (Assumption \ref{a:ignore} below).
In particular, the \emph{interaction} between the covariates and the first reader score plays an important role in the overall selection mechanism, as applicants who received a score of ``No'' or ``Yes'' from the first reader could still have been asked to submit an LOR based on their individual and school information.
Figure \ref{fig:std_diff} shows covariate imbalance for several key covariates --- measured as the absolute difference in means divided by the pooled standard deviation --- for applicants who submitted LORs versus those who did not.\footnote{The full set of student-level variables we include in our analysis are: weighted and unweighted GPA, GPA percentile within school, parental income and education, SAT composite score and math score, the number of honors courses and percentage out of the total available, number of AP courses, ethnic group, first generation college student status, and fee waiver status. The school level variables we control for are: average SAT reading, writing, and math scores, average ACT score, average parental income, percent of students taking AP classes, and the school Academic Performance Index (API) evaluated through California's accountability tests. For students that did not submit an SAT score but did submit an ACT score, we imputed the SAT score via the College Board's SAT to ACT concordance table. For the 992 applicants with neither an SAT nor an ACT score, we impute the SAT score as the average among applicants from the school.}
For the purposes of this study, we follow the university in defining  a URM applicant as one who is a low-income student, a student in a low-performing high school, a first-generation college student, or from an underrepresented racial or ethnic group. 
We see that there are large imbalances in observable applicant characteristics, most notably average school income, GPA, the number of honors and AP classes taken, and SAT score.
There were also large imbalances in first reader scores (not shown in Figure \ref{fig:std_diff}): 70\% of applicants that submitted LORs had ``Possible'' scores, compared to only 4\% of those who did not. There is a smaller imbalance in URM status, with 61\% of those submitting LORs classified as URMs, versus 53\% of those who did not submit.

\subsubsection{Heterogeneity across \emph{a priori} probability of admission}
\label{sec:ai}
\begin{figure}[tb]
  \centering \includegraphics[width=0.85\maxwidth]{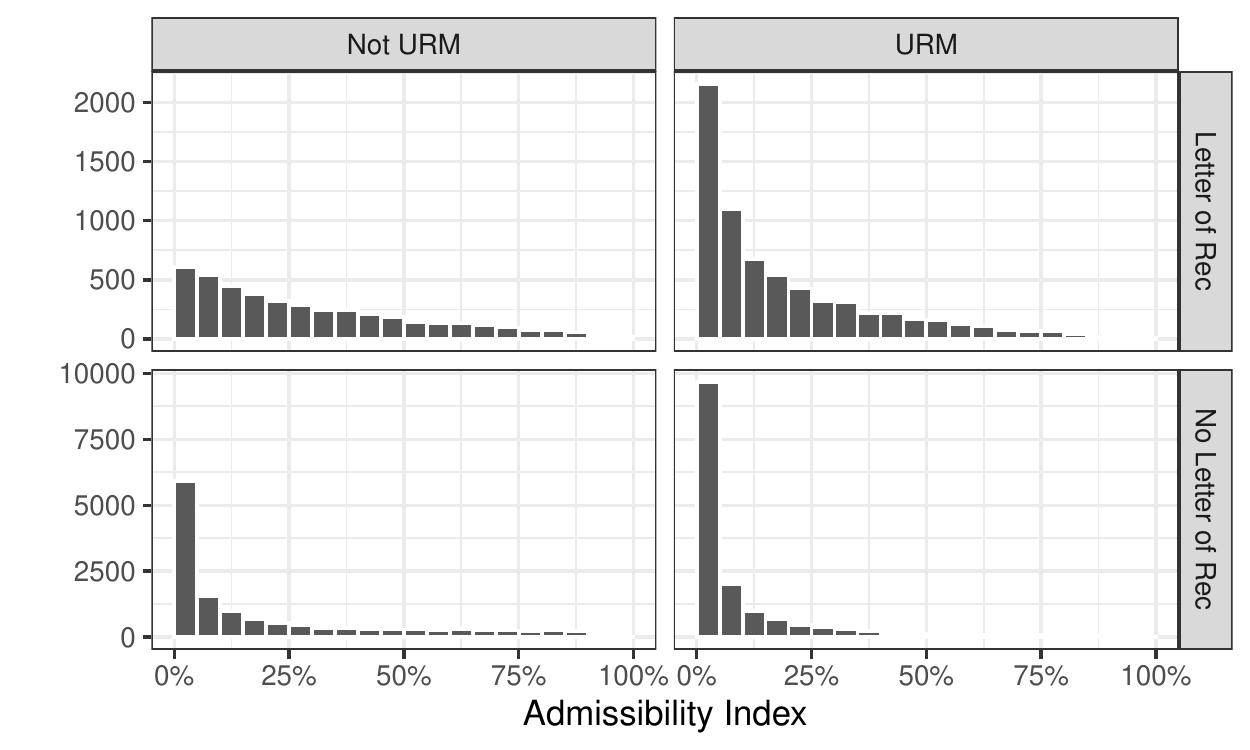}
\caption{Distribution of the ``admissibility index'' --- an estimate of the \emph{a priori} probability of acceptance --- for the 2016 UC Berkeley application cohort, separated into URM and non-URM and those that submitted a letter versus those that did not.}
\label{fig:ai_hist}
\end{figure}

To better understand who was invited to submit LORs and any differential impacts between URM and non-URM applicants, we construct a univariate summary of applicant- and school-level characteristics. We use logistic regression to estimate the probability of admission given observable characteristics using the \emph{prior year} (2015) admissions data. 
We then use this model to predict \emph{a priori} admissions probabilities for the applicants of interest in 2016; we refer to these predicted probabilities as the Admissibility Index (AI).
Overall, the AI is a good predictor of admission in 2016: for applicants who do not submit LORs, the overall Area Under the Curve (AUC) in predicting 2016 admissions is 0.92 and the mean square error is 7\% (see Appendix Table \ref{tab:ai_roc_auc}). However, the predictive accuracy decreases for higher AI applicants, for whom the AI under-predicts admissions rates (see Appendix Figure \ref{fig:ai_performance}).

Figure \ref{fig:ai_hist} shows the AI distribution for the 2016 applicant cohort, broken out by URM status and LOR submission. There are several features of this distribution that have important implications for our analysis. First, although the probability of admission is quite low overall, 
applicants across nearly the full support of probabilities submitted LORs. This is primarily because applicants who received ``Possible'' scores from the first readers come from a wide range of admissibility levels. This will allow us to estimate heterogeneous effects across the full distribution, with more precision for applicants with lower AIs.
Second, because the admissions model disproportionately predicted that URM students had high chances of receiving ``Possible'' scores, many more URM applicants were invited to submit letters than non-URM applicants, and so our estimates for URM applicants will be more precise than those for non-URM applicants. 
Third, at higher AI levels large shares of applicants submitted LORs, leaving few comparison observations. This will make it challenging to form balanced comparison groups for high-AI URM applicants who submit letters.

From Figure \ref{fig:ai_hist} we know that the distribution of AI varies between URM and non-URM applicants, and so apparent differences in estimated effects between the two groups may be due to compositional differences.
Therefore, in the subsequent sections we will focus on estimating effects within subgroups defined by both URM status and admissibility. 
To do this, we define subgroups by creating four (non-equally-sized) strata of the AI: $<5\%$, $5\%-10\%$, $10\%-20\%$ and $> 20\%$. Interacting with URM status, this leads to eight non-overlapping subgroups; we will marginalize over these to estimate the other subgroup effects above. Table \ref{tab:subgroup_counts} shows the total number of applicants in each of the eight groups, along with the proportion submitting letters of recommendation. As we discuss in Section \ref{sec:results}, we will further divide each of these subgroups by first reader score and college, to ensure exact balance on these important covariates.

\begin{table}[tbp]   
  \centering
  \begin{tabular}{@{}llrrr@{}}
    \toprule
    AI Range & URM & Number of Applicants & Number Submitting LOR & Proportion Treated\\
    \midrule
    \multirow{2}{*}{$< 5\%$} & URM & 11,832 & 2,157 & 18\%\\
    & Not URM & 6,529 & 607 & 9\%\\
    \midrule
    \multirow{2}{*}{5\% - 10\%} & URM & 3,106 & 1,099 & 35\%\\
    & Not URM & 2,099 & 536 & 25\%\\
    \midrule
    \multirow{2}{*}{10\% - 20\%} & URM & 2,876 & 1,212 & 42\%\\
    & Not URM & 2,495 & 828 & 33\%\\
    \midrule
    \multirow{2}{*}{$>20\%$} & URM & 4,645 & 2,345 & 50\%\\
    & Not URM & 6,959 & 2,359 & 34\%\\
    \bottomrule
  \end{tabular}
  \caption{Number of applicants and proportion treated by subgroup.}
  \label{tab:subgroup_counts}
\end{table}

%%%
%%% SETUP
%%%
% \clearpage
\section{Treatment effect variation in observational studies}
\label{sec:prelim}
\subsection{Setup and estimands}
We now describe the letter of recommendation study as an observational study where for each applicant $i=1,\ldots,n$, we observe applicant and school level-covariates $X_i \in \calX$; a group indicator $G_i \in \{1,\ldots,K\}$ denoting a pre-defined subgroup of interest;
% e.g., URM status or coarsened AI; 
a binary indicator for submitting a letter of recommendation $W_i \in \{0,1\}$;  and whether the applicant is admitted, which we denote as $Y_i \in \{0,1\}$.
Let $n_{1g}$ and $n_{0g}$ represent the number of treated and control units in subgroup $G_i = g$, respectively. We assume that for each applicant, $(X_i, G_i, W_i, Y_i)$ are sampled i.i.d. from some distribution $\calP(\cdot)$. Following the potential outcomes framework \citep{neyman1923,Holland1986}, we assume SUTVA \citep{rubin1980} and posit two potential outcomes $Y_i(0)$ and $Y_i(1)$ for each applicant $i$, corresponding to $i$'s outcome if that applicant submits a letter of recommendation or not, respectively; the observed outcome is $Y_i = W_i Y_i(1) + (1-W_i)Y_i(0)$.\footnote{There is a possibility of interference induced by the number of admitted applicants being capped. With 6874 admitted students, we consider the potential interference to be negligible} In this study we are interested in estimating two types of effects. First, we wish to estimate the overall Average Treatment Effect on the Treated (ATT), the treatment effect for applicants who submit a letter,
 $$\tau = \E[Y(1) - Y(0) \mid W = 1],$$
 where we denote $\mu_1 = \E[Y(1) \mid W = 1]$ and $\mu_0 = \E[Y(0) \mid W = 1]$. 
Second, for each subgroup $G_i = g$, we would like to estimate the Conditional ATT (CATT),
\begin{equation}
    \label{eq:catt}
    \tau_g = \E[Y(1) - Y(0) \mid G = g, W = 1],
\end{equation}
where similarly we denote $\mu_{1g} = \E[Y(1) \mid G = g, W = 1]$ and $\mu_{0g} = \E[Y(0) \mid G = g, W = 1]$.

Estimating $\mu_{1g}$ is relatively straightforward: we can simply use the average outcome for treated units in group $g$, $\hat{\mu}_{1g} \equiv \frac{1}{n_{1g}} \sum_{G_i = g} W_i Y_i$. However, estimating $\mu_{0g}$ is more difficult due to confounding; we focus much of our discussion on imputing this counterfactual mean for the group of applicants who submitted letters of recommendation. To do this, we rely on two key assumptions that together form the usual \emph{strong ignorability} assumption \citep{Rosenbaum1983}.
\begin{assumption}[Ignorability]
    \label{a:ignore}
    The potential outcomes are independent of treatment given the covariates and subgroup:
    \begin{equation}
        \label{eq:ignore}
        Y(1), Y(0) \indep W \mid X, G.
    \end{equation}
\end{assumption}
\begin{assumption}[One Sided Overlap]
    \label{a:overlap}
    The \emph{propensity score} $e(x, g) \equiv P(W = 1 \mid X = x, G = g)$ is less than 1:
    \begin{equation}
        \label{eq:overlap}
        e(X, G) < 1.
    \end{equation}
\end{assumption}
\noindent In our context, Assumption \ref{a:ignore} says that conditioned on the first reader score and applicant- and school-level covariates, submission of LORs is independent of the potential admissions outcomes. Due to the selection mechanism we describe in Section \ref{sec:selection}, we believe that this is a reasonable starting point for estimating these impacts; see \citet{rothstein_lor2017} and Appendix \ref{sec:within} for alternatives. In Appendix \ref{sec:sensitivity}, we assess the sensitivity of our conclusions to violations of this assumption.

Assumption \ref{a:overlap} corresponds to assuming that no applicant would have been guaranteed to submit a letter of recommendation. Although some applicants were guaranteed to be \emph{invited} to submit an LOR, we believe that this is a reasonable assumption for actually submitting a letter. In Section \ref{sec:diagnostic} we assess overlap empirically.

With this setup, let $m_0(x, g) = \E[Y(0) \mid X = x, G = g]$ be the \emph{prognostic score}, the expected control outcome conditioned on covariates $X$ and group membership $G$. 
Under Assumptions \ref{a:ignore} and \ref{a:overlap}, we have the standard identification result:
\begin{equation}
    \label{eq:identify}
    \mu_{0g} = \E[m_0(X, G) \mid W = 1] = \E\left[\frac{e(X, G)}{1 - e(X,G)} Y \mid W = 0 \right].
\end{equation}
Therefore we can obtain a plug-in estimate for $\mu_{0g}$ with an estimate of the prognostic score, $m_0(\cdot, \cdot)$, an estimate of propensity score, $e(\cdot, \cdot)$, or an estimate of the treatment odds themselves, $\frac{e(\cdot, \cdot)}{1 - e(\cdot, \cdot)}$. 
We next review existing methods for such estimation, turning to our proposed weighting approach in the following section.

\subsection{Related work: methods to estimate subgroup treatment effects}
\label{sec:related}

There is an extensive literature on estimating varying treatment effects in observational studies; see \citet{anoke2019approaches} and \citet{Carvalho2019} for recent discussions. This is an active area of research, and we narrow our discussion here to methods that assess heterogeneity across pre-defined, discrete subgroups. In particular, we will focus on linear weighting estimators that take a set of weights  $\hat{\gamma} \in \R^n$, and estimate $\mu_{0g}$ as a weighted average of the control outcomes in the subgroup:
\begin{equation}
    \label{eq:mu0g_hat}
    \hat{\mu}_{0g} \equiv \frac{1}{n_{1g}}\sum_{G_i = g} \hat{\gamma}_i(1-W_i)Y_i.
\end{equation}
Many estimators take this form; we focus on design-based approaches that do not use outcome information in constructing the estimators \citep{rubin2008objective}. See \citet{hill2011bayesian, Kunzel2019, Carvalho2019, Nie2019, Hahn2020} for discussions of approaches that instead focus on outcome modeling.

\paragraph{Methods based on estimated propensity scores.}
A canonical approach in this setting is Inverse Propensity Weighting (IPW) estimators for $\mu_{0g}$ \citep[see][]{green2014examining}. Traditionally, this proceeds in two steps: first estimate the propensity score $\hat{e}(x, g)$, e.g. via logistic regression; second, estimate $\mu_{0g}$ as in Equation \eqref{eq:mu0g_hat}, with weights $\hat{\gamma}_i = \frac{\hat{e}(X_i, G_i)}{1 - \hat{e}(X_i, G_i)}$:
\begin{equation}
    \label{eq:ipw}
    \hat{\mu}_{0g} = \frac{1}{n_{1g}}\sum_{W_i = 0, G_i = g} \frac{\hat{e}(X_i, G_i)}{1 - \hat{e}(X_i, G_i)} Y_i
\end{equation}
where these are ``odds of treatment'' weights to target the ATT.
A natural approach to estimating $\hat{e}(X_i, G_i)$, recognizing that $G_i$ is discrete,
is to estimate a logistic model for treatment separately for each group or, equivalently, with full interactions between $G_i$ and (possibly transformed) covariates $\phi(X_i) \in \R^p$:
\begin{equation}
    \label{eq:ipw_interact}
    \logit(e(x, g)) = \alpha_g + \beta_g \cdot \phi(x).
\end{equation}
Due to the high-dimensional nature of the problem, it is often infeasible to estimate Equation \eqref{eq:ipw_interact} without any regularization: the treated and control units might be completely separated, particularly when some groups are small.
Classical propensity score modeling with random effects is one common solution, but can be numerically unstable in settings similar to this \citep{zubizarreta2017optimal}.
Other possible solutions in high dimensions include $L^1$ penalization \citep{Lee2019}, hierarchical Bayesian modeling \citep{Li2013}, and generalized boosted models \citep{mccaffrey2004propensity}.
In addition, \citet{dong2020subgroup} propose a stochastic search algorithm to estimate a similar model when the number of subgroups is large, and \citet{Li2017} and \citet{Yang2020_overlap} propose \emph{overlap weights}, which upweight regions of greater overlap. 
% We explore overlap weights further in Section \ref{sec:results}.

Under suitable assumptions and conditions, methods utilizing the estimated propensity score will converge to the true ATT asymptotically.
However, in high dimensional settings with a moderate number of subgroups these methods can often fail to achieve good covariate balance in the sample of interest; as we show in Section \ref{sec:diagnostic}, these methods fail to balance covariates in the UC Berkeley LOR study.
The key issue is that traditional IPW methods focus on estimating the propensity score itself (i.e., the conditional probability of treatment) rather than finding weights that achieve good in-sample covariate balance.

\paragraph{Balancing weights.}

Unlike traditional IPW, balancing weights estimators instead find weights that directly target in-sample balance. One example is the Stable Balancing Weights (SBW) proposal from \citet{Zubizarreta2015}, which finds the minimum variance weights that achieve a user-defined level of covariate balance in $\phi(X_i) \in \R^p$:
\begin{equation}
    \label{eq:sbw}
    \begin{aligned}
    \min_{\gamma} \;\;\;\; & \|\gamma\|_2^2\\
    \text{subject to} \;\;\;\; & \max_j \left| \frac{1}{n_1} \sum_{W_i = 1}\phi_j(X_i) - \frac{1}{n_1} \sum_{W_i = 0}\gamma_i \phi_j(X_i)\right| \leq \delta,
    \end{aligned}
\end{equation}
for weights $\gamma$, typically constrained to the simplex, and for allowable covariate imbalance $\delta$.
These methods have a long history in calibrated survey weighting \citep[see, e.g.][]{Deming1940,Deville1993}, and have recently been extensively studied in the observational study context \citep[e.g.][]{Hainmueller2011, Zubizarreta2015,Athey2018a,Hirshberg2019, hazlett2018kernel}. They have also been shown to estimate the propensity score with a loss function designed to achieve good balance \citep{Zhao2016a,Wang2019,Chattopadhyay2019}. 

While balancing weights achieve better balance than the traditional IPW methods above, we must take special care to use them appropriately when estimating subgroup treatment effects. As we will show in Section \ref{sec:diagnostic}, designing balancing weights estimators without explicitly incorporating the subgroup structure also fails to balance covariates within subgroups in the LOR study. We turn to designing such weights in the next section.

%%%
%%% APPROXIMATE BALANCING WEIGHTS
%%%
% \clearpage
\section{Approximate balancing weights for treatment effect variation}
\label{sec:approx_weights}

Now we describe a specialization of balancing weights that minimizes the bias for subgroup treatment effect estimates. This approach incorporates the subgroup structure into the balance measure and optimizes for the ``local balance" within each subgroup. First we show that the error for the subgroup treatment effect estimate is bounded by the level of local imbalance within the subgroup. Furthermore, the error for estimating the overall ATT depends on both the global balance and the local balance within each subgroup. We then describe a convex optimization problem to minimize the level of imbalance within each subgroup while ensuring exact global balance in the full sample. Next, we connect the procedure to IPW with a hierarchical propensity score model, using the procedure's Lagrangian dual formulation. We conclude by describing how to augment the weighting estimate with an outcome model.

\subsection{Local balance, global balance, and estimation error}
\label{sec:local_balance}

\subsubsection{Subgroup effects}
We initially consider the role of local imbalance in estimating subgroup treatment effects. This is the subgroup-specific specialization of standard results in balancing weights; see \citet{benmichael_balancing_review} for a recent review.
We will compare the estimate $\hat{\mu}_{0g}$ to $\tilde{\mu}_{0g} \equiv \frac{1}{n_{1g}}\sum_{G_i = g}W_i m_0(X_i, g)$, our best approximation to $\mu_{0g}$ if we knew the true prognostic score.
Defining the residual $\varepsilon_i = Y_i - m_0(X_i, G_i)$, the error is 
\begin{equation}
    \label{eq:error_general}
    \hat{\mu}_{0g} - \tilde{\mu}_{0g} = \underbrace{\frac{1}{n_{1g}}\sum_{G_i = g} \hat{\gamma}_i (1-W_i) m_0(X_i, g) - \frac{1}{n_{1g}}\sum_{G_i = g} W_i m_0(X_i, g)}_{\text{bias}_g} + \underbrace{\frac{1}{n_{1g}}\sum_{G_i = g}(1-W_i) \hat{\gamma}_i\varepsilon_i}_\text{noise} .
\end{equation}
\noindent Since the weights $\hat{\gamma}$ are \emph{design-based}, they will be independent of the outcomes, and the noise term will be mean-zero and have variance proportional to the sum of the squared weights $\frac{1}{n_{1g}^2}\sum_{G_i = g}(1-W_i) \hat{\gamma}_i^2$.\footnote{In the general case with heteroskedastic errors, the variance of the noise term is $\frac{1}{n_{1g}^2}\sum_{G_i = g}\hat{\gamma}_i^2 \Var(\varepsilon_i) \leq  \max_i \{\Var(\varepsilon_i) \}\frac{1}{n_{1g}^2}\sum_{G_i = g}\hat{\gamma}_i^2$.}
At the same time, the conditional bias term, $\text{bias}_g$, depends on the imbalance in the true prognostic score $m_0(X_i, G_i)$. The idea is to bound this imbalance by the worst-case imbalance in all functions $m$ in a model class $\calM$. 
While the setup is general,\footnote{See \citet{Wang2019} for the case where the prognostic score can only be approximated by a linear function; see \citet{hazlett2018kernel} for a kernel representation and \citet{Hirshberg2019} for a general nonparametric treatment.} we describe the approach assuming that the prognostic score within each subgroup is a linear function of transformed covariates $\phi(X_i) \in \R^p$ with $L^2$-bounded coefficients; i.e., $\calM = \{m_0(x, g) = \eta_g \cdot \phi(x) \mid \|\eta_g\|_2 \leq C\}$. We can then bound the bias by the level of \emph{local imbalance} within the subgroup via the Cauchy-Schwarz inequality:
\begin{equation}
    \label{eq:bias_bound}
    \left|\text{bias}_g\right| \leq C \underbrace{\left\| \frac{1}{n_{1g}}\sum_{G_i = g} \hat{\gamma}_i (1-W_i)\phi(X_i) - \frac{1}{n_{1g}}\sum_{G_i = g}W_i \phi(X_i)\right\|_2}_{\text{local imbalance}}.
\end{equation}

\noindent  Based on Equation \eqref{eq:bias_bound}, we could control local bias solely by controlling local imbalance. This approach would be reasonable if we were solely interested in subgroup impacts. In practice, however, we are also interested in the overall effect, as well as in aggregated subgroup effects.

\subsubsection{Overall treatment effect}
% such as the impact for all URM applicants, not just the specific URM $\times$ AI stratum. 
We can estimate aggregated effects by taking a weighted average of the subgroup-specific estimates, e.g. we estimate $\mu_{0}$ as $\hat{\mu}_0 = \sum_{g=1}^K \frac{n_{1g}}{n_1}\hat{\mu}_{0g} = \frac{1}{n_1}\sum_{W_i = 0} n_{1G_i} \hat{\gamma}_i Y_i$.
The imbalance within each subgroup continues to play a key role in estimating this overall treatment effect, alongside global balance. To see this, we again compare to our best estimate if we knew the prognostic score, $\tilde{\mu}_0 = \frac{1}{n_1}\sum_{g=1}^K n_{1g} \tilde{\mu}_{0g}$, and see that the local imbalance plays a part. The error is
\begin{equation}
    \label{eq:error_global}
    \begin{aligned}
        \hat{\mu}_0 - \tilde{\mu}_0  & = \bar{\eta} \cdot \left(\frac{1}{n_1} \sum_{i=1}^nn_{1G_i} \hat{\gamma}_i (1-W_i)\phi(X_i) - \frac{1}{n_1}\sum_{i=1}^n W_i \phi(X_i)\right) \;\; + \\
        & \quad\quad \frac{1}{n_1}\sum_{g=1}^k n_{1g} \left(\eta_g - \bar{\eta}\right) \cdot \left(\sum_{G_i = g} \hat{\gamma}_i(1-W_i) \phi(X_i) - \frac{1}{n_{1g}}\sum_{G_i=g} W_i \phi(X_i)\right) \;\; +\\
        & \quad\quad \frac{1}{n_1}\sum_{i=1}^n \hat{\gamma}_i (1-W_i)\varepsilon_i,
    \end{aligned}
\end{equation}
where $\bar{\eta} \equiv \frac{1}{K}\sum_{g=1}^K \eta_g$ is the average of the model parameters across all subgroups. Again using Cauchy-Schwarz we see that the overall bias is controlled by the \emph{local imbalance} within each subgroup as well as the \emph{global balance} across subgroups:
\begin{equation}
    \label{eq:bias_bound_global}
    \begin{aligned}
        \left|\text{bias}\right| & \leq \|\bar{\eta}\|_2 \underbrace{\left\|\frac{1}{n_1} \sum_{i=1}^nn_{1G_i} \hat{\gamma}_i (1-W_i)\phi(X_i) - \frac{1}{n_1}\sum_{i=1}^n W_i \phi(X_i)\right\|_2}_{\text{global balance}} + \\
        & \quad\quad \sum_{g=1}^G \frac{n_{1g}}{n_1} \|\eta_g - \bar{\eta}\|_2  \underbrace{\left\|\sum_{G_i = g} \hat{\gamma}_i(1-W_i) \phi(X_i) - \frac{1}{n_{1g}}\sum_{G_i=g} W_i \phi(X_i)\right\|_2}_{\text{local balance}}.
    \end{aligned}
\end{equation}
In general, we will want to achieve \emph{both} good local balance within each subgroup and good global balance across subgroups.
Ignoring local balance can incur bias by ignoring heterogeneity in the outcome model across subgroups, while ignoring global balance leaves potential bias reduction on the table.
Equation \eqref{eq:bias_bound_global} shows that the relative importance of local and global balance for estimating the overall ATT is controlled by the level of similarity in the outcome process across groups. In the extreme case where the outcome process does not vary across groups --- i.e., $\eta_g = \bar{\eta}$ for all $g$ --- then controlling the global balance is sufficient to control the bias. In the other extreme where the outcome model varies substantially across subgroups --- e.g., $\|\eta_g - \bar{\eta}\|_2$ is large for all $g$ --- we will primarily seek to control the local imbalance within each subgroup in order to control the bias for the ATT. Typically, we expect that interaction terms are weaker than ``main effects,'' i.e., $\|\eta_g - \bar{\eta}\|_2 < \|\bar{\eta}\|_2$ \citep[see][]{cox1984interaction, feller2015hierarchical}. As a result, our goal is to find weights that prioritize global balance while still achieving good local balance.

\subsection{Optimizing for both local and global balance}
\label{sec:opt_problem}
We now describe a convex optimization procedure to find weights that optimize for local balance while ensuring exact global balance across the sample. The idea is to stratify across subgroups and find approximate balancing weights within each stratum, while still constraining the overall level of balance. 
% In our setting, we stratify on first reader score, URM status, the coarsened AI measure, and the college that the applicant is applying to; see Section \ref{sec:results}.
To do this, we find weights $\hat{\gamma}$ that solve the following optimization problem:
\begin{equation}
    \label{eq:primal}
    \begin{aligned}
        \min_{\gamma} \;\;\;\;\;\;\; & \sum_{g=1}^K \left\|\sum_{G_i = g, W_i = 0} \gamma_i \phi(X_i) - \sum_{G_i = g, W_i = 1} \phi(X_i)\right\|_2^2 \;\; + \;\; \frac{\lambda_g}{2}\sum_{G_i=G,W_i=0} \gamma_i^2\\[1.2em]
        \text{subject to } & \sum_{W_i = 0} \gamma_i \phi(X_i) = \sum_{W_i = 1}\phi(X_i)\\[1.2em]
        & \sum_{G_i = G, W_i = 0} \gamma_i = n_{1g}\\[1.2em]
        & \gamma_i  \geq 0 \;\;\;\;\; \forall i=1,\ldots,n
    \end{aligned}
\end{equation}

The optimization problem \eqref{eq:primal} has several key components. First, following Equation \eqref{eq:bias_bound} we try to find weights that minimize the local imbalance for each stratum defined by $G$; this is a proxy for the stratum-specific bias. 
We also constrain the weights to \emph{exactly balance} the covariates globally over the entire sample. 
Equivalently, this finds weights that achieve exact balance marginally on the covariates $\phi(X_i)$ and only approximate balance for the interaction terms $\phi(X_i) \times \mathbbm{1}_{G_i}$, placing greater priority on main effects than interaction terms.
Taken together, this ensures that we are minimizing the overall bias as well as the bias within each stratum.
In principle, weights that exactly balance the covariates within each stratum would also yield exact balance globally. Typically, however, the sample sizes are too small to achieve exact balance within each stratum, and so this combined approach at least guarantees global balance.\footnote{This constraint induces a dependence across the strata, so that the optimization problem does not decompose into $J$ sub-problems.}

From Equation \eqref{eq:bias_bound_global}, we can see that if there is a limited amount of heterogeneity in the baseline outcome process across groups, the global exact balance constraint will limit the estimation error when estimating the ATT, even if local balance is relatively poor. 
In principle, incorporating the global balance constraint could lead to worse local balance. However, we show in both the simulations in Section \ref{sec:sims} and the analysis of the LOR pilot study in Section \ref{sec:results} that the global constraint leads to negligible changes in the level of local balance and the performance of the subgroup estimators, but can lead to large improvements in the global balance and the performance of the overall estimate. 
Thus, there seems to be little downside in terms of subgroup estimates from an approach that controls both local and global imbalance --- but large gains for overall estimates.
Note that while we choose to enforce exact global balance, we could also limit to \emph{approximate} global balance, with the relative importance of local and global balance controlled by an additional hyperparameter set by the analyst.

Second, we include an $L^2$ regularization term that penalizes the sum of the squared weights in the stratum; from Equation \eqref{eq:error_general}, we see that this is a proxy for the variance of the weighting estimator. For each stratum, the optimization problem includes a hyper-parameter $\lambda_g$ that negotiates the bias-variance tradeoff within that stratum. 
When $\lambda_g$ is small, the optimization prioritizes minimizing the bias through the local imbalance, and when $\lambda$ is large it prioritizes minimizing the variance through the sum of the squared weights. As a heuristic, we 
limit the number of hyperparameters by choosing
$\lambda_g = \frac{\lambda}{n_g}$ for a common choice of $\lambda$. For larger strata where better balance is possible, this heuristic will prioritize balance --- and thus bias --- over variance; for smaller strata, by contrast, this will prioritize lower variance. We discuss selecting $\lambda$ in the letters of recommendation study in Section \ref{sec:diagnostic}.

Next, we incorporate two additional constraints on the weights.
We include a fine balance constraint \citep{Rosenbaum2007_finebalance}: within each stratum the weights sum up to the number of treated units in that stratum, $n_{1g}$. 
Since each stratum maps to only one subgroup, this guarantees that the weights sum to the number of treated units in each subgroup. 
We also restrict the weights to be non-negative, which stops the estimates from extrapolating outside of the support of the control units \citep{king2006dangers}. Together, these induce several stability properties, including that the estimates are sample bounded.

In addition, we could extend the optimization problem in Equation \eqref{eq:primal} to balance intermediate levels between global balance and local balance. % within the strata.
Incorporating additional balance constraints for each intermediate level is unwieldy in practice due to the proliferation of hyperparameters.
Instead, we can expand $\phi(x)$ to include additional interaction terms between covariates and levels of the hierarchy. We discuss this this choice in the letters of recommendation study in Section \ref{sec:results}.

Finally, we compute the variance of our estimator conditioned on the design $(X_1,Z_1,W_1),\ldots,$ $(X_n,Z_n,W_n)$ or, equivalently, conditioned on the weights. The conditional variance is
\begin{equation}
  \label{eq:var_mu0g}
  \Var(\hat{\mu}_{0g} \mid \hat{\gamma}) = \frac{1}{n_{1g}^2}\sum_{G_i = g} (1 - W_i)\hat{\gamma}_i^2 \Var(Y_i).
\end{equation}
Using the $i$\super{th} residual to estimate $\Var(Y_i)$ yields the empirical sandwich estimator for the treatment effect
\begin{equation}
    \label{eq:sandwich}
    \widehat{\Var}(\hat{\mu}_{1g} - \hat{\mu}_{0g} \mid \hat{\gamma}) = \frac{1}{n_{1g}^2} \sum_{G_i = g} W_i (Y_i - \hat{\mu}_{1g})^2 +  \frac{1}{n_{1g}^2}\sum_{G_i = g}(1-W_i) \hat{\gamma}_i^2 (Y_i - \hat{\mu}_{0g})^2,
\end{equation}
where, as above, $\hat{\mu}_{1g}$ is the average outcome for applicants in subgroup $g$ who submit an LOR.
This is the fixed-design Huber-White heteroskedastic robust standard error for the weighted average. See \citet{Hirshberg2019} for discussion on asymptotic normality and semi-parametric efficiency for estimators of this form.

\subsection{Dual relation to partially pooled propensity score estimation}
\label{sec:dual}
Thus far, we have motivated the approximate balancing weights approach by appealing to the connection between local bias and local balance. We now draw on recent connections between approximate balancing weights and (calibrated) propensity score estimation through the Lagrangian dual problem. The weights that solve optimization problem \eqref{eq:primal} correspond 
to estimating the inverse propensity weights with a (truncated) linear odds function with the stratum $G$ interacted with the covariates $\phi(X)$,\footnote{The truncation arises from constraining weights to be non-negative, and the linear odds form arises from penalizing the $L^2$ norm of the weights. We can consider other penalties that will lead to different forms. See \citet{benmichael_balancing_review} for a review of the different choices.}
\begin{equation}
    \label{eq:linear_odds}
    \frac{P(W = 1 \mid X = x, G = g)}{1 - P(W = 1 \mid X = x, G = g)} = \left[\alpha_g + \beta_g \cdot \phi(x)\right]_+,
\end{equation}
where the coefficients $\beta_g$ are \emph{partially pooled} towards a global model.

To show this, we first derive the Lagrangian dual. For each stratum $g$, the sum-to-$n_{1g}$ constraint induces a dual variable $\alpha_g \in \R$, and the local balance measure induces a dual variable $\beta_g \in \R^p$. 
These dual variables are part of the \emph{balancing loss function} for stratum $z$:
\begin{equation}
    \label{eq:dual_loss}
    \calL_g(\alpha_g, \beta_g) \equiv \sum_{W_i = 0, G_i = g} \left[\alpha_g + \beta_g \cdot \phi(X_i)\right]_+^2 - \sum_{W_i = 1, G_i = g} \left(\alpha_g + \beta_g\cdot \phi(X_i)\right),
\end{equation}
where $[x]_+ = \max\{0, x\}$. With this definition we can now state the Lagrangian dual.
\begin{proposition}
    \label{prop:dual}
    With $\lambda_g > 0$, if a feasible solution to \eqref{eq:primal} exists, the Lagrangian dual is 
    \begin{equation}
        \label{eq:dual}
        \min_{\alpha, \beta_1,\ldots,\beta_J, \mu_\beta} \sum_{g = 1}^K\underbrace{\calL_g(\alpha_g, \beta_g)}_{\text{balancing loss}} \;\;+\;\; \underbrace{\sum_{z=1}^J \frac{\lambda_g}{2}\|\beta_g - \mu_\beta\|_2^2}_{\text{shrinkage to global variable}}.
    \end{equation}
    If $\hat{\alpha}, \hat{\beta}_1,\ldots,\hat{\beta}_J$ are the solutions to the dual problem, then the solution to the primal problem \eqref{eq:primal} is
    \begin{equation}
        \label{eq:primal_sol}
        \hat{\gamma}_i = \left[\hat{\alpha}_{Z_i} +  \hat{\beta}_{Z_i} \cdot \phi(X_i)\right]_+ .
    \end{equation}
\end{proposition}
The Lagrangian dual formulation sheds additional light on the approximate balancing weights estimator. First, we apply results on the connection between approximate balancing weights and propensity score estimation 
\citep[e.g.,][]{Zhao2016a, Wang2019, Hirshberg2019_amle, Chattopadhyay2019}.
We see that this approach estimates propensity scores of the form \eqref{eq:linear_odds}, which corresponds to a fully interacted propensity score model where the coefficients on observed covariates vary across strata.
Recall that we find \emph{approximate} balancing weights for each stratum because the number of units per stratum might be relatively small; therefore we should not expect to be able to estimate this fully interacted propensity score well.

The dual problem in Equation \eqref{eq:dual} also includes a global dual variable $\mu_\beta$ induced by the global balance constraint in the primal problem \eqref{eq:primal}. 
Because we enforce \emph{exact} global balance, this global model is not regularized.
However, by penalizing the deviations between the stratum-specific variables and the global variables via the $L^2$ norm, $\|\beta_g - \mu_\beta\|_2^2$, the dual problem \emph{partially pools} the stratum-specific parameters towards a global model. 
Thus, we see that the approximate balancing weights problem in Equation \eqref{eq:primal} corresponds to
a hierarchical propensity score model \citep[see, e.g.][]{Li2013},
as in Section \ref{sec:related}, fit with a loss function designed to induce covariate balance.

Excluding the global constraint removes the global dual variable $\mu_\beta$, and the dual problem shrinks the stratum-specific variables $\beta_g$ towards zero without any pooling. In contrast, ignoring the local balance measure by setting $\lambda_g \to \infty$ constrains the stratum-specific variables $\beta_g$ to all be \emph{equal} to the global variable $\mu_\beta$, resulting in a fully pooled estimator.
For intermediate values, 
$\lambda_g$
controls the level of partial pooling. When $\lambda_g$ is large the dual parameters are heavily pooled towards the global model, and when $\lambda_g$ is small the level of pooling is reduced. By setting $\lambda_g = \frac{\lambda}{n_g}$ as above, larger strata will be pooled less than smaller strata.\footnote{It is also possible to have covariate-specific shrinkage by measuring imbalance in the primal problem \eqref{eq:primal} with a \emph{weighted} $L^2$ norm, leading to an additional $p$ hyper-parameters. We leave exploring this extension and hyper-parameter selection methods to future work.}

\subsection{Augmentation with an outcome estimator}
\label{sec:augment}

The balancing weights we obtain via the methods above may not achieve perfect balance, leaving the potential for bias.
We can augment the balancing weights estimator with an outcome model, following similar proposals in a variety of settings \citep[see, e.g.][]{Athey2018a, Hirshberg2019_amle,benmichael2019_augsynth}. Analogous to bias correction for matching \citep{rubin1973bias} or model-assisted estimation in survey sampling \citep{sarndal2003model}, the essential idea is to adjust the weighting estimator using an estimate of the bias.
Specifically, we can estimate the prognostic score $m_0(x, g)$ with a working model $\hat{m}_0(x, g)$, e.g., with a flexible regression model. An estimate of the bias in group $g$ is then:
\begin{equation}
    \label{eq:bias_est}
    \widehat{\text{bias}}_g = \frac{1}{n_{1g}}\sum_{W_i = 1, G_i = g} \hat{m}_0(X_i, g) - \frac{1}{n_{1g}} \sum_{W_i = 0, G_i = g} \hat{\gamma}_i \hat{m}_0(X_i, g).
\end{equation}
This is the bias due to imbalance in estimated prognostic score in group $g$ \emph{after} weighting.
With this estimate of the bias, we can explicitly bias-correct our weighting estimator, estimating $\mu_{0g}$ as 
\begin{equation}
    \label{eq:mu0g_hat_aug}
    \begin{aligned}
        \hat{\mu}_{0g}^{\text{aug}} &\equiv \hat{\mu}_{0g} \;\;+\;\; \widehat{\text{bias}}_g \\
        &= \frac{1}{n_{1g}}\sum_{W_i = 0, G_i = g} \hat{\gamma}_iY_i \;\;+\;\; \left[\frac{1}{n_{1g}}\sum_{W_i = 1, G_i = g} \hat{m}_0(X_i, g) - \frac{1}{n_{1g}} \sum_{W_i = 0, G_i = g} \hat{\gamma}_i \hat{m}_0(X_i, g)\right].
    \end{aligned}
\end{equation}
\noindent Thus, if the balancing weights fail to achieve good covariate balance in a given subgroup, the working outcome model, $ \hat{m}_0(X_i, g)$, can further adjust for any differences. See \citet{benmichael_balancing_review} for further discussion.

%%%
%%% SIM STUDY
%%%
% \clearpage
\section{Simulation study}
\label{sec:sims}

Before estimating the differential impacts of letters of recommendation, we first present simulations assessing the performance of our proposed approach versus traditional inverse propensity score weights fit via regularized logistic regression as well as outcome modelling with machine learning approaches.
To better reflect real-world data, we generate correlated covariates and include binary and skewed covariates.
For each simulation run, 
with $d = 50$ covariates,
we begin with a diagonal covariance matrix $\Sigma$ where $\Sigma_{jj} = \frac{(d - j + 1)^5}{d^5}$ and sample a random orthogonal $d\times d$ matrix $Q$ to create a new covariance matrix $\tilde{\Sigma} = Q\Sigma$ with substantial correlation. For $n = 10,000$ units, we draw covariates from a multivariate normal distribution $X_{i} \overset{iid}{\sim} N(0, \tilde{\Sigma})$.
We then transform some of these covariates. For $j =1,11,21,32,41$ we dichotomize the variable and define $\tilde{X}_{ij} = \bbone\{X_{ij} \geq q_{.8}(X_{\cdot j})\}$, where $q_{.8}(X_{\cdot j})$ is the 80th percentile of $X_j$ among the $n$ units.
For $j = 2,7,12,...,47$ we create a skewed covariate $\tilde{X}_{ij} = \exp(X_{ij})$.
To match our study, we create discrete subgroup indicators from the continuous variable $X_{id}$. To do this, we create a grid over $X_{id}$ with grid size $\frac{n}{G}$, and sample $G - 1$ points from this grid. We then create subgroup indicators $G_i$ 
by binning $X_{id}$ according to the $G-1$ points.
We consider $G \in \{10, 50\}$ groups.

With these covariates, we generate treatment assignment and outcomes. We use a separate logistic propensity score model for each group following Equation \eqref{eq:ipw_interact},\footnote{The logistic specification differs from the truncated linear odds in Equation \ref{eq:linear_odds}. If the transformed covariates $\phi(X_i)$ include a flexible basis expansion, the particular form of the link function will be less important.}
\begin{equation}
    \label{eq:sim_pscore}
    \text{logit} \;e(X_i, G_i) = \alpha_{G_i} + (\mu_\beta + U_g^\beta \odot B_g^\beta)\cdot X_i,
\end{equation}
and also use a separate linear outcome model for each group,
\begin{equation}
    \label{eq:sim_outcome}
    Y_i(0) = \eta_{0G_i} + (\mu_\eta + U_g^\eta \odot B_g^\eta) \cdot X_i + \varepsilon_i,
\end{equation}
where $\varepsilon_i \sim N(0,1)$ and $\odot$ denotes element-wise multiplication. 
We draw the fixed effects and varying slopes for each group according to a hierarchical model with sparsity. We draw the fixed effects as $\alpha_g \overset{\text{iid}}{\sim} N(0,1)$ and $\eta_{0g} \overset{\text{iid}}{\sim} N(0,1)$. For the slopes, we first start with a mean slope vector $\mu_\beta, \mu_\eta \in \{-\frac{3}{\sqrt{d}},\frac{3}{\sqrt{d}}\}^K$, where each element is chosen independently with uniform probability. Then we draw isotropic multivariate normal random variables $U_g^\beta, U_g^\eta \overset{\text{iid}}{\sim} MVN(0, I_d)$. Finally, we draw a set of $d$ binary variables $B_{gj}^\beta, B_{gj}^\eta$ that are Bernoulli with probability $p = 0.25$. The slope is then constructed as a set of sparse deviations from the mean vector, which is $\mu_\beta + U_g^\beta \odot B_g^\beta$ for the propensity score and $\mu_\eta + U_g^\eta \odot B_g^\eta$ for the outcome model.

To incorporate the possibility that treatment effects vary with additional covariates that are not the focus of our analysis, we generate the treatment effect for unit $i$ as $\tau_i = X_{id} - X_{i3} + 0.3 X_{id} X_{i3}$ and set the treated potential outcome as $Y_i(1) = Y_i(0) + \tau_{i} W_i$. Note that the effect varies with the underlying continuous variable $X_{id}$ that we use to form groups, as well as the additional variable $X_{i3}$.
The true ATT for group $g$ in simulation $j$ is thus $\tau_{gj} = \frac{1}{n_{1g}}\sum_{G_i = g} W_i (Y_i(1) - Y_i(0))$, and the overall ATT is $\tau_j = \frac{1}{n_1}\sum_{i=1}^n W_i (Y_i(1) - Y_i(0))$.

\begin{figure}[tb]
  \centering \includegraphics[width=\maxwidth]{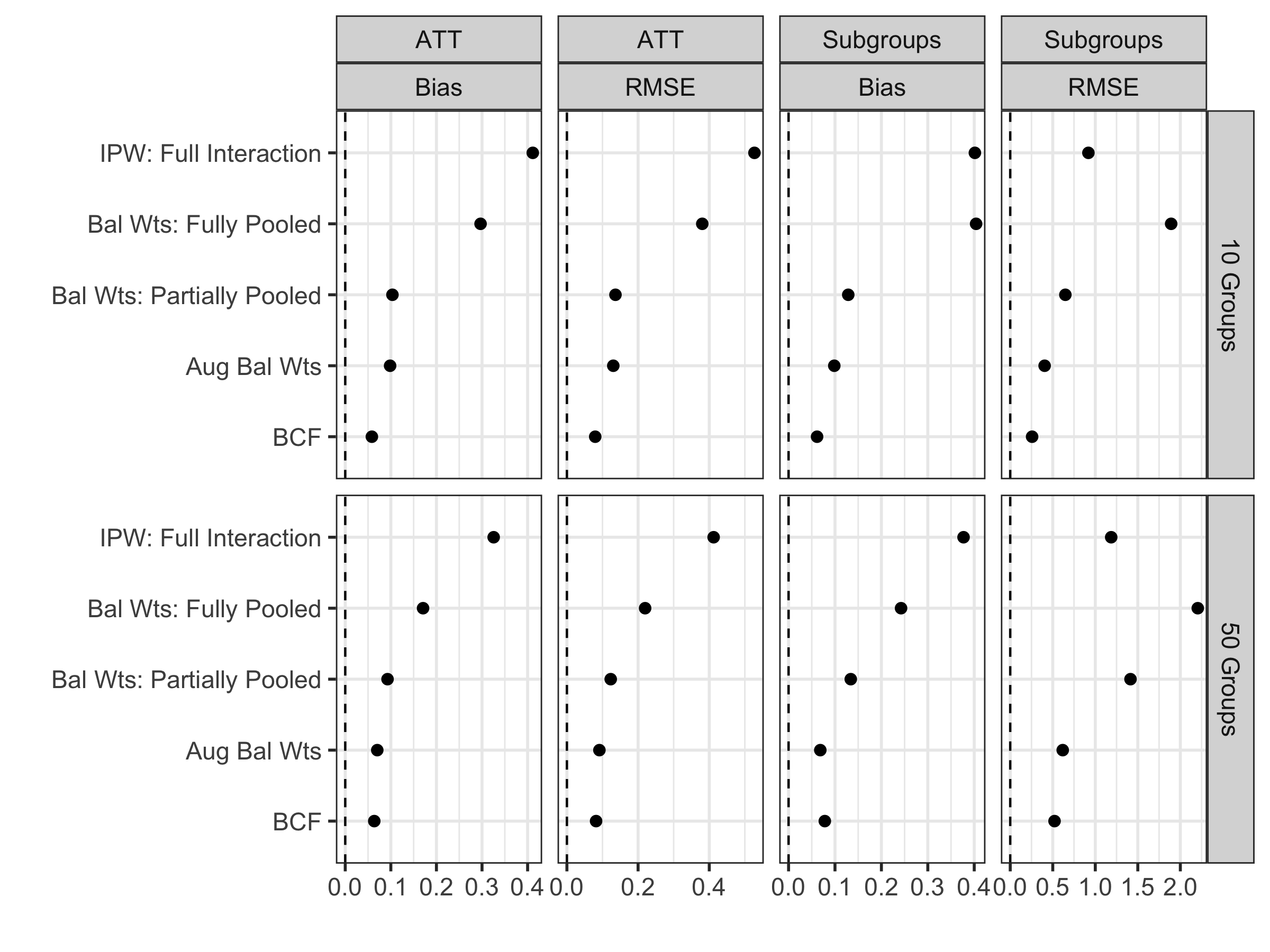}
\caption{Performance of approximate balancing weights, traditional IPW with logistic regression, and outcome modelling for estimating subgroup treatment effects.}
\label{fig:main_sim_results}
\end{figure}

For $j=1,\ldots,m$ with $m = 500$ Monte Carlo samples, we estimate the treatment effects for group $g$, $\hat{\tau}_{gj}$, and the overall ATT, $\hat{\tau}_j$, and compute a variety of metrics. Following the metrics studied by \citet{dong2020subgroup}, for subgroup treatment effects we compute (a) the mean absolute bias across the $G$ treatment effects, $\frac{1}{m}\sum_{j=1}^m \left|\frac{1}{g}\sum_{g=1}^G \hat{\tau}_{gj} -\tau_g \right|$,
and (b) the mean root mean square error $\sqrt{\frac{1}{mG}\sum_{j=1}^m\sum_{g=1}^G (\hat{\tau}_{gj} - \tau_g)^2}$. For the overall ATT we measure (a) the absolute bias $\left|\frac{1}{m}\sum_{j=1}^m \hat{\tau}_j - \tau_j\right|$ and (b) the root mean square error $\sqrt{\frac{1}{m}\sum_{j=1}^m (\hat{\tau}_j - \tau_j)^2}$. 

Here we focus on five estimators:
 \begin{itemize}

  \item \emph{Full interaction IPW:} traditional IPW with a fully interacted, ridge-penalized model that estimates a separate propensity score within each stratum as in Equation \eqref{eq:ipw_interact}.

  \item \emph{Fully pooled balancing weights:} approximate balancing weights that solve Equation \eqref{eq:primal}, but ignore local balance by setting $\lambda \to \infty$, thus fully pooling towards the global model. This is equivalent to stable balancing weights in Equation \eqref{eq:sbw} with an exact global balance constraint $\delta = 0$.   

  \item \emph{Partially pooled balancing weights:} approximate balancing weights that solve Equation \eqref{eq:primal}, using $G$ as the stratifying variable and prioritizing local balance by setting $\lambda_g = \frac{1}{n_{1g}}$.
  
  \item \emph{Augmented balancing weights:} augmenting the partially pooled balancing weights as in Equation \eqref{eq:mu0g_hat_aug} where $\hat{m}_0(x, g)$ is fit via ridge regression with all interactions.

  \item \emph{Bayesian Causal Forests (BCF)}: Estimating $\tau_g$ as $\frac{1}{n_{1g}} \sum_{G_i = g} W_i \hat{\tau}_i$, where $\hat{\tau}_i$ are the posterior predictive means from a Bayesian Causal Forest estimator \citep{Hahn2020}.
  
\end{itemize}
In Appendix \ref{sec:sim_appendix} we also consider various other estimators.
 For the fully interacted specification of the logistic regression in IPW and the ridge regression in the augmented balancing weights, we include a set of global parameters $\mu_\beta$ so that the slope for group $g$ is $\mu_\beta + \Delta_g$, with a squared $L^2$ penalty for each component. 
 These are correctly specified for the models above.
 We estimate the models with \texttt{glmnet} \citep{Friedman2010} with the hyperparameter chosen through 5-fold cross validation.
 
Figure \ref{fig:main_sim_results} shows the results for the overall ATT and for subgroup effects. 
In most cases, the partially pooled approximate balancing approach has much lower bias than RMSE than the logistic regression-based IPW estimator; however, the RMSEs are comparable with 50 subgroups.
Furthermore, prioritizing local balance with partial pooling yields lower bias and RMSE than ignoring local balance entirely with the fully pooled approach.
In Appendix Figure \ref{fig:plot_all_weights}, we also consider excluding the global balance constraint, and find that the constraint yields much lower bias for the ATT in some settings, with relatively little cost to the subgroup estimates.

The partially-pooled balancing weights estimator, which is transparent and design-based,
also performs nearly as well as the black-box BCF method.
Augmenting the partially-pooled weights provides some small improvements to the bias, indicating that the weights alone are able to achieve good balance in these simulations, and a larger improvement in the RMSE in the setting with many groups where the weighting estimator alone have larger variance.
In Appendix Figure \ref{fig:plot_all_weights} we also compare to ridge regression, finding similarly comparable performance.
In addition, Appendix Figure \ref{fig:sim_coverage} shows the coverage of 95\% intervals for the different approaches. 
We see that the weighting estimator, both with and without augmentation, has reasonable uncertainty quantification, with much better coverage than either of the two model-based approaches.

%%%
%%% RESULTS
%%%
% \clearpage
\section{Differential impacts of letters of recommendation}
\label{sec:results}
We now turn to estimating the differential impacts of letters of recommendation on 
admissions decisions. 
We focus on the eight subgroups defined in Table \ref{tab:subgroup_counts}, based on the interaction between URM status (2 levels) and admissibility index (4 levels).
Due to the selection mechanism described in Section \ref{sec:lor}, however, it is useful to create even more fine-grained strata and then aggregate to these eight subgroups. 
Specifically, we define $G = 41$ fine-grained strata based on URM status, AI grouping, first reader score, and college applied to.\footnote{Of the 48 possible strata, we drop 7 strata where no applicants submitted a letter of recommendation. These are non-URM applicants in both colleges in the two lowest AI strata but where the first reader assigned a ``Yes'' or ``No''. This accounts for $\sim 2\%$ of applicants. The remaining 41 strata have a wide range of sizes with a few very large strata. Min: 15, p25: 195, median: 987, p75: 1038, max: 8000.} 
While we are not necessarily interested in treatment effect heterogeneity across all 41 strata, this allows us to exactly match on key covariates and then aggregate to obtain the primary subgroup effects. 

Another key component in the analysis is the choice of transformation of the covariates $\phi(\cdot)$.
Because we have divided the applicants into many highly informative strata, we choose $\phi(\cdot)$ to include all of the raw covariates. Additionally, because of the importance of the admissibility index, we also include a natural cubic spline
for AI with knots at the sample quantiles. Finally, we include the output of the admissions model and a binary indicator for whether the predicted probability of a ``Possible'' score is greater than 50\%. 
We further prioritize local balance in the admissibility index by 
including in $\phi(x)$ the interaction between the AI, URM status, and an indicator for admissibility subgroup.
As we discuss above, this ensures local balance in the admissibility index at an intermediate level of the hierarchy between global balance and local balance. Finally, we standardize each component of $\phi(X)$ to have mean zero and variance one.
If desired, we could also consider other transformations such as a higher-order polynomial transformation, using a series of basis functions for all covariates, or computing inner products via the kernel trick to allow for an infinite dimensional basis \citep[see, e.g.][]{hazlett2018kernel,Wang2019,Hirshberg2019_amle}.

\subsection{Diagnostics: local balance checks and assessing overlap}
\label{sec:diagnostic}

\begin{figure}[tbp]
  \centering
    \begin{subfigure}[t]{0.5\textwidth}  
  {\centering \includegraphics[width=0.8\textwidth]{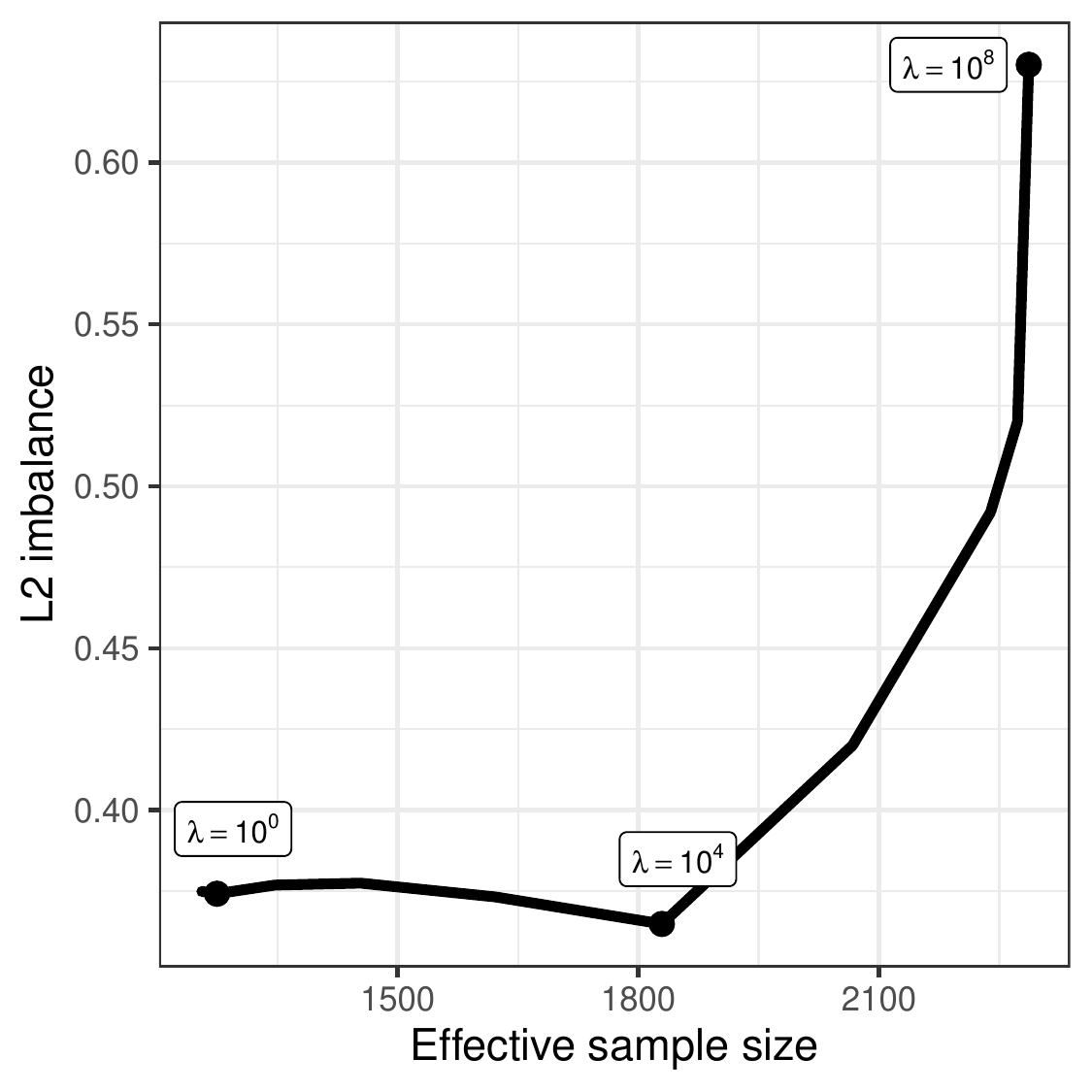} 
  }
  \caption{Imbalance vs effective sample size.\\$\lambda = 1,10^4,10^8$ noted.}
    \label{fig:lambda_plot}
    \end{subfigure}%
    ~
    \begin{subfigure}[t]{0.5\textwidth}  
    {\centering \includegraphics[width=0.8\textwidth]{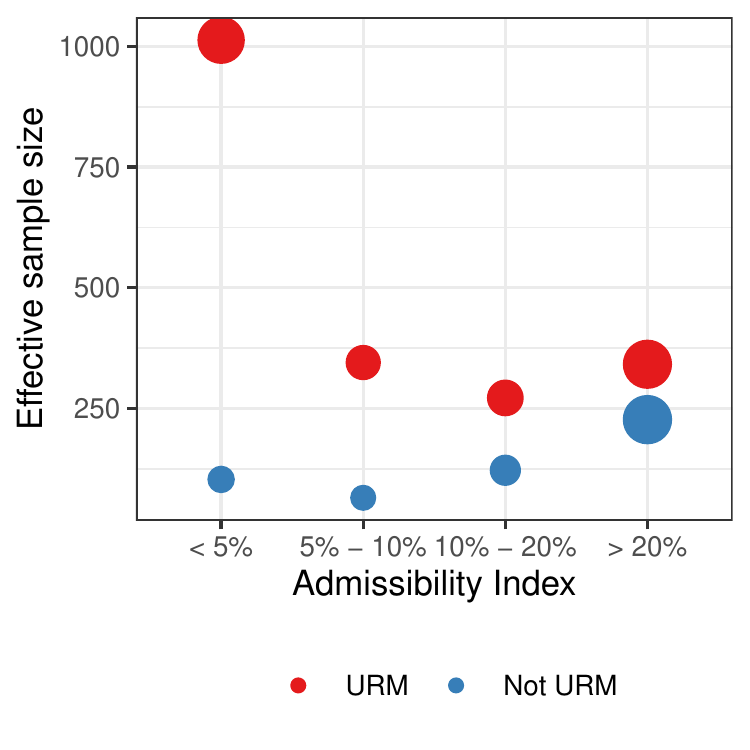} 
    }
    \caption{Effective sample sizes, area proportional to number of treated units}
      \label{fig:eff_sample_size}
      \end{subfigure}
\caption{(a) Imbalance measured as the square root of the objective in \eqref{eq:primal} plotted against the effective sample size of the overall control group. (b) Effective sample size of the control group for each subgroup, with weights solving the approximate balancing weights problem \eqref{eq:primal} with $\lambda_g = \frac{10^4}{n_g}$.} 
\label{fig:love_plot_main}
\end{figure}

In order to estimate effects, we must first choose 
values of the common hyperparameter $\lambda$ in the optimization problem \eqref{eq:primal}, where we set $\lambda_g = \frac{\lambda}{n_g}$. 
Recall that this hyperparameter negotiates the bias-variance tradeoff: small values of $\lambda$ will prioritize bias by reducing local balance while higher values will prioritize variance by increasing the effective sample size. Figure \ref{fig:lambda_plot} shows this tradeoff. We plot the square root of the local balance measure in \eqref{eq:primal} against the \emph{effective sample size} for the re-weighted control group, 
$n_1 \big/ \left(\sum_{W_i = 0}\hat{\gamma}_i^2\right)$. 
Between $\lambda = 10^0$ and $10^4$, we see that the imbalance is relatively flat while the overall effective sample size increases, after which the imbalance increases quickly with $\lambda$. We therefore select $\lambda = 10^4$ for the results we present.

Figure \ref{fig:eff_sample_size} shows the effective control group sample size for each of the primary URM and AI subgroups, scaled by the number of applicants in the group submitting LORs.
Across the board, the URM subgroups have larger effective sample sizes than the non-URM subgroups, with particularly stark differences for the lower AI subgroups. 
Furthermore, for all non-URM subgroups the effective sample size is less than 250.
Comparing to the sample sizes in Table \ref{tab:subgroup_counts}, we see that the weighting approach leads to a large design effect: many applicants who did not submit LORs are not comparable to those who did. However, lower admissibility non-URM applicants also submitted letters at lower rates.
This design effect, combined with the smaller percentage of non-URM applicants submitting LORs, means that we should expect to have greater precision in the estimates for URM applicants than non-URM applicants.

\begin{figure}[tbp]
  \centering
    \begin{subfigure}[t]{0.45\textwidth}  
  {\centering \includegraphics[width=0.8\textwidth]{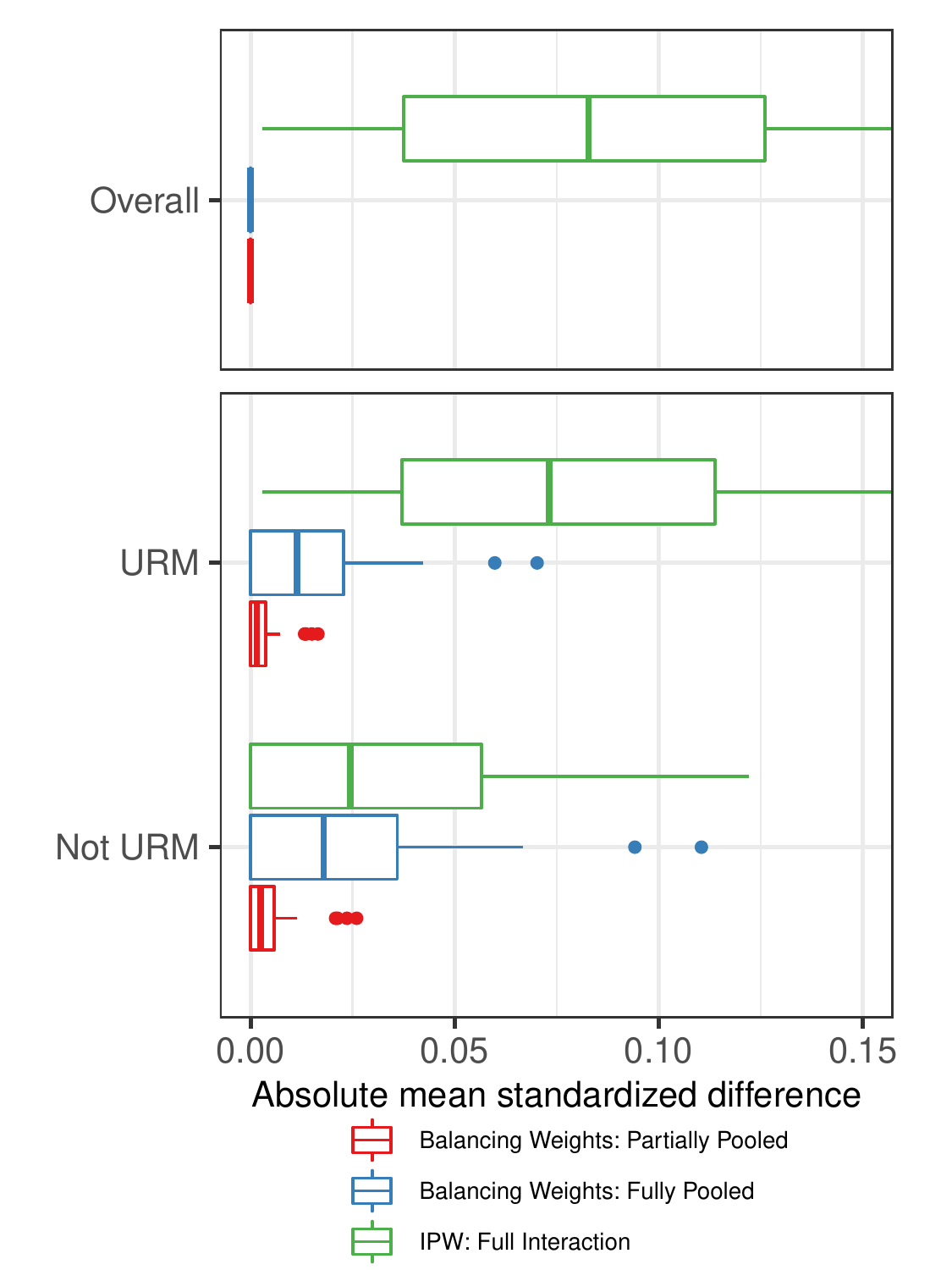} 
  }
  \caption{Overall and by URM status and AI.} 
    \label{fig:love_plot_marginal}
    \end{subfigure}%
    ~
    \begin{subfigure}[t]{0.45\textwidth}  
    {\centering \includegraphics[width=0.8\textwidth]{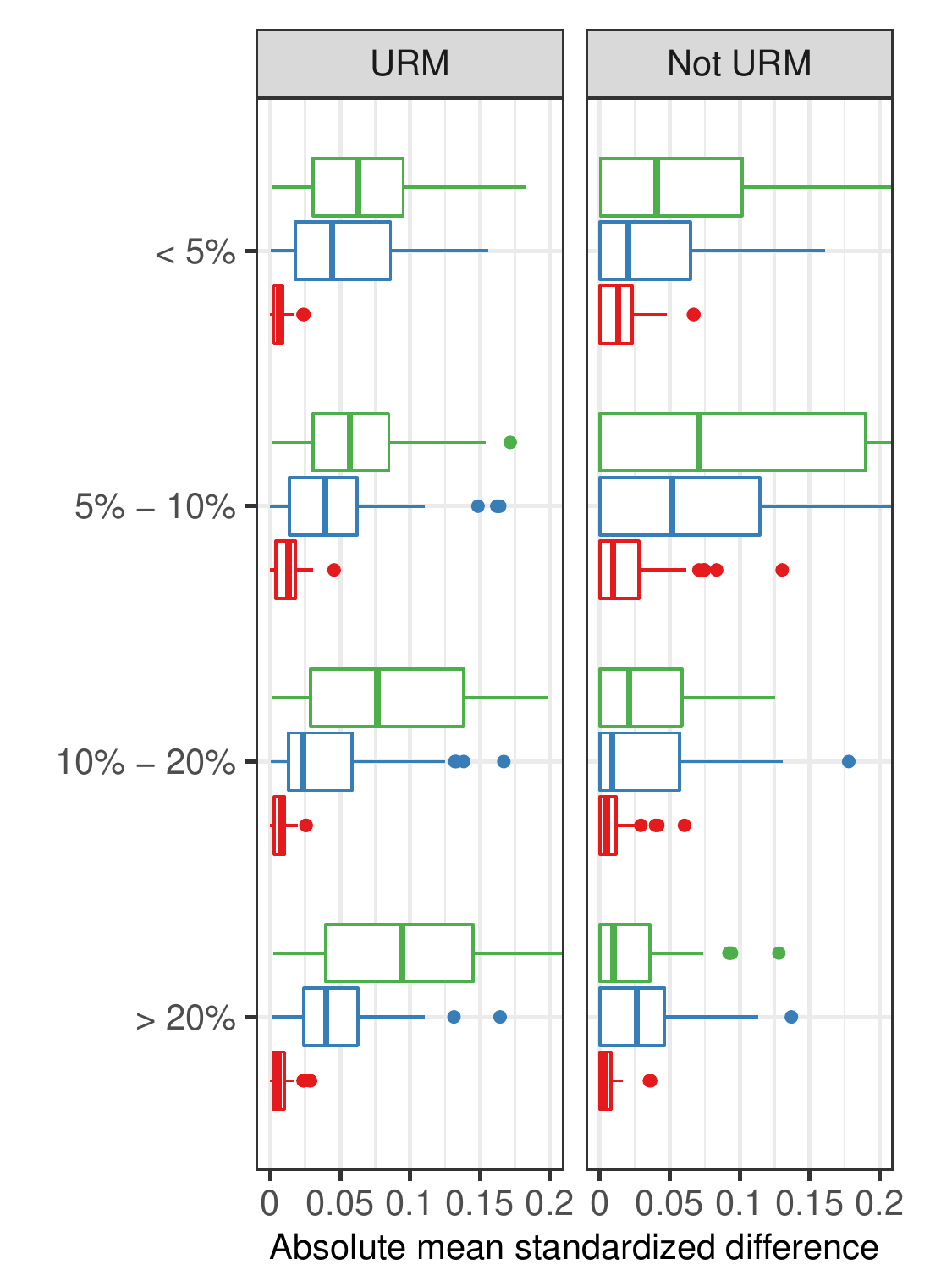} 
    }
    \caption{By URM status interacted with AI.}
      \label{fig:love_plot_interact}
      \end{subfigure}
\caption{The distribution of  imbalance in each component of $\phi(X)$
after weighting with both the partially- and fully-pooled balancing weights estimators, as well as the fully interacted IPW estimator. } 
\label{fig:love_plot_main}
\end{figure}

We now assess the level of local balance within each subgroup, following the discussion in Section \ref{sec:local_balance}.
We focus on three of the estimators described in Section \ref{sec:sims}: fully- and partially-pooled balancing weights and the full interaction IPW estimator.
Figure \ref{fig:love_plot_main} shows the distribution of the imbalance in each of the 51 (standardized) components of $\phi(X)$.
The fully interacted IPW approach has very poor balance overall, due in part to the difficulty of estimating the high-dimensional propensity score model.
As expected, both the fully- and partially-pooled balancing weights achieve perfect balance overall; 
however, only the partially pooled balancing weights achieve excellent local balance.
Appendix Figure \ref{fig:love_plot_box} shows these same metrics for the no-pooled balancing weights and fixed effects IPW estimators we discuss in Appendix \ref{sec:sim_appendix}, as well as subgroup overlap weights \citep{Yang2020_overlap}.
The partially- and no-pooled approaches have similar global and local balance overall, but the partially-pooled approach sacrifices a small amount of local balance for an improvement in global balance.
In contrast, both the fixed effects IPW and overlap weights approaches yield poor local balance.

\begin{figure}[tbp]
  \centering
    {\centering \includegraphics[width=0.95\textwidth]{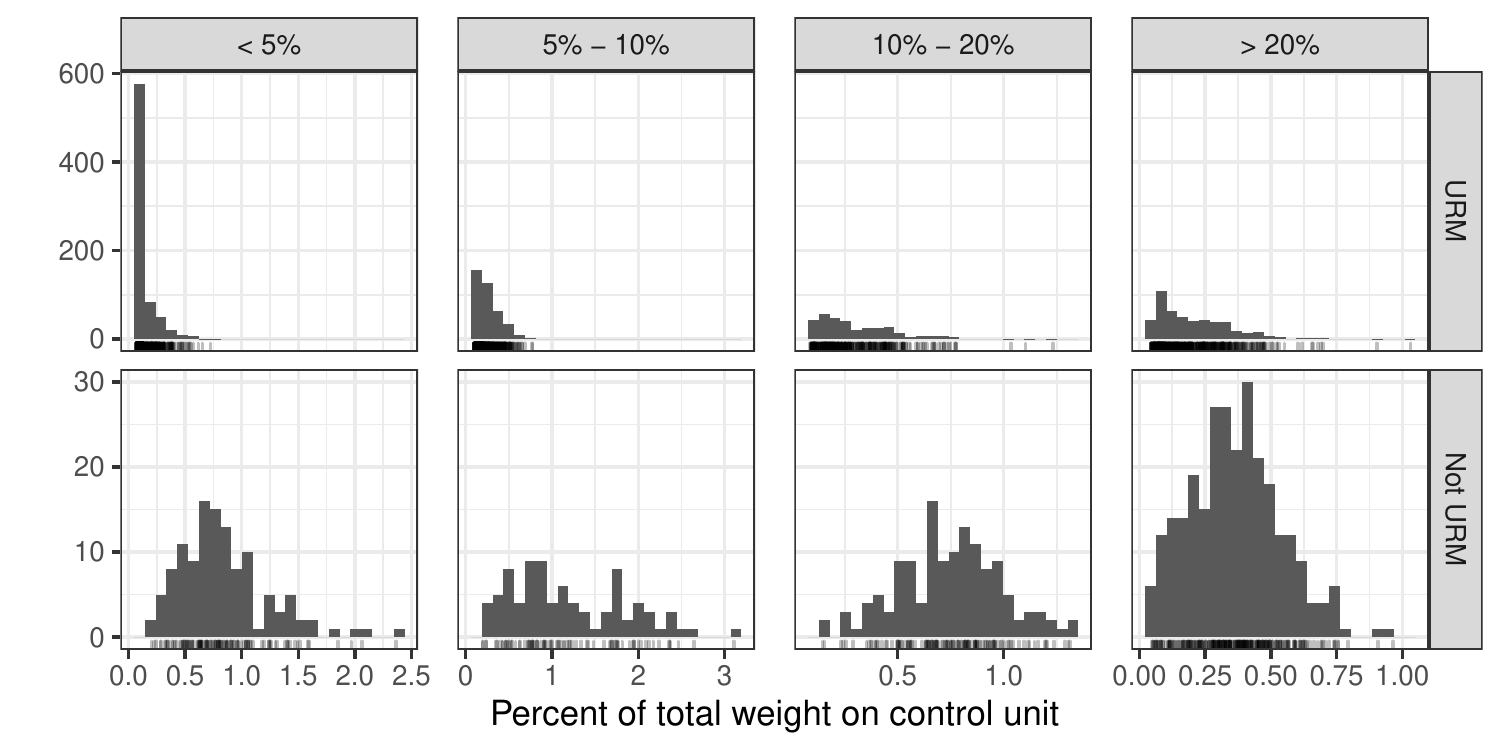}}
    \caption{Weights on control units from solving the approximate balancing weights problem \eqref{eq:primal}. \emph{Not pictured:} the 66\% of control units that receive zero weight.}
    \label{fig:weight_hist}
  \end{figure}

Finally, we assess overlap within each subgroup.
A key benefit of weighting approaches is that overlap issues manifest in the distribution of our weights $\hat{\gamma}$. 
Figure \ref{fig:weight_hist} plots the distribution of the weights over the comparison applicants by URM status and AI group, normalized by the number of treated applicants in the subgroup.
The vast majority of control units receive zero weight and are excluded from the figure. Of the 28,556 applicants who did not submit an LOR, only 9,834 (34\%) receive a weight larger than 0.001. This is indicative of a lack of ``left-sided'' overlap: very many applicants who did not submit a letter of recommendation had nearly zero odds of doing so in the pilot program. 
This is problematic for estimating the overall average treatment effect, but is less of a concern when we focus on estimating the average treatment effect on the treated.

For each AI subgroup we also see that the distribution of weights is skewed more positively for the non-URM applicants. In particular, for the lower AI, non-URM subgroups we see a non-trivial number of comparison applicants that ``count for'' over 1\% of the re-weighted sample, with a handful of outliers that count for more than 2\%. 
While large weights do not necessarily affect the validity of the estimator, large weights decrease the effective sample size, reducing the precision of our final estimates, as we see in Figure \ref{fig:eff_sample_size}.

\subsection{Treatment effect estimates}

\begin{figure}[tbp]
    \centering
      \begin{subfigure}[t]{0.45\textwidth}  
    {\centering \includegraphics[width=\textwidth]{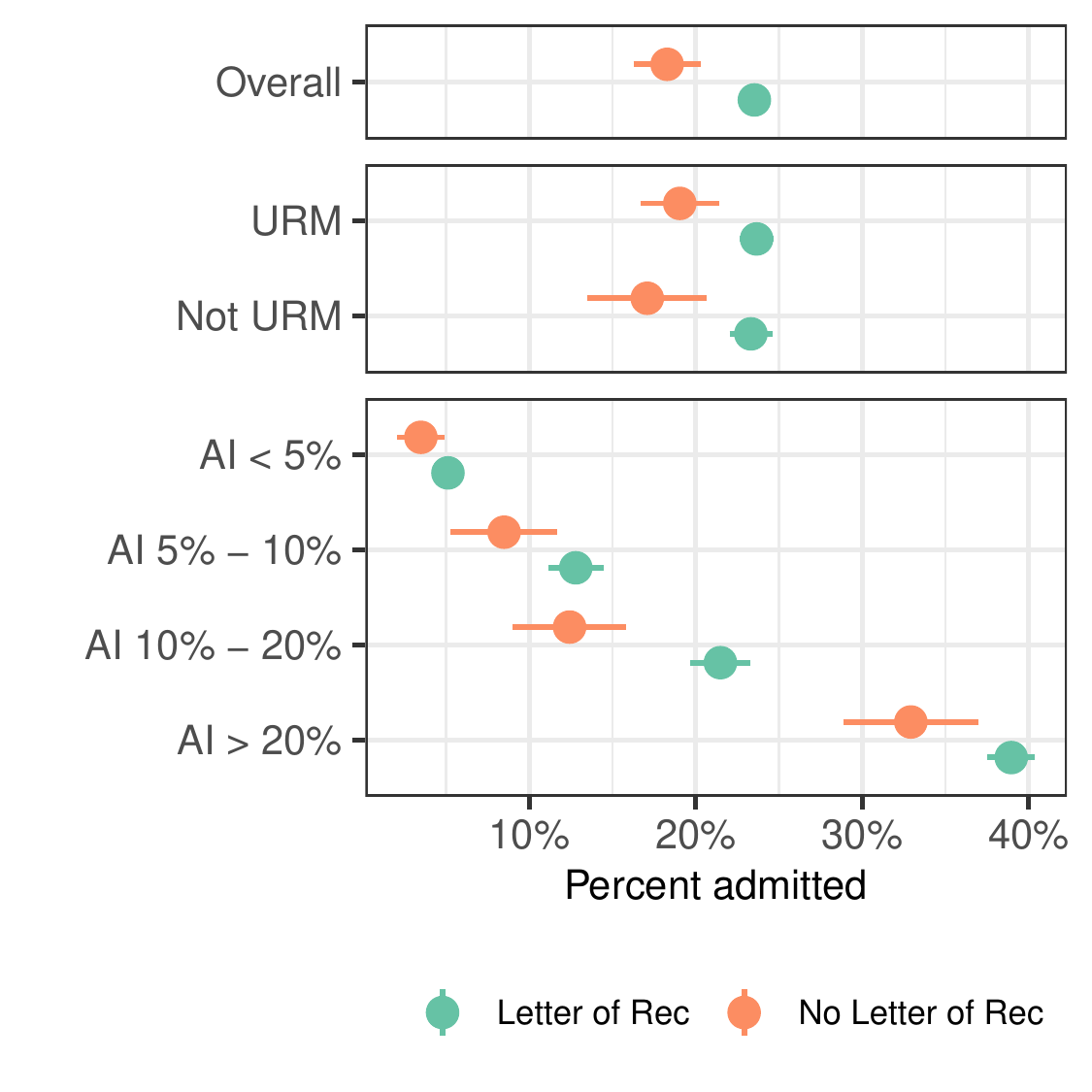} 
    }
    \caption{Treated and re-weighted control percent admitted.} 
      \label{fig:main_estimates_marginal}
      \end{subfigure}%
      ~
      \begin{subfigure}[t]{0.45\textwidth}  
        {\centering \includegraphics[width=\textwidth]{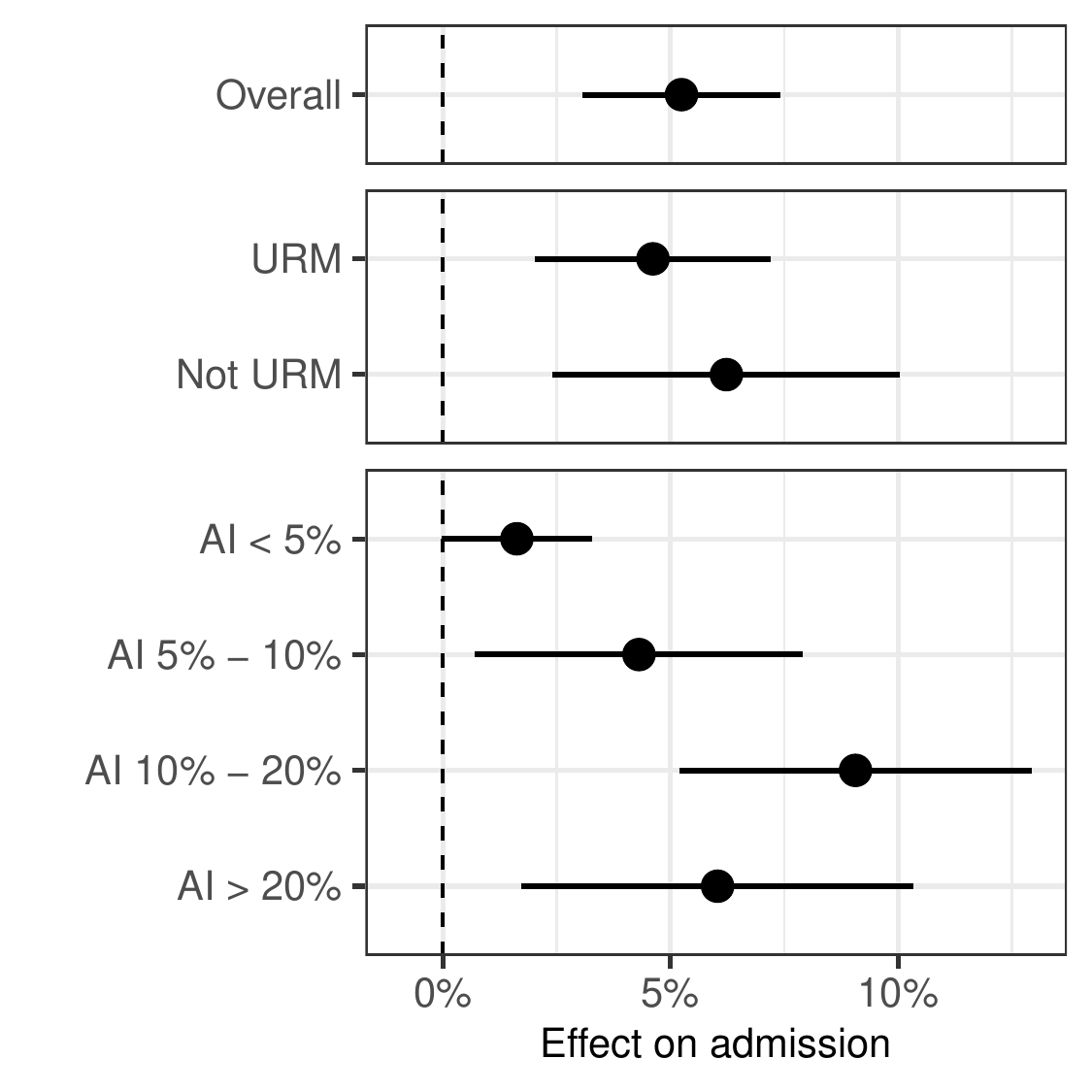} 
        }
        \caption{Estimated effects on admission.} 
          \label{fig:main_estimates_marginal_admit}
          \end{subfigure}%
      \caption{Estimated treated and control means and treatment effect of letters of recommendation on admission $\pm$ two standard errors, overall and by URM status and Admissibility Index.} 
      \label{fig:marginal_estimates}
    \end{figure}

After assessing local balance and overlap, we can now turn to estimating the differential impacts of letters of recommendation.
Figure \ref{fig:marginal_estimates} shows (1) the percent of applicants who submitted LORs who were accepted, $\hat{\mu}_{1g}$; (2) the imputed counterfactual mean, $\hat{\mu}_{0g}$; and (3) the ATT, $\hat{\mu}_{1g} - \hat{\mu}_{0g}$. The standard errors are computed via the sandwich estimator in Equation \eqref{eq:sandwich}.
Overall, we estimate that LORs increased admission rates by 5 percentage points (pp).
While we estimate a larger effect for non-URM applicants (6.2 pp) than URM applicants (4.5 pp), there is insufficient evidence to distinguish between the two effects. 
Second, we see a roughly positive trend between treatment effects and the AI, potentially with a peak for the 10\%-20\% group. This is driven by the very small estimated effect for applicants with AI $< 5\%$, who are very unlikely to be accepted with or without LORs. LORs seem to have a larger effect for applicants closer to the cusp of acceptance.

\begin{figure}[htbp]
  \centering
  \begin{subfigure}[t]{0.45\textwidth}  
    {\centering \includegraphics[width=\textwidth]{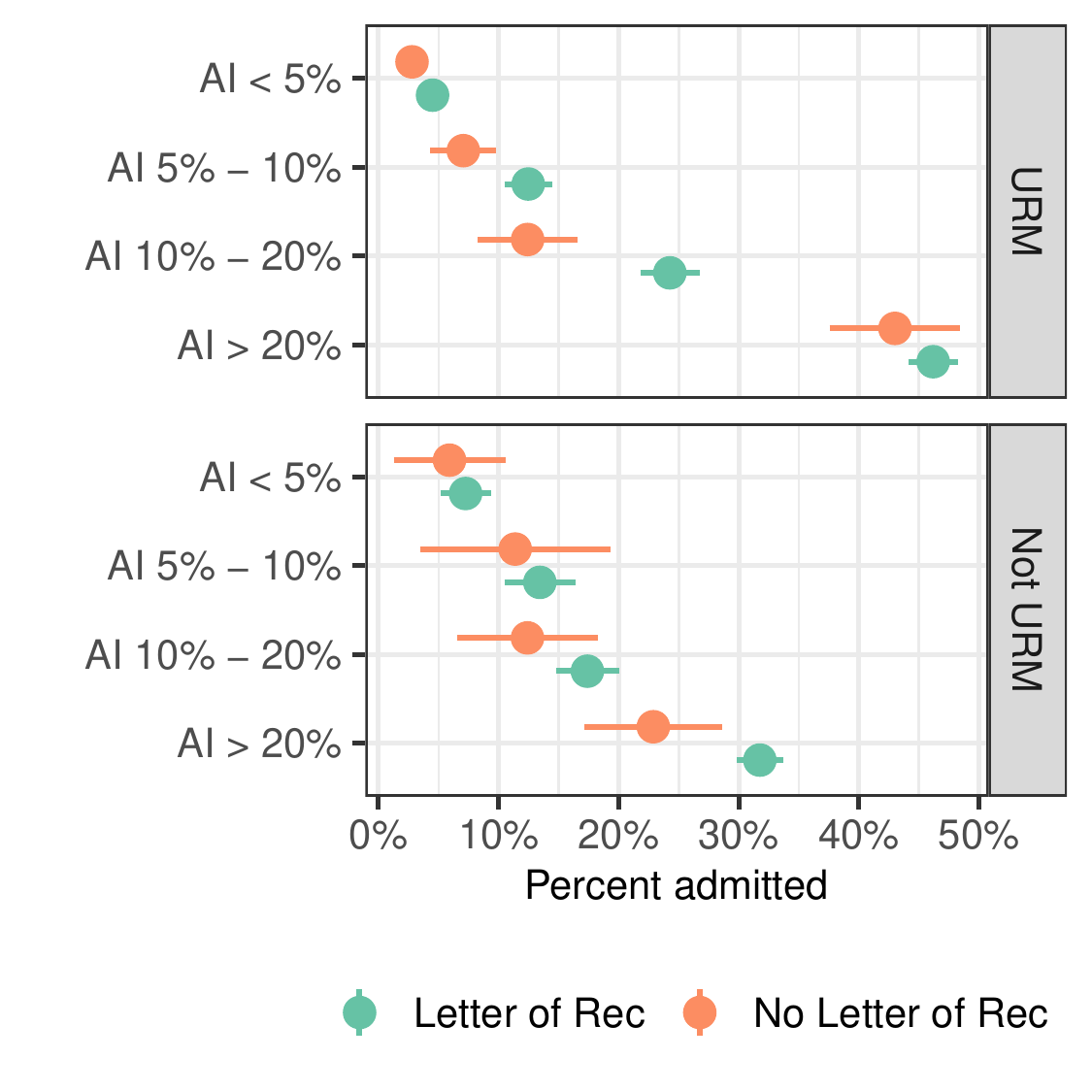} 
    }
    \caption{Treated and re-weighted control percent admitted.}
      \label{fig:main_estimates_interact_admit_mu}
      \end{subfigure}%
    ~
    \begin{subfigure}[t]{0.45\textwidth}  
    {\centering \includegraphics[width=\textwidth]{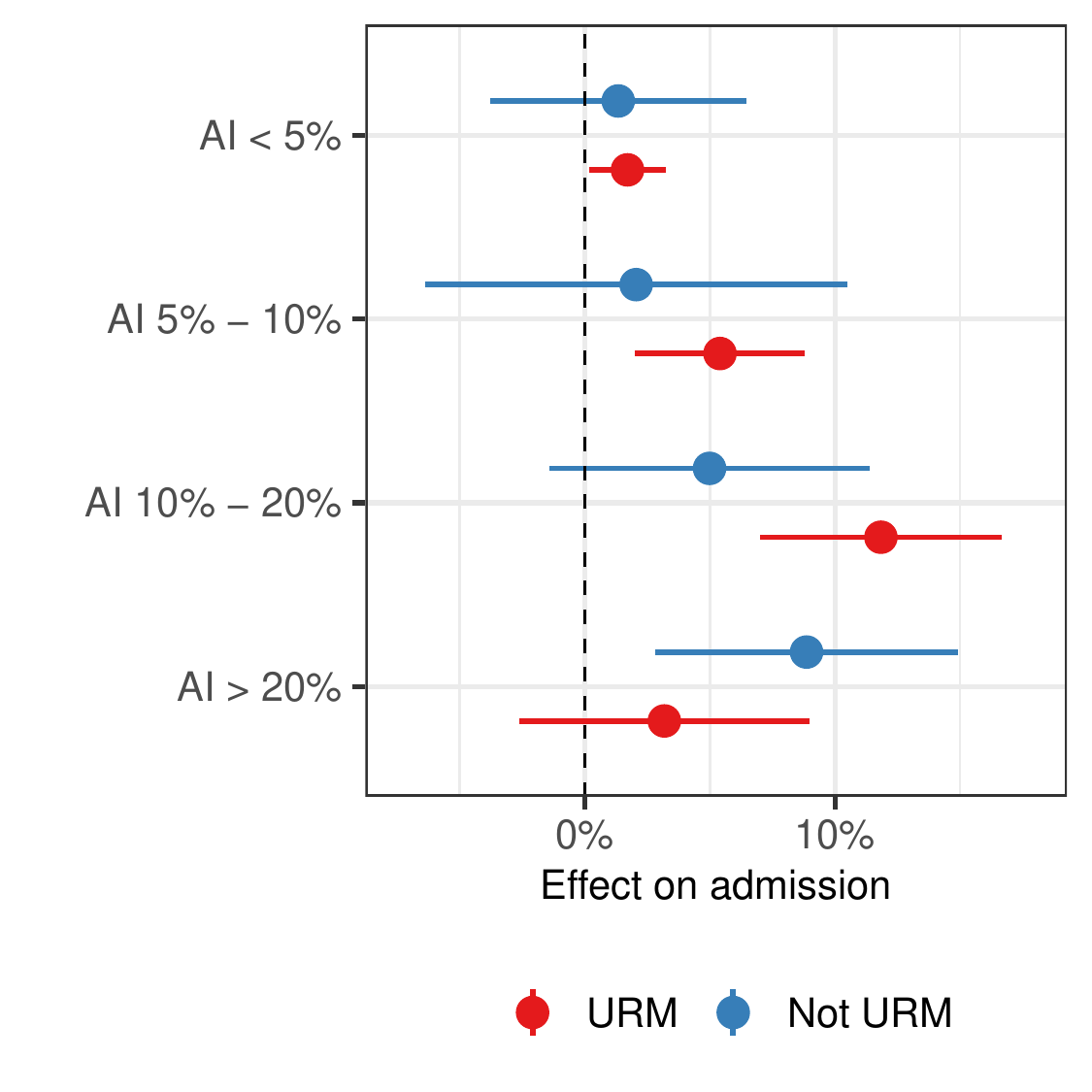} 
    }
    \caption{Estimated effects on admission.}
      \label{fig:main_estimates_interact_admit}
      \end{subfigure}
    \caption{Estimated treated and control means and treatment effect of letters of recommendation on admission $\pm$ two standard errors, further broken down by URM status interacted with the Admissibility Index. }
    \label{fig:interaction_estimates}
  \end{figure}

Figure \ref{fig:interaction_estimates} further stratifies the subgroups, showing the effects jointly by URM status and AI. While the point estimate for the overall increase in admission rates is slightly larger for non-URM applicants than for URM applicants, this is mainly a composition effect. For applicants very unlikely to be admitted (AI $< 5$\%) the point estimates are nearly identical for URM and non-URM applicants, although the URM subgroup is estimated much more precisely.
For the next two levels of the admissibility index (AI between 5\% and 20\%), URM applicants have a higher estimated impact, with imprecise estimates for non-URM applicants.
For the highest admissibility groups (AI $> 20$\%),  non-URM applicants have larger positive effects, though again these estimates are noisy. 
Since URM applicants have lower AI on average, the overall estimate is also lower for URM applicants.
Furthermore, the peak in the effect for middle-tier applicants is more pronounced for URM applicants than non-URM applicants. 
From Figure \ref{fig:main_estimates_interact_admit_mu} we see that this is primarily because high admissibility URM applicants have very high imputed admission rates.

Appendix \ref{sec:robustness_appendix} includes extensive robustness checks and other sensitivity analyses. 
We first compare our main results with simple estimates based on linear regression and based on plugging in the AI.
Both approaches diverge from our main results for the highest admissibility non-URM applicants, a group with limited overlap.
For other subgroups, the estimates are similar for the AI-based estimates, but differ for the linear regression estimates, which has poor control over local imbalance. 
% Next, we assess the sensitivity of our results to different definitions of the sample and choice of outcome.
We also explore an alternative approach that instead leverages unique features of the UC Berkeley pilot study, which included an additional review without the letters of recommendation from a sample of 10,000 applicants. 
These results are broadly similar to the estimates from the observational study, again differing mainly in regions with relatively poor overlap.
Finally, we conduct a formal sensitivity analysis for violations of the ignorability assumption (Assumption \ref{a:ignore}), adapting a recent proposal from \citet{soriano2019sensitivity}. 
Using this approach we conclude that there would need to be substantial unmeasured confounding, of roughly the same predictive power as the AI, to qualitatively change our conclusions.

Taken together, our main results and accompanying sensitivity checks paint a relatively clear picture of differential impact of letters of recommendation across applicants' \emph{a priori} probability of admission. Treatment effects are low for applicants who are unlikely to be accepted and high for applicants on the margin for whom letters provide useful context,  with some evidence of a dip for the highest admissibility applicants.
Our estimates of differential impacts between URM and non-URM students are more muddled, due to large sampling errors, and do not support strong conclusions. Point estimates indicate that LORs benefit URM applicants more than they do non-URM applicants at all but the highest academic indexes. Because non-URM applicants are overrepresented in the high-AI category, the point estimate for the average treatment effect is larger for non-URMs; however, there is insufficient precision to distinguish between the two groups.

\subsection{Augmented and machine learning estimates}
We now consider augmenting the weighting estimator with an estimate of the prognostic score, $\hat{m}(x, g)$. In Appendix Figure \ref{fig:augmented_estimates_ridge} we show estimates after augmenting with ridge regression, fully interacting $\phi(X)$ with the strata indicators; we compute standard errors via Equation \eqref{eq:sandwich}, replacing $Y_i - \hat{\mu}_{0g}$ with the empirical residual $Y_i - \hat{m}(X_i, g)$. Because the partially pooled balancing weights achieve excellent local balance for $\phi(X)$, augmenting with a model that is also linear in $\phi(X)$ results in minimal adjustment.
We therefore augment with random forests, a nonlinear outcome estimator. 
Tree-based estimators are a natural choice, creating ``data-dependent strata'' similar in structure to the strata we define for $G$. 
For groups where the weights $\hat{\gamma}$ have good balance across the estimates $\hat{m}(x, g)$, there will be little adjustment due to the outcome model.
Conversely, if the raw and bias-corrected estimate disagree for a subgroup, then the weights have poor local balance across important substantive data-defined strata. For these subgroups we should be more cautious of our estimates.

\begin{figure}[tbp]
  \centering
      \begin{subfigure}[t]{0.45\textwidth}  
  {\centering \includegraphics[width=\textwidth]{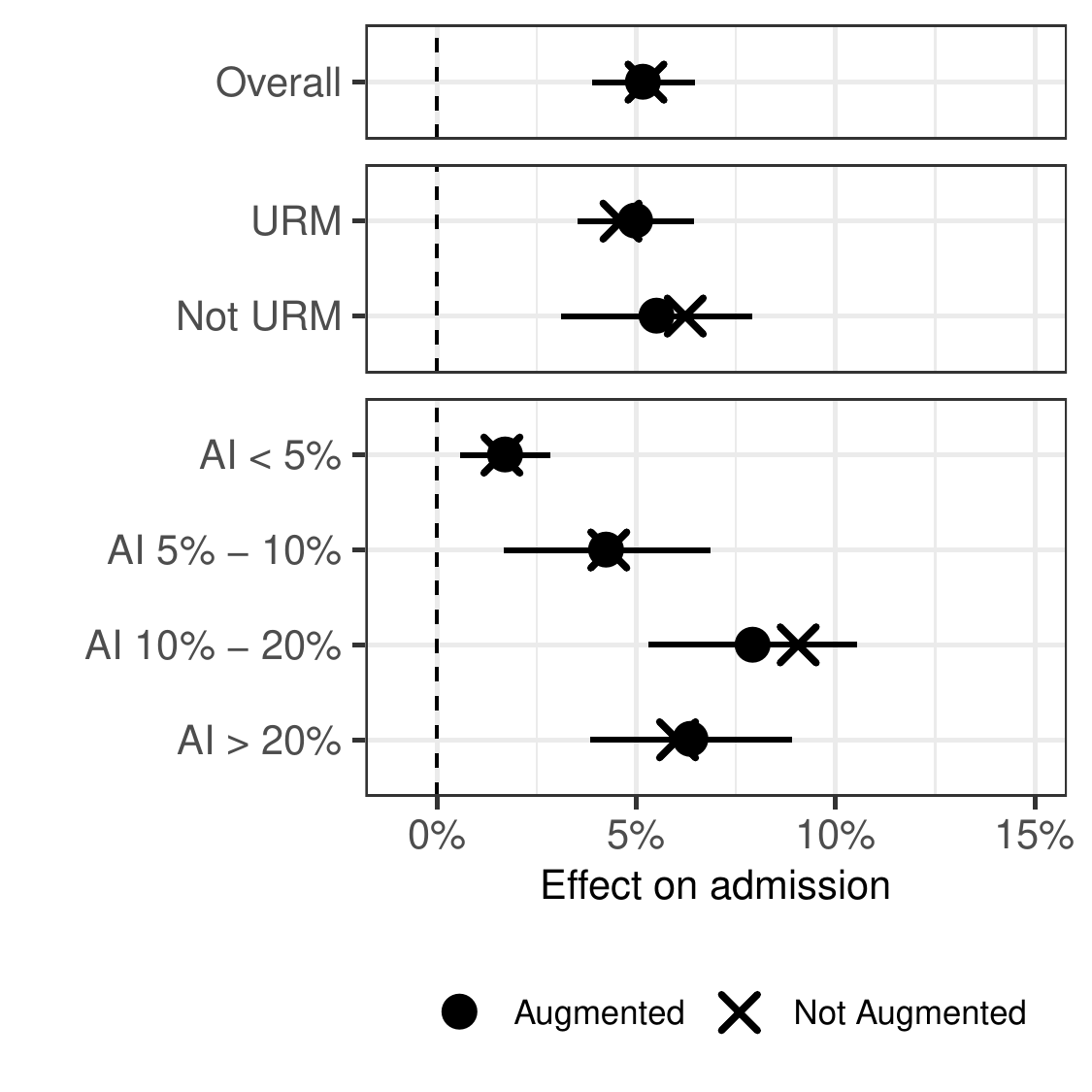} 
  }
  \caption{Overall and by URM status and AI.} 
      \label{fig:augmented_estimates_marginal}
      \end{subfigure}%
      ~
      \begin{subfigure}[t]{0.45\textwidth}  
      {\centering \includegraphics[width=\textwidth]{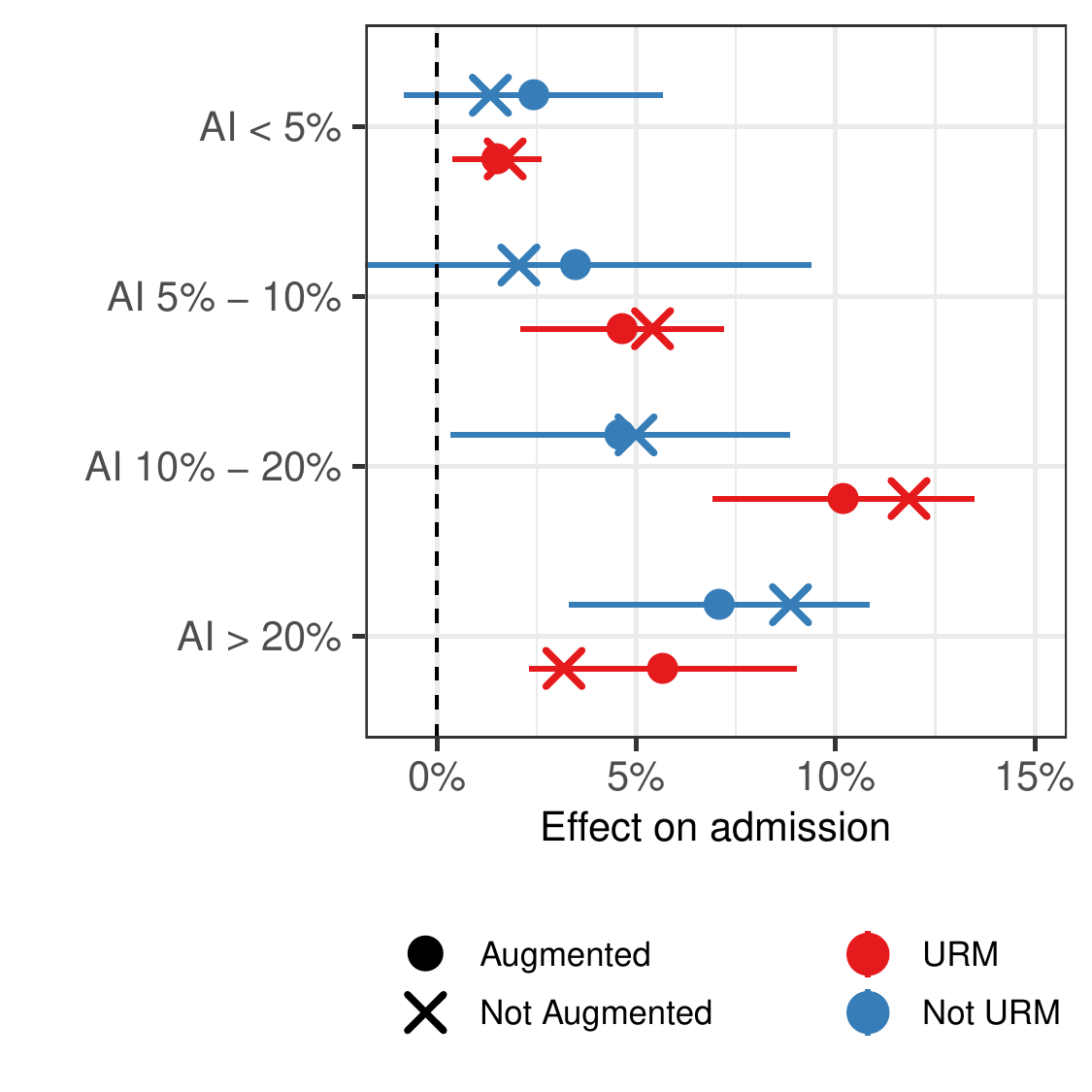} 
      }
      \caption{By URM status interacted with AI.}
      \label{fig:augmented_estimates_interact}
      \end{subfigure}
      \caption{Estimated effect of letters of recommendation on admission rates with and without augmentation via a random forest outcome model.} 
      \label{fig:augmented_estimates}
  \end{figure}

Figure \ref{fig:augmented_estimates} shows the random forest-augmented effect estimates relative to the un-augmented estimates; the difference between the two is the estimated bias. Overall, the random forest estimate of the bias is negligible and, as a result, the un-adjusted and adjusted estimators largely coincide. Augmentation, however, does seem to stabilize the higher-order interaction between AI and URM status, with particularly large adjustments for the highest AI group (AI $\geq 20$\%). This suggests that we should be wary of over-interpreting any change in the relative impacts for URM and non-URM applicants as AI increases.

We also compare to treatment effects estimated via automatic, flexible machine learning approaches that give no special precedence to our pre-defined subgroups. First, Appendix Figure \ref{fig:simple_estimates} shows the result of using the ridge regression model above to impute the counterfactual means $\hat{\mu}_{0g}$ for each subgroup. 
Because the partially-pooled weights achieve excellent local balance, the ridge estimates broadly comport with both the weighting and augmented estimates. However, ridge regression leads to more precise estimates for the subgroups with lower effective sample sizes and fewer comparable control units. While the augmented balancing weights limit the amount of extrapolation, the ridge estimator nonetheless extrapolates away from the support of the control units an arbitrary amount in order to find a lower-variance estimate \citep{benmichael2019_augsynth}.

Finally, we use Bayesian causal forests \citep[BCF;][]{Hahn2020} to estimate the conditional average treatment effect given the covariates and subgroup, $\hat{\tau}(x,g)$, then aggregate over the treated units in the group to estimate the CATT, $\hat{\tau}_g = \frac{1}{n_{1g}}\sum_{G_i = g} W_i \hat{\tau}(X_i, G_i)$.
This approach gives no special consideration to the subgroups of interest: $G$ enters symmetrically with the other covariates $X$.
Appendix Figure \ref{fig:bcf_estimates} shows the results. The BCF estimate of the overall ATT is nearly the same as our main estimates and similarly finds no variation between URM and non-URM applicants. However, the BCF estimates find less heterogeneity across admissibility levels and little to no heterogeneity across URM and admissibility subgroups, in part because this approach regularizes higher-order interactions in the CATE function. While this can be beneficial in many cases, with pre-defined subgroups this can lead to over-regularization of the effects of interest, as happens here. Furthermore, the BCF estimates are extremely precise, even in regions with limited overlap \citep[see discussion in][]{Hahn2020}. Overall, we find this approach less credible in our setting with pre-defined subgroups of interest.

%%%
%%% DISCUSSION
%%%
\section{Discussion}
\label{sec:discussion}

Estimating heterogeneous treatment effects and assessing treatment effect variation in observational studies is a challenge, even for pre-specified subgroups. Focusing on weighting estimators that estimate subgroup treatment effects by re-weighting control outcomes, we show that the estimation error depends on the level of \emph{local imbalance} between the treated and control groups after weighting. We then present a convex optimization problem that finds approximate balancing weights that directly target the level of local imbalance within each subgroup, while ensuring exact global balance to also estimate the overall effect. Using this method to estimate heterogeneous effects in the UC Berkeley letters of recommendation pilot study, we find evidence that letters of recommendation lead to better admissions outcomes for stronger applicants, with mixed evidence of differences between URM and non-URM applicants. 

There are several directions for future methodological work. First, hyperparameter selection for balancing weights estimators is an important question in practice but remains an open problem. We elect to choose the hyperparameter by explicitly tracing out the level of balance and effective sample size as the hyper-parameter changes. However, cross validation approaches such as that proposed by \cite{Wang2019} may have better properties. This an an important avenue for future work. 

Second, we directly estimate the effect of submitting an LOR among those who submit. However, we could instead frame the question in terms of non-compliance and use the \emph{invitation} to submit an LOR as an instrument for submission. Using the approximate balancing weights procedure described above we could adjust for unequal invitation probabilities, and estimate the effect on compliers via weighted two-stage least squares.

Finally, we could consider applying this approach to subgroup effects in randomized trials, adapting recent proposals from, among others, \citet{zeng2021propensity} and \citet{yang2021covariate}. We expect that the overlap will be substantially improved in this setting, leading to larger effective sample size, even if the randomized trials themselves are smaller. We also anticipate that the design-based approach we advocate here will help researchers navigate Simpsons' Paradox-type challenges in subgroup effects \citep{vanderweele2011interpretation}.

%%%
%%% BIBLIOGRAPHY
%%%
\clearpage
\bibliography{citations.bib}
\bibliographystyle{chicago}

\clearpage
\section*{Supplementary materials}
\appendix
\renewcommand\thefigure{\thesection.\arabic{figure}} 
\renewcommand\thetable{\thesection.\arabic{table}}       
\setcounter{figure}{0}    

\singlespacing

\section{Robustness checks and sensitivity analysis}
\label{sec:robustness_appendix}

We now assess how much these conclusions change with different estimates and under violations of the key ignorability assumption.

\subsection{Alternative estimators and sample definitions}
First, we can see how our main estimates contrast with simple comparisons between those who submitted letters and those who did not. Appendix Figure \ref{fig:simple_estimates} shows effects estimated by comparing the admission rate for the treatment group to the admission rate expected by the AI.
This comparison shows a large difference between the effects for URM and non-URM applicants, with a \emph{negative} estimated effect on non-URM applicants.
This is primarily driven by a large negative estimate for the highest admissibility non-URM applicants, with the other estimates roughly in line with our main estimates. As we discuss in Section \ref{sec:ai}, while the AI is very predictive overall it has less predictive power for higher admissibility applicants, resulting in unreliable estimates of the effect. 

Appendix Figure \ref{fig:simple_estimates} also shows simple linear regression estimates of the effects, regressing admission on treatment and an additional set of terms from the transformed covariates $\phi(X)$. We estimate effects first with treatment alone, then treatment interacted by URM status, interacted by AI group, and interacted by both. This approach does not directly control for local imbalances and relies on a correctly specified linear additive specification. Because of this, off-the-shelf linear regression disagrees with our main estimates, estimating a smaller overall effect and no effect for non-URM applicants. This is driven by a negative effect for low admissibility non-URM applicants, a region of limited overlap as specified above.

Next, we consider the sensitivity of our results to a different definition of the sample.
Recall that an applicant may not have submitted an LOR for one of two reasons: (i) they were not invited to do so, and (ii) they did not submit even though they were invited.
We assess the sensitivity of our results to excluding this first group 
when using the weighting approach.
Appendix Figure \ref{fig:subset_estimates} shows the overall estimated effect and the effect for URM and non-URM applicants with the full sample and restricted to applicants who were invited to submit an LOR.
The point estimates are similar, 
and although the number of control units is much smaller in the restricted sample --- 3,452 vs 29,398 ---
the standard errors are only slightly larger.
This reflects the fact the weighting approach finds that few of the no-invitation control units are adequate comparisons to the treated units.

\subsection{Effect on second reader scores and within-subject comparison}
\label{sec:within}

% \eli{Moved this paragraph to appendix}
We now consider effects on an intermediate outcome: whether the second reader --- who has access to the LORs --- gives a ``Yes'' score. 
Because these are \emph{design-based} weights, we use the same set of weights to estimate effects on both second reader scores and admissions decisions. We find a similar pattern of heterogeneity overall.

With this outcome we can also make use of a within-study design to estimate treatment effects, leveraging scores from additional third readers who did not have access to the letters of recommendation.
After the admissions process concluded, 10,000 applicants who submitted letters were randomly sampled and the admissions office recruited several readers to conduct additional evaluations of the applicants \citep{rothstein_lor2017}. During this supplemental review cycle, the readers were \emph{not} given access to the letters of recommendation, but otherwise the evaluations were designed to be as similar as possible to the second reads that were part of the regular admissions cycle; in particular, readers had access to the first readers' scores.

With these third reads we can estimate the treatment effect by taking the average difference between the second read (with the letters) and the third read (without the letters). One major issue with this design is that readers might have applied different standards during the supplemental review cycle. Regardless, if the third readers applied a different standard consistently across URM and admissibility status, we can distinguish between treatment effects within these subgroups.

Appendix Figures \ref{fig:reader2_marginal}  and \ref{fig:reader2_interact} show the results for both approaches. Overall for second reader scores we see a similar structure of heterogeneity as for admission rates, although there does not appear to be an appreciable decline in the treatment effect for the highest admissibility non-URM applicants.
The two distinct approaches yield similar patterns of estimates overall, with the largest discrepancy for applicants with a predicted probability of admission between 5\% and 10\%, particularly for non-URM applicants. However, this group has a very low effective sample size, and so the weighting estimates are very imprecise.

\subsection{Formal sensitivity analysis}
\label{sec:sensitivity}

We assess sensitivity to the key assumption
underlying our estimates, Assumption \ref{a:ignore}: an applicant's LOR submission is conditionally independent of that applicant's potential admission decision.
Since we observe all the information leading to an invitation to submit an LOR, we believe that Assumption \ref{a:ignore} is plausible for this step in the process. However, applicants' decisions to submit LORs given that invitation might vary in unobserved ways that are correlated with admission.

To understand the potential impact of such unmeasured confounding, we perform a formal sensitivity analysis. Following the approach in \citet{Zhao2019_sensitivity, soriano2019sensitivity}, we allow the true propensity score conditioned on the control potential outcome $e(x, g, y) \equiv P(W = 1 \mid X = x, G = g, Y(0) = y)$ to differ from the probability of treatment given covariates $x$ and group membership $g$, $e(x,g)$, by a factor of $\Lambda$ in the odds ratio: 

\begin{equation}
  \label{eq:sens}
  \Lambda^{-1} \leq \frac{\nicefrac{e(x,g)}{1 - e(x,g)}}{\nicefrac{e(x,g,y)}{1 - e(x,g,y)}} \leq \Lambda.
\end{equation}

\noindent This generalizes Assumption \ref{a:ignore} to allow for a pre-specified level of unmeasured confounding, where $\Lambda = 1$ corresponds to the case with no unmeasured confounding. The goal is then to find the smallest and largest ATT, $[\tau^{\text{min}}, \tau^{\text{max}}]$, consistent with a given $\Lambda$ for the marginal sensitivity model in Equation \eqref{eq:sens}. Following \citet{soriano2019sensitivity}, we use the percentile bootstrap to construct a 95\% confidence for this bound, $[L, U]$.
In particular, we focus on the largest value of $\Lambda$ for which the overall ATT remains statistically significant at the 95\% level (i.e., $L > 0$), which we compute as $\Lambda = 1.1$.

We then modify this approach to focus on subgroup differences; we focus on differences between URM and non-URM applicants but can apply this to other subgroups as well. First, we use the same procedure to find bounds on the effect for URM applicants, $[L^{\text{urm}}, U^{\text{urm}}]$, and non-URM applicants, $[L^{\text{non}}, U^{\text{non}}]$. We then construct worst-case bounds on their \emph{difference}, $[L^{\text{urm}} - U^{\text{non}}, U^{\text{urm}} - L^{\text{non}}]$.
Although we fail to detect a difference between the two groups, there may be unmeasured variables that confound a true difference in effects. To understand how large the difference could be, we use the sensitivity value that nullifies the overall effect, $\Lambda = 1.1$ and construct a 95\% confidence interval for the difference. % $\tau_{\text{urm}} - \tau_{\text{non}}$. 
We find that we cannot rule out a true difference as large as 12 pp, with 95\% confidence interval $(-12\%, 8.2\%)$.

To understand this number, Appendix Figure \ref{fig:sens} shows the strength required of an unmeasured confounder in predicting the admission outcome (measured as the magnitude of a regression of the outcome on the unmeasured confounder) to produce enough error to correspond to $\Lambda = 1.1$, for a given level of imbalance in the unmeasured confounder between applicants who did and did not submit LORs.\footnote{Letting $\delta$ denote this imbalance, the absolute regression coefficient must be larger than 
$\max\{|L^{\text{urm}} - U^{\text{non}}|, |U^{\text{urm}} - L^{\text{non}}|\} / \delta = 0.592 / \delta$.}
It compares these values to the imbalance in the components of $\phi(X)$ \emph{before weighting} and the regression coefficient for each component where we regress the outcome on $\phi(X)$ for the control units. We find that the unmeasured confounder would have to have a higher level of predictive power or imbalance than any of our transformed covariates, except for the AI. Thus, an unmeasured confounder would have to be relatively strong and imbalanced in order to mask a substantial difference between URM and non-URM applicants.

\clearpage
\section{Additional simulation estimators}
\label{sec:sim_appendix}
 We compute treatment effects for eight weighting estimators:
 \begin{itemize}
  \item \emph{Partially pooled balancing weights:} approximate balancing weights that solve Equation \eqref{eq:primal}, using $G$ as the stratifying variable and prioritizing local balance by setting $\lambda_g = \frac{1}{n_{1g}}$.
  
  \item \emph{Augmented balancing weights:} augmenting the partially pooled balancing weights as in Equation \eqref{eq:mu0g_hat_aug} where $\hat{m}_0(x, g)$ is fit via ridge regression with all interactions.
  
  \item \emph{Fully pooled balancing weights:} approximate balancing weights that solve Equation \eqref{eq:primal}, but ignore local balance by setting $\lambda \to \infty$, thus fully pooling towards the global model. This is equivalent to stable balancing weights in Equation \eqref{eq:sbw} with an exact global balance constraint $\delta = 0$. 
  
  \item \emph{No pooled balancing weights:} approximate balancing weights that solve Equation \eqref{eq:primal}, but without the exact global balance constraint.
  
  \item \emph{Full interaction IPW:} traditional IPW with a fully interacted model that estimates a separate propensity score within each stratum as in Equation \eqref{eq:ipw_interact}.
  
  \item \emph{Fixed effects IPW:} full interaction IPW with stratum-specific coefficients constrained to be equal to a global parameter $\beta_g = \beta$ for all $g$.
  
  \item \emph{Full interaction ridge regression outcome model:} Estimating $\mu_{0g}$ via $\hat{\mu}_{0g} = \frac{1}{n_1}\sum_{G_i = g}W_i \hat{m}_0(X_i, g)$, where $\hat{\mu}_0(x, g)$ is the same ridge regression predictor as for augmented balancing weights
  
  \item \emph{Bayesian Causal Forests (BCF)}: Estimating $\tau_g$ as $\frac{1}{n_{1g}} \sum_{G_i = g} W_i \hat{\tau}_i$, where $\hat{\tau}_i$ are the posterior predictive means from a Bayesian Causal Forest estimator \citep{Hahn2020}.
  
\end{itemize}

%%%
\clearpage
\section{Additional figures and tables}
\setcounter{figure}{0}    

% \clearpage

\begin{figure}[!htb]
  \centering 
  \begin{subfigure}[t]{0.45\textwidth}  
    {\centering \includegraphics[width=\textwidth]{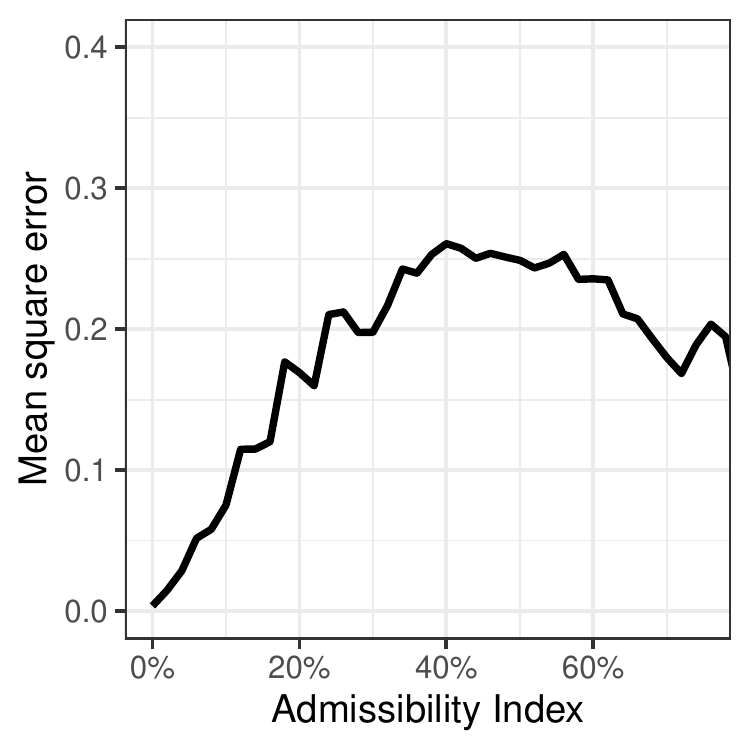} 
    }
    \caption{Brier score.} 
      \label{fig:ai_mse}
      \end{subfigure}%
      ~
      \begin{subfigure}[t]{0.45\textwidth}  
      {\centering \includegraphics[width=\textwidth]{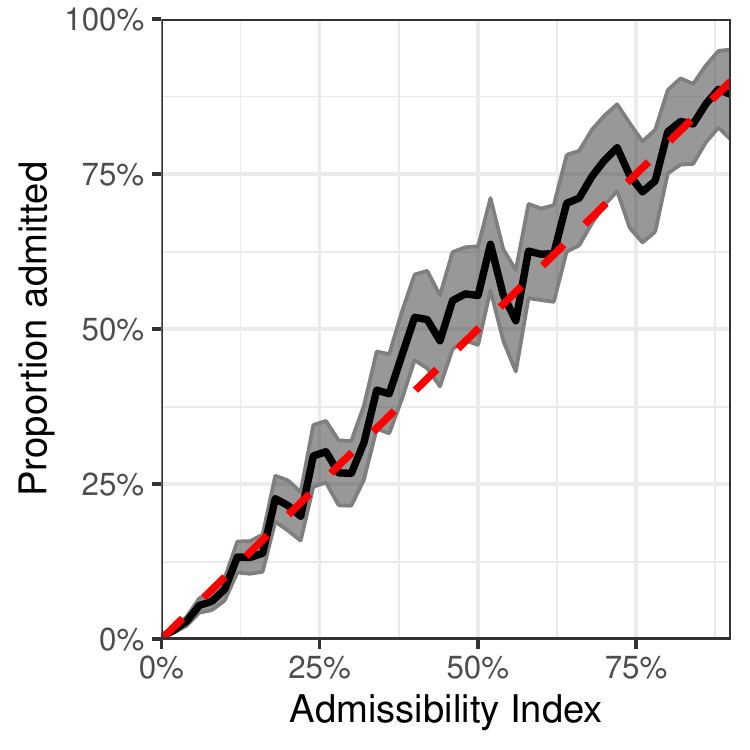} 
      }
      \caption{Admission rates}
        \label{fig:ai_avg}
  \end{subfigure}
\caption{(a) Mean square error (Brier score) and (b) admission rates for the Admissibility Index predicting the 2016-2017 cycle admissions results, computed in 2\% groups.}
\label{fig:ai_performance}
\end{figure}

\begin{table}[!htbp]   
    \centering
    \begin{tabular}{@{}llrr@{}}
      \toprule
      College & URM & AUC & Brier Score\\
      \midrule
      \multirow{2}{*}{Letters and Science} & URM & 91.7\% & 6\%\\
      & Not URM & 91.7\% & 9\%\\
      \midrule
      \multirow{2}{*}{Engineering} & URM & 91.6\% & 4\%\\
      & Not URM & 91.8\% & 9\%\\
      \bottomrule
    \end{tabular}
    \caption{AUC and Brier score for the Admissibility Index predicting the 2016-2017 cycle admissions results.}
    \label{tab:ai_roc_auc}
  \end{table}

\begin{figure}[!htb]
  \centering \includegraphics[width=\maxwidth]{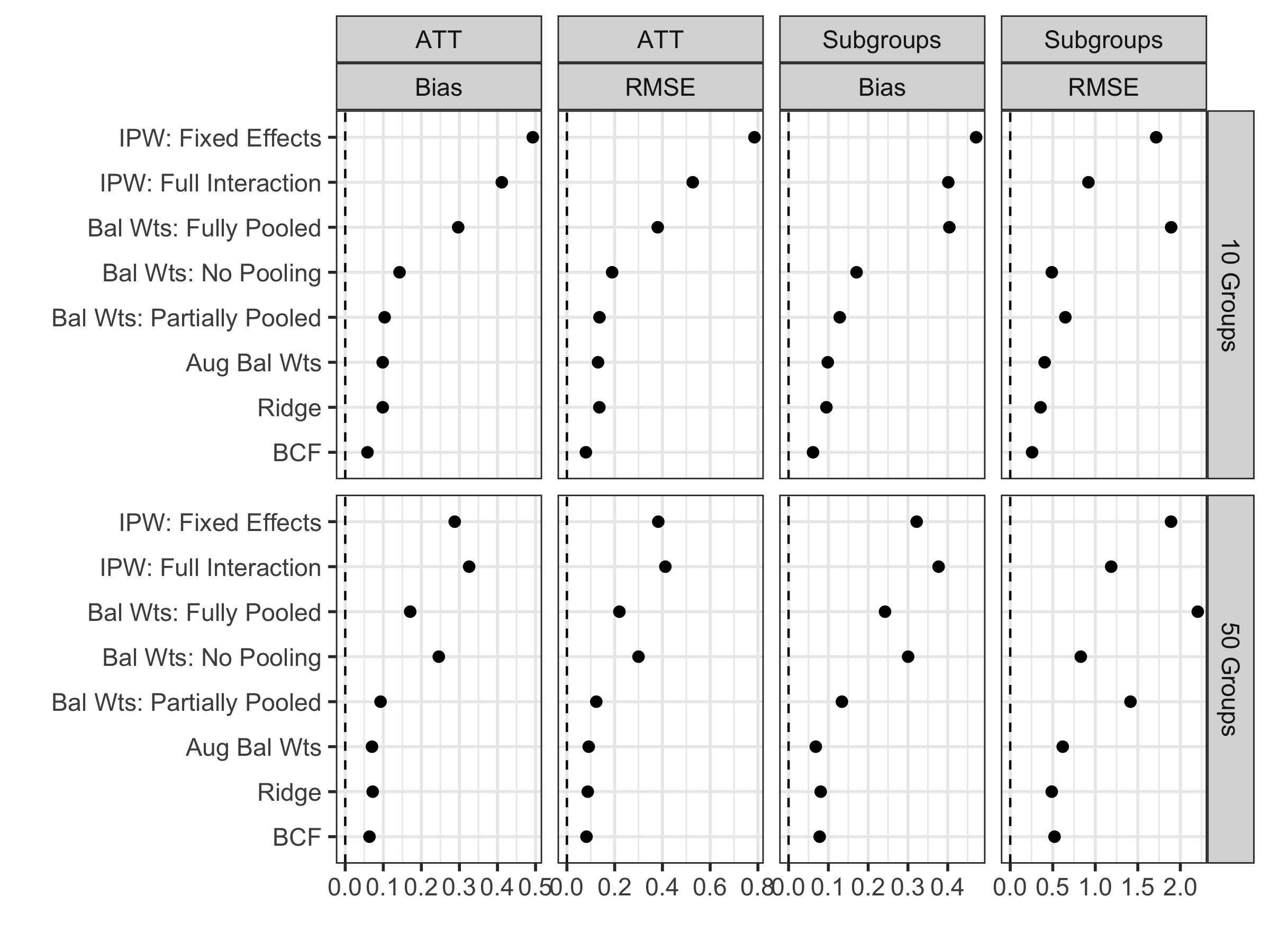}
\caption{Performance of approximate balancing weights, traditional IPW with logistic regression, and outcome modelling for estimating subgroup treatment effects.}
\label{fig:plot_all_weights}
\end{figure}

\begin{figure}[tb]
  \centering \includegraphics[width=\maxwidth]{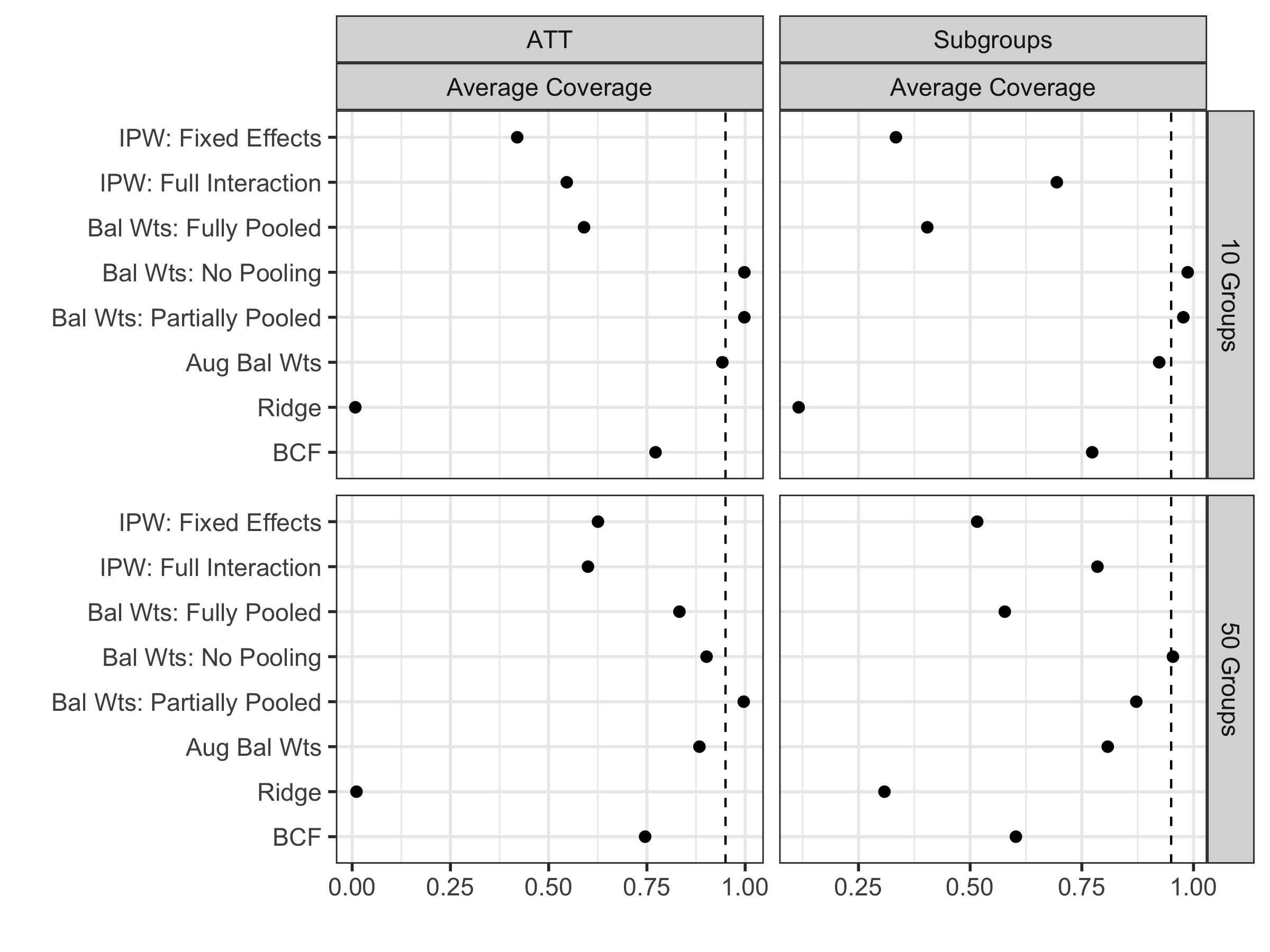}
\caption{Coverage for approximate balancing weights, traditional IPW with logistic regression, and outcome modelling for estimating subgroup treatment effects.}
\label{fig:sim_coverage}
\end{figure}

  \begin{figure}[tbp]
    \centering \includegraphics[width=\maxwidth]{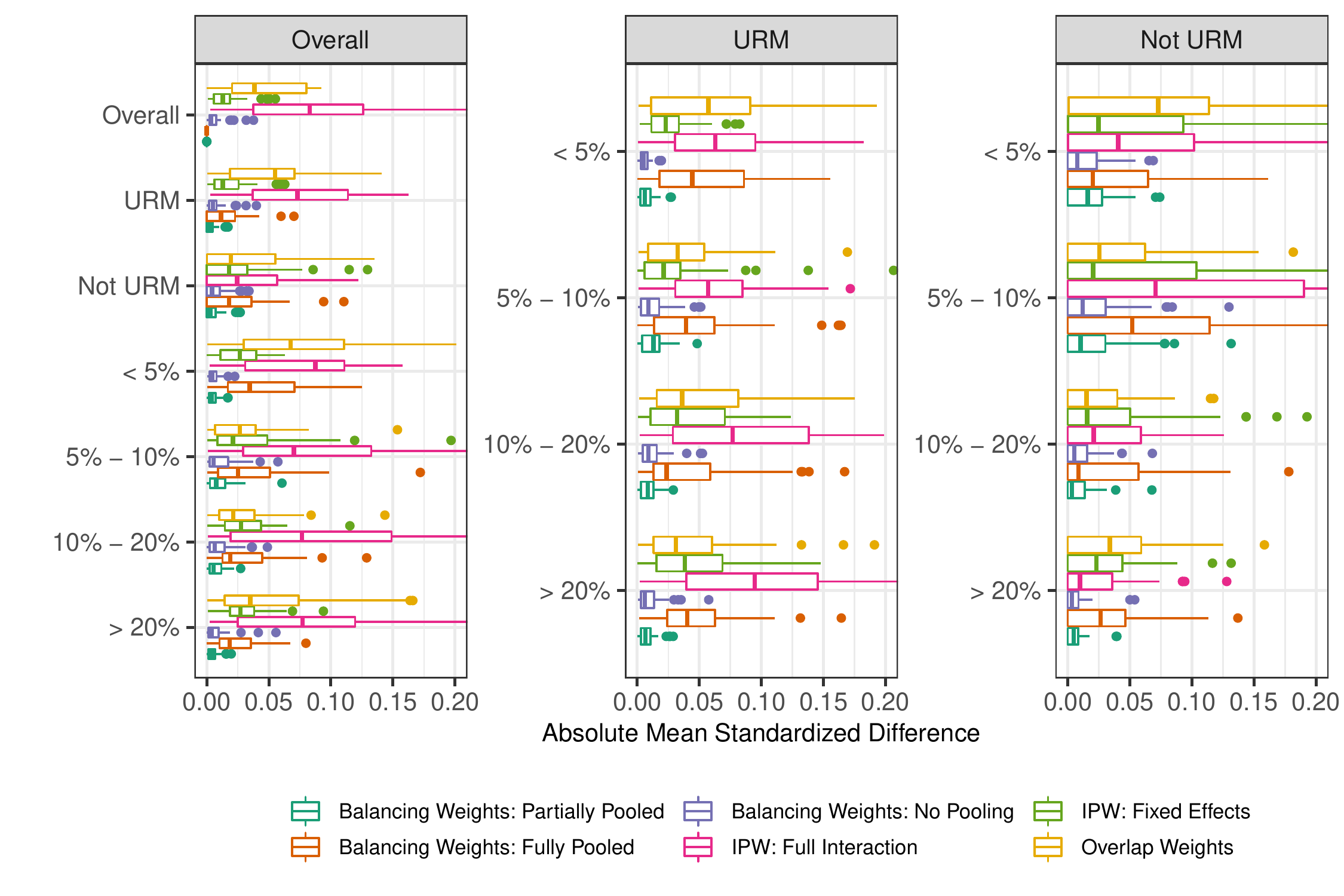}
  \caption{Distribution of covariate balance measured by the mean standardized difference for different weighting methods.}
  \label{fig:love_plot_box}
\end{figure}

\begin{figure}[tbp]

  \centering
    \begin{subfigure}[t]{0.45\textwidth}  
  {\centering \includegraphics[width=\textwidth]{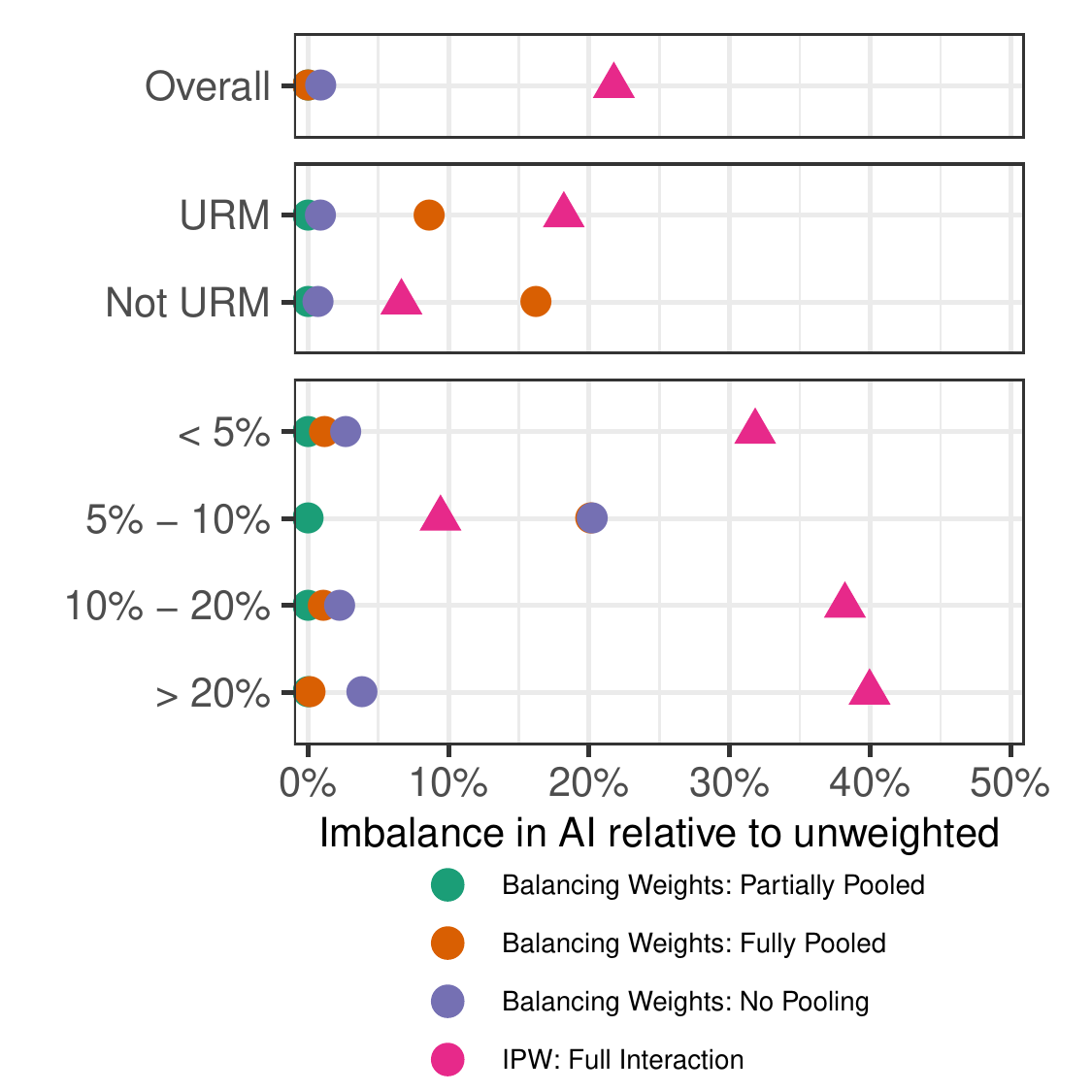} 
  }
  \caption{Overall and by URM status and AI.} 
    \label{fig:pct_bias_reduce_marginal}
    \end{subfigure}%
    ~
    \begin{subfigure}[t]{0.45\textwidth}  
    {\centering \includegraphics[width=\textwidth]{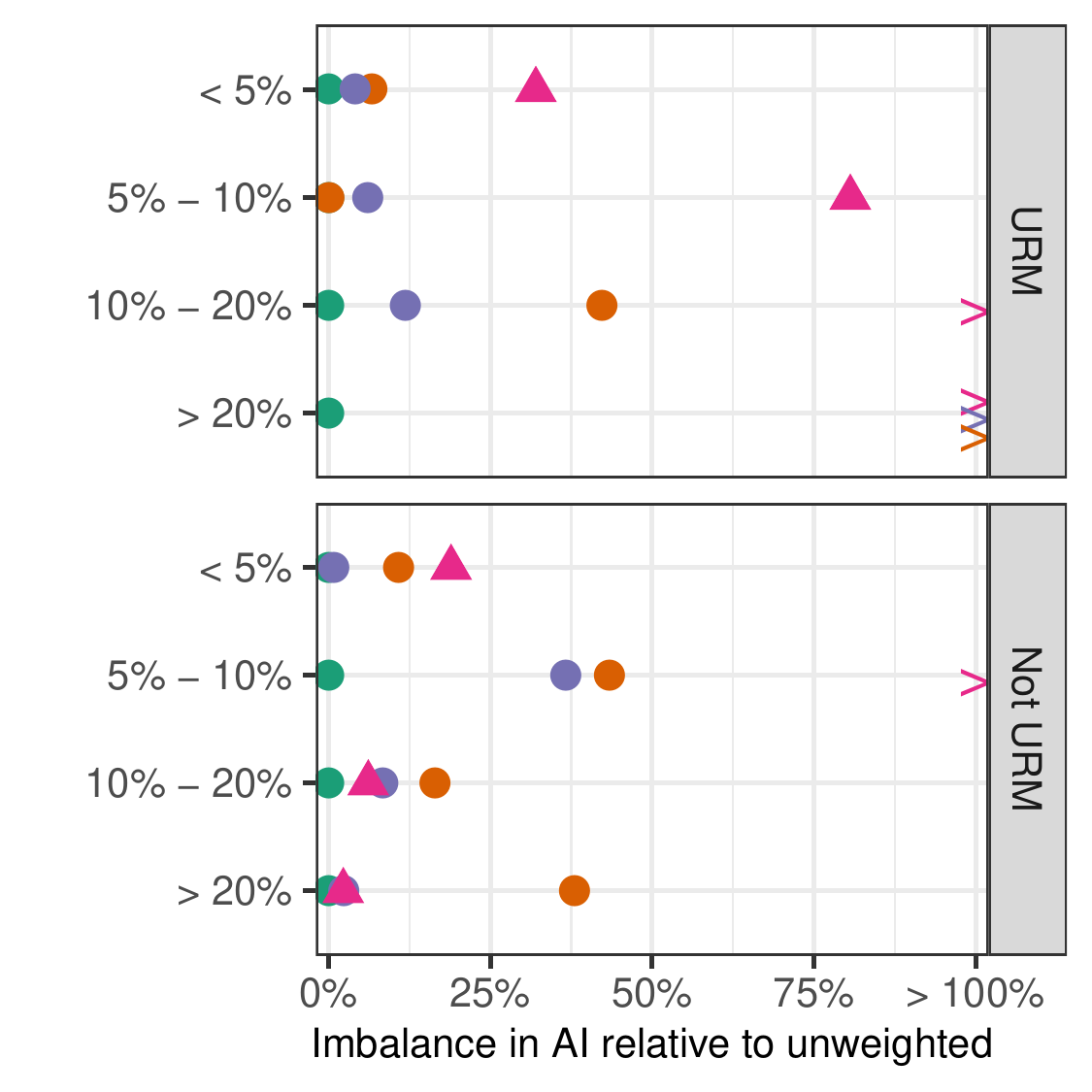} 
    }
    \caption{By URM status interacted with AI.}
      \label{fig:pct_bias_reduce_interact}
      \end{subfigure}
\caption{Imbalance in the admissibility index after weighting relative to before weighting, overall and within each subgroup.
For several subgroups, the fully pooled balancing weights procedure results in \emph{increased} imbalance in the admissibility index, denoted by an arrow.} 
\label{fig:pct_bias_reduce}
\end{figure}

  \begin{figure}[tbp]
    \centering
        \begin{subfigure}[t]{0.45\textwidth}  
    {\centering \includegraphics[width=\textwidth]{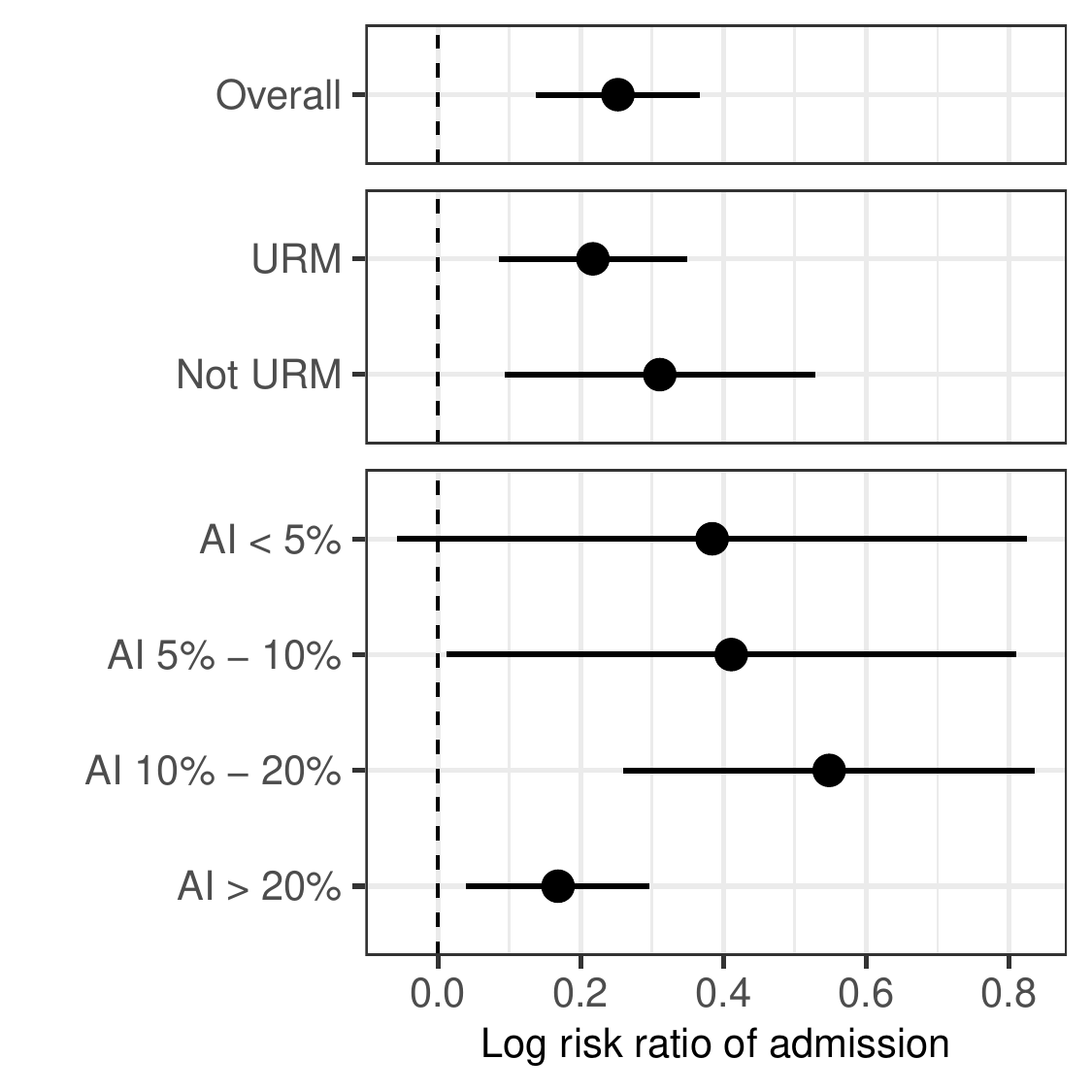} 
    }
    \caption{Overall and by URM status and AI.} 
        \label{fig:risk_ratio_estimates_marginal.pdf}
        \end{subfigure}%
        ~
        \begin{subfigure}[t]{0.45\textwidth}  
        {\centering \includegraphics[width=\textwidth]{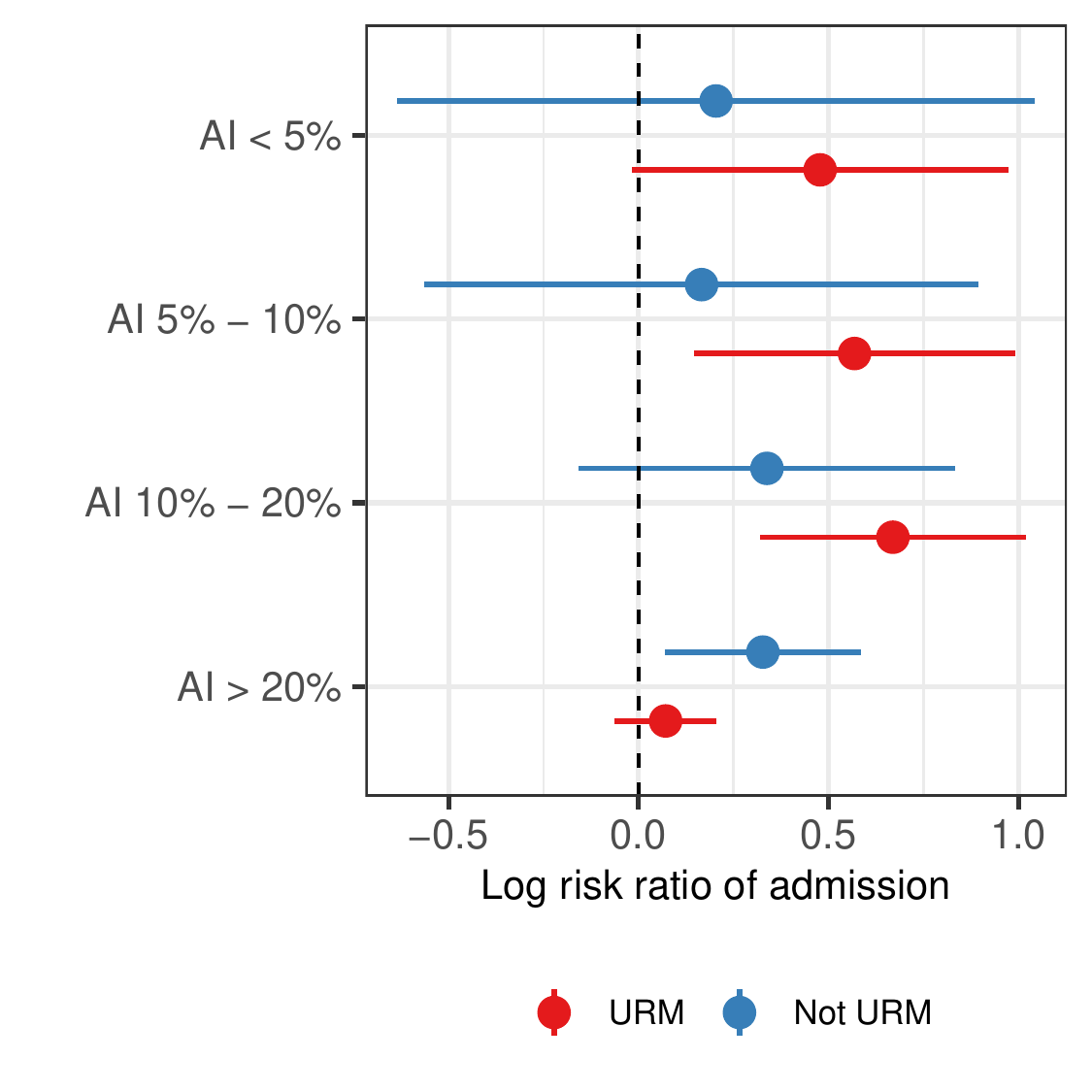} 
        }
        \caption{By URM status interacted with AI.}
        \label{fig:risk_ratio_estimates_interact}
        \end{subfigure}
        \caption{Estimated log risk ratio of admission with and without letters of recommendation $\pm$ two standard errors computed via the delta method, overall and by URM status and AI.} 
        \label{fig:risk_ratio_estimates}
    \end{figure}

    \begin{figure}[tbp]
      \centering
      \begin{subfigure}[t]{0.45\textwidth}  
        {\centering \includegraphics[width=\textwidth]{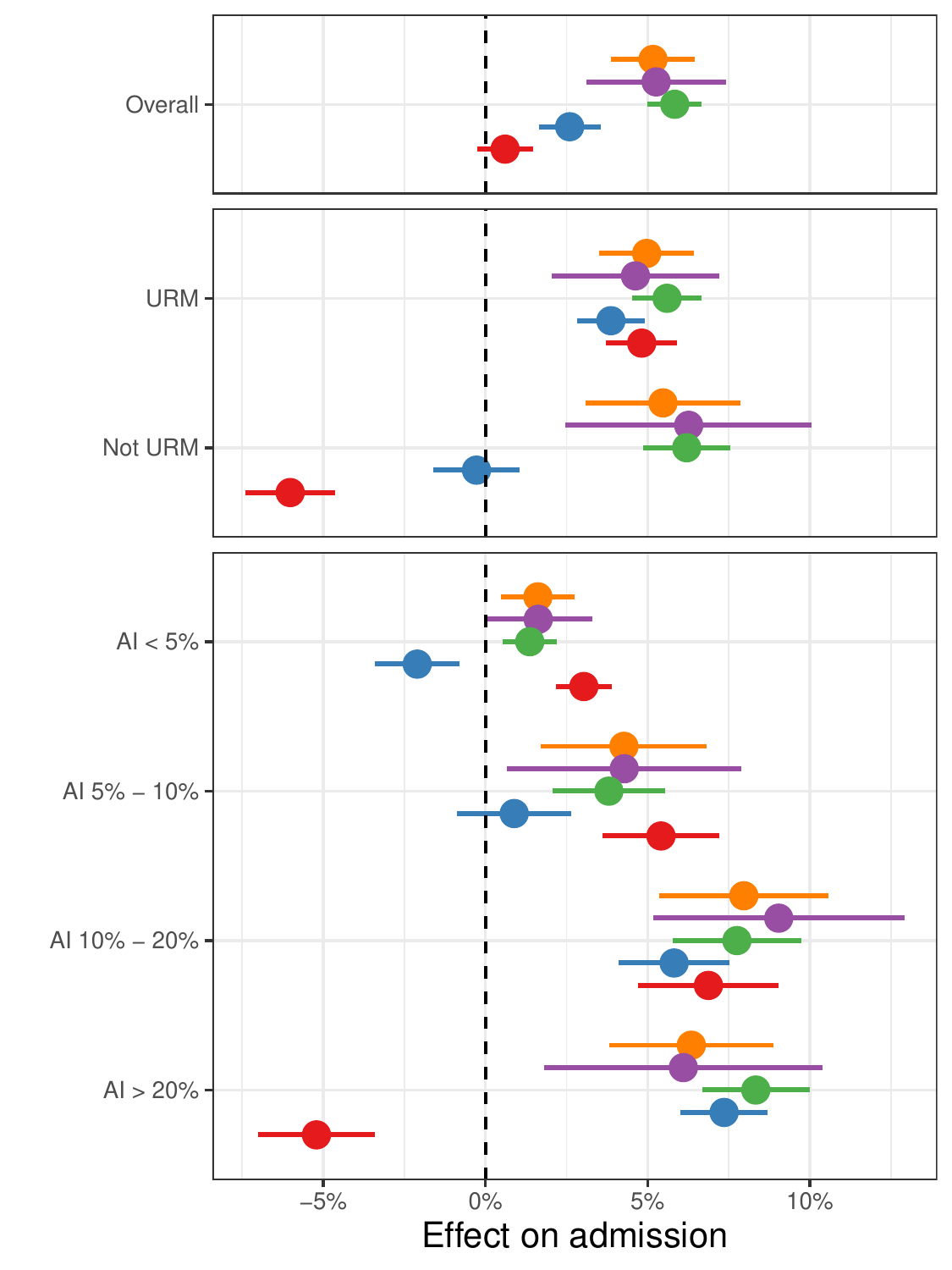} 
        }
        \caption{Overall and by URM status and AI}
          \label{fig:simple_estimates_interact_admit_mu}
          \end{subfigure}%
        ~
        \begin{subfigure}[t]{0.45\textwidth}  
        {\centering \includegraphics[width=\textwidth]{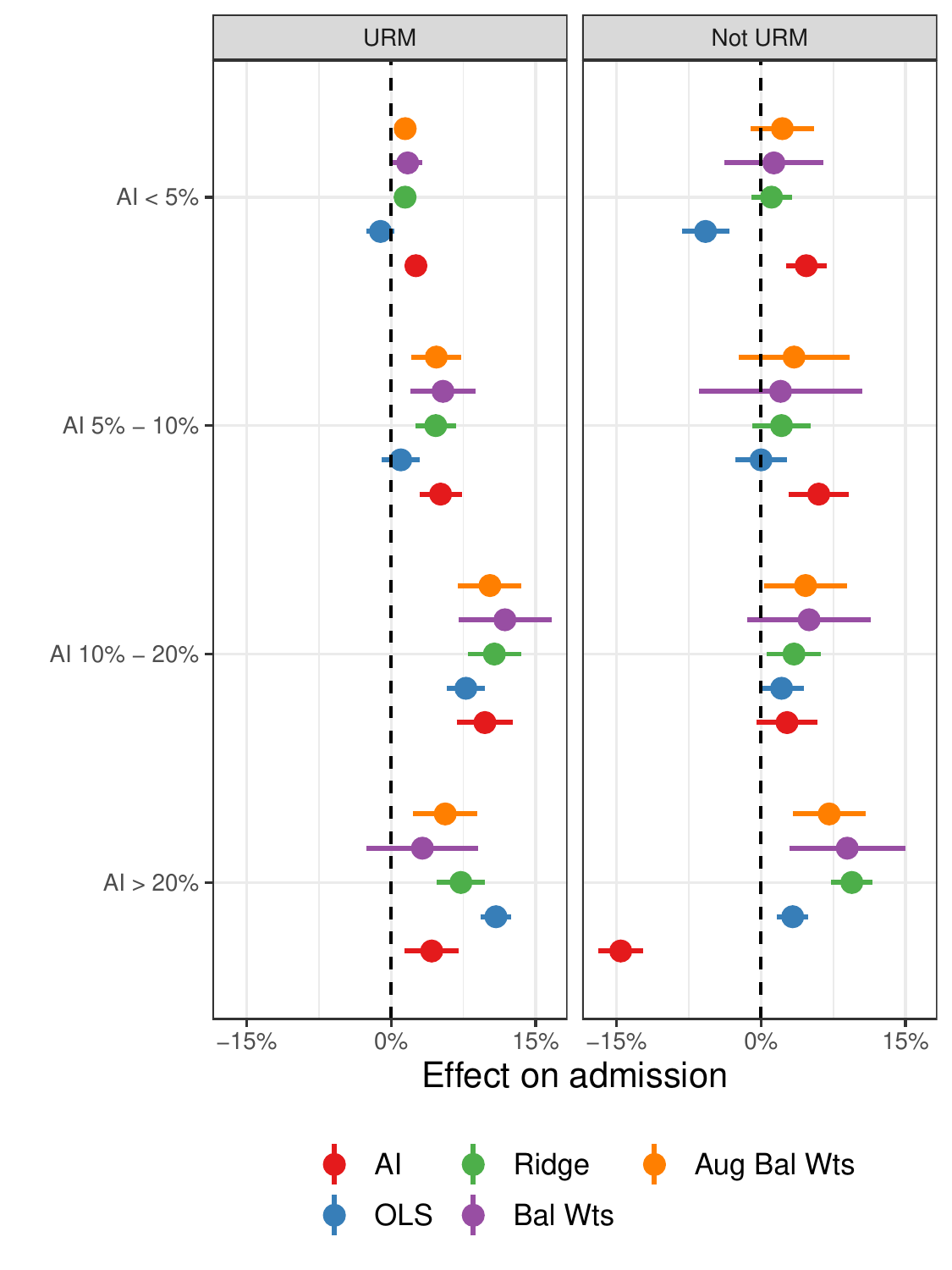} 
        }
        \caption{By URM status interacted with AI}
          \label{fig:simple_estimates_interact_admit}
          \end{subfigure}
        \caption{Estimated effect of letters of recommendation on admission rates \emph{without} adjusting for selection, and comparing to the expected admission rate from the AI score and regression.} 
        \label{fig:simple_estimates}
      \end{figure}

    \begin{figure}[tbp]
      \centering
          \begin{subfigure}[t]{0.45\textwidth}  
      {\centering \includegraphics[width=\textwidth]{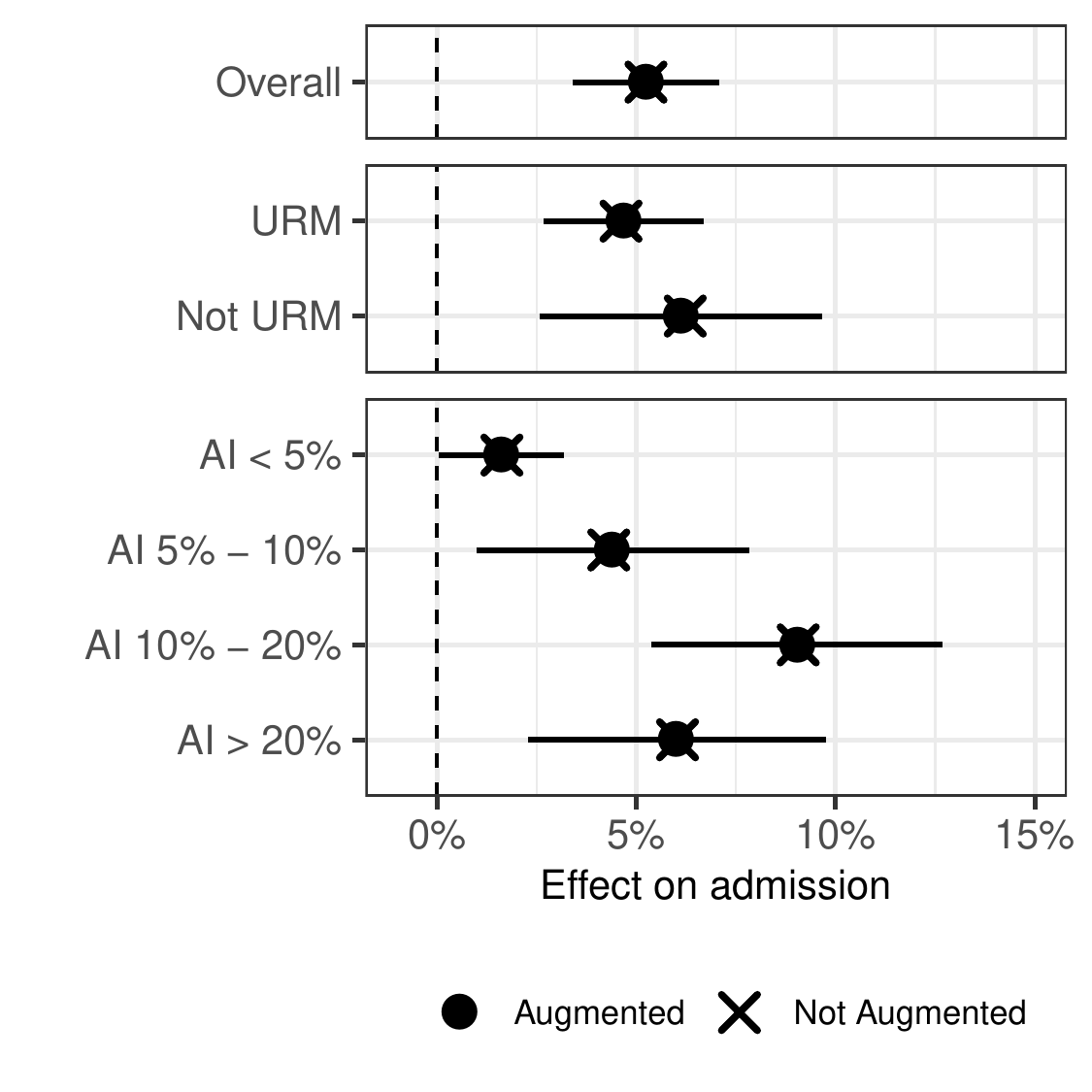} 
      }
      \caption{Overall and by URM status and AI.} 
          \label{fig:augmented_estimates_marginal_ridge}
          \end{subfigure}%
          ~
          \begin{subfigure}[t]{0.45\textwidth}  
          {\centering \includegraphics[width=\textwidth]{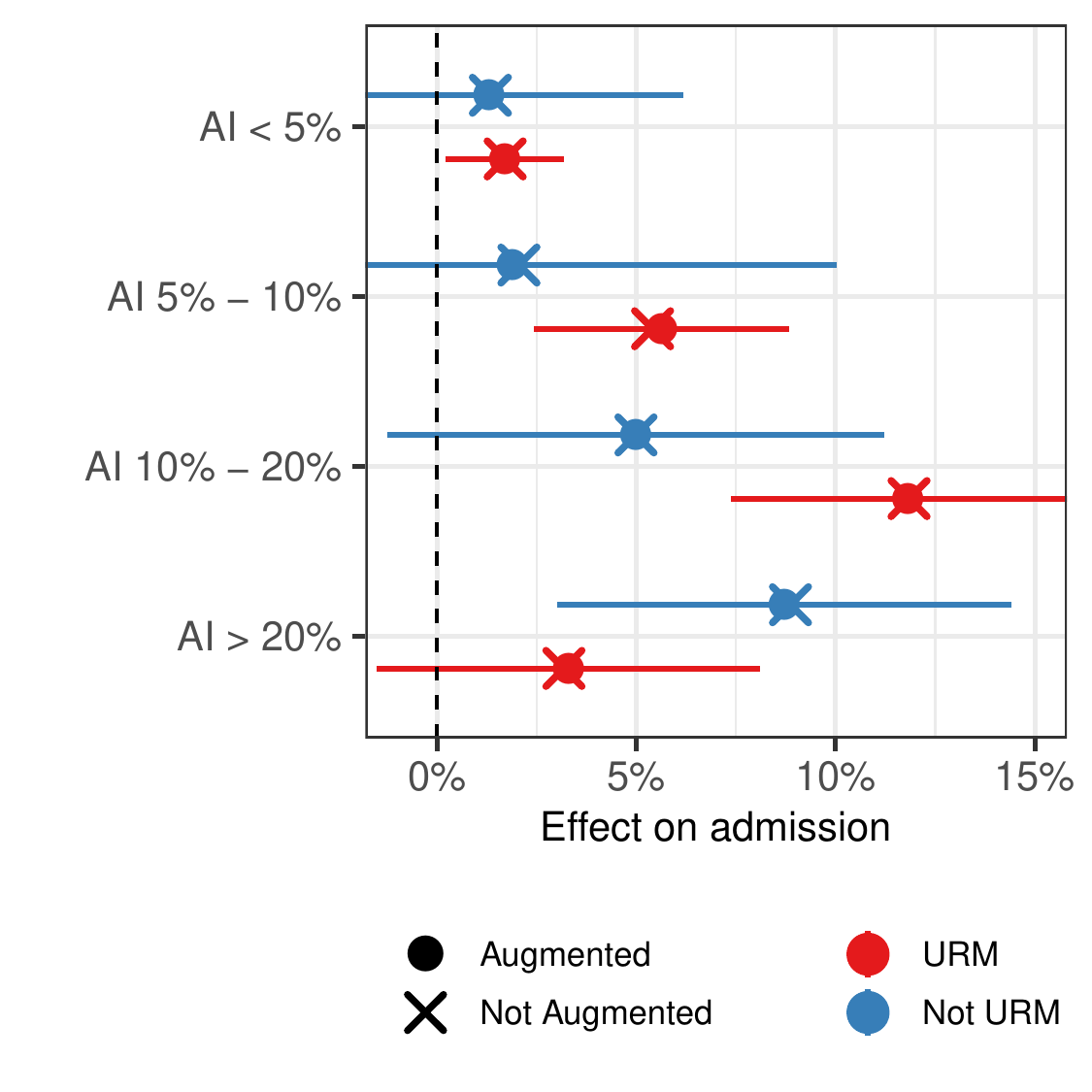} 
          }
          \caption{By URM status interacted with AI.}
          \label{fig:augmented_estimates_interact_ridge}
          \end{subfigure}
          \caption{Estimated effect of letters of recommendation on admission rates with and without augmentation via ridge regression with 5-fold cross validation.} 
          \label{fig:augmented_estimates_ridge}
      \end{figure}

\begin{figure}[tbp]
    \centering
      \begin{subfigure}[t]{0.45\textwidth}  
    {\centering \includegraphics[width=\textwidth]{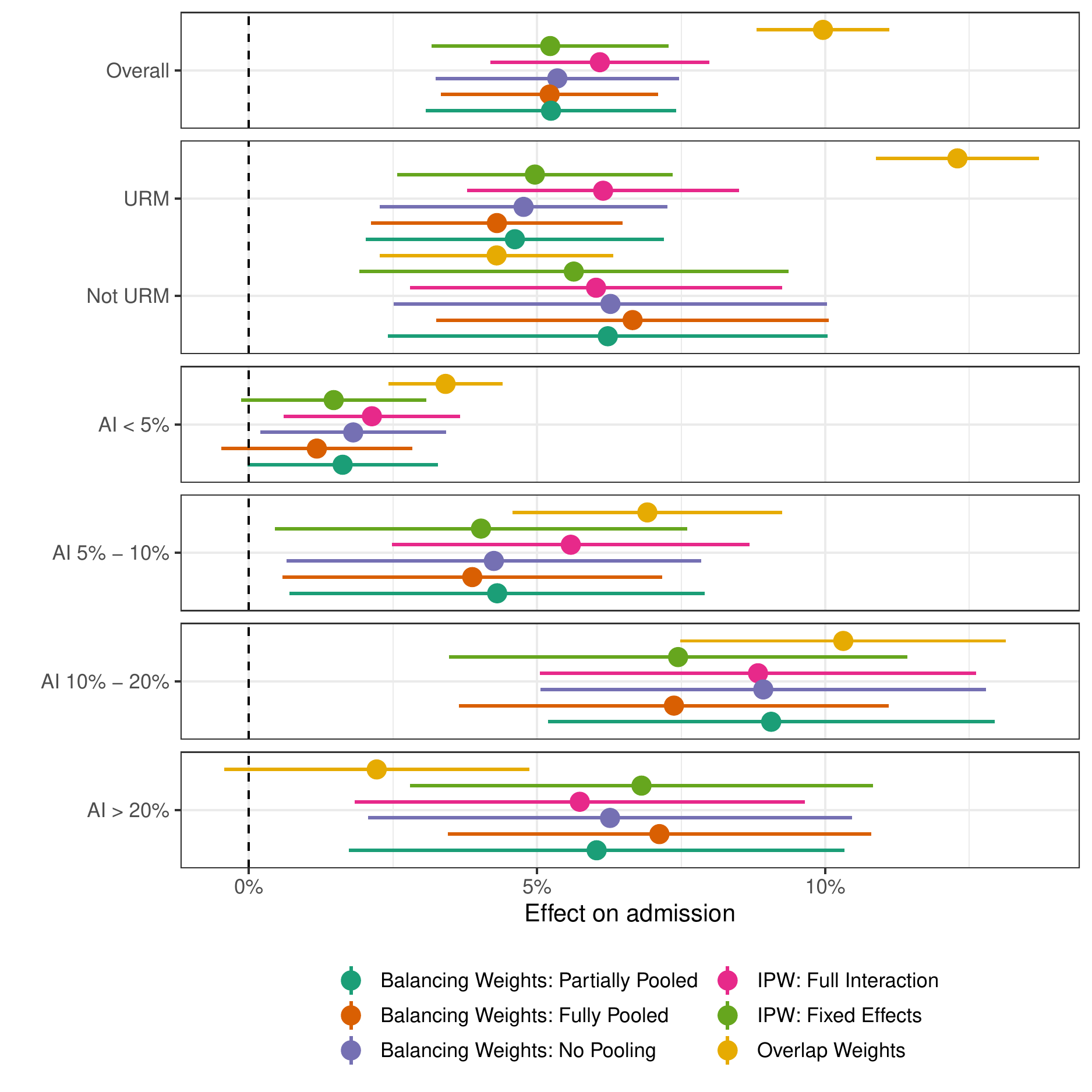} 
    }
    \caption{Overall and by URM status and AI.} 
      \label{fig:weighting_estimates_marginal}
      \end{subfigure}%
      ~
      \begin{subfigure}[t]{0.45\textwidth}  
      {\centering \includegraphics[width=\textwidth]{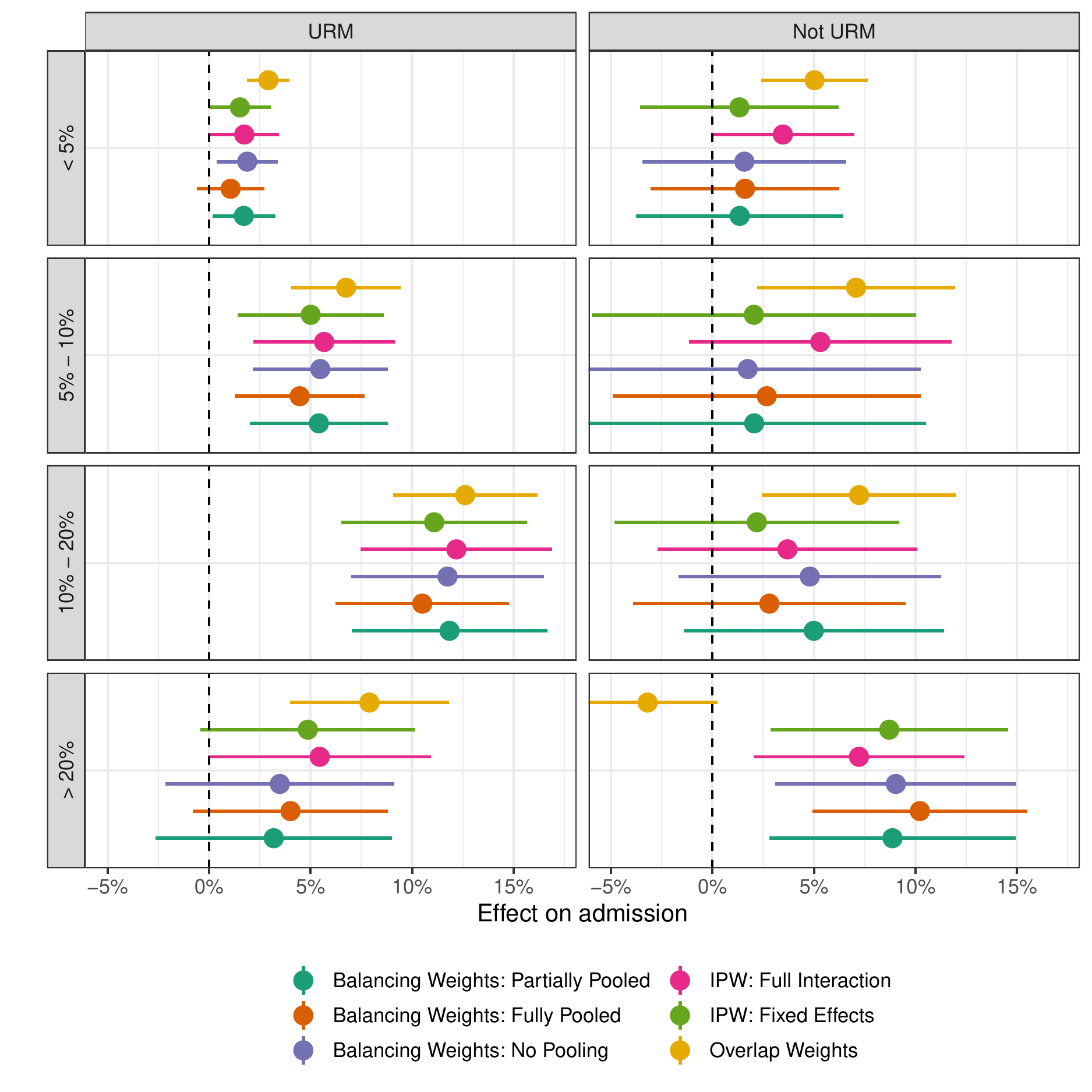} 
      }
      \caption{By URM status interacted with AI.}
        \label{fig:weighting_estimates_interact}
        \end{subfigure}
      \caption{Estimated effect of letters of recommendation on admission rates for comparable weighting estimators.} 
      \label{fig:weighting_estimates}
    \end{figure}

\begin{figure}[tbp]
  \centering
      \begin{subfigure}[t]{0.45\textwidth}  
  {\centering \includegraphics[width=\textwidth]{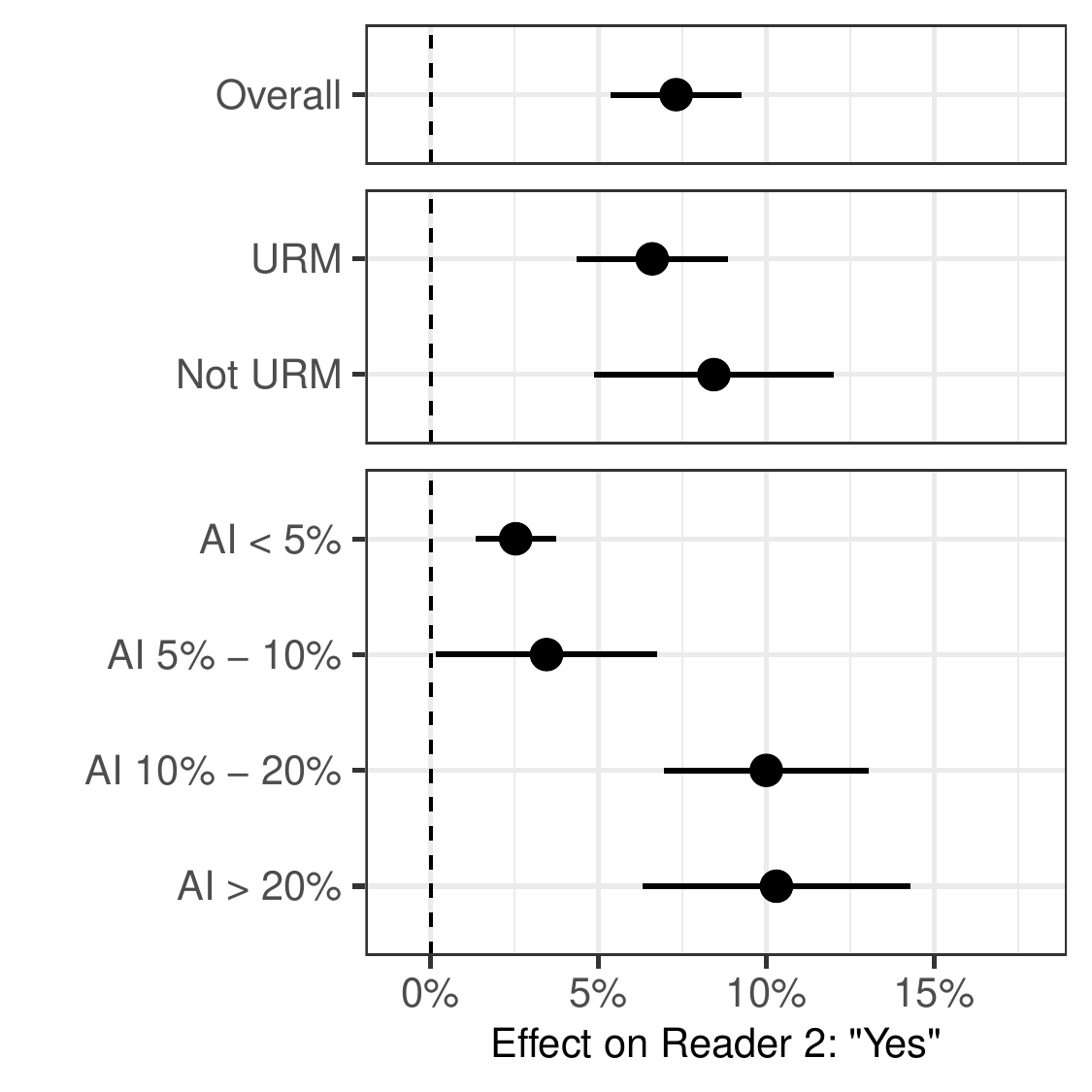} 
  }
  \caption{Partially pooled balancing weights} 
      \label{fig:reader2_estimates_marginal}
      \end{subfigure}%
      ~
      \begin{subfigure}[t]{0.45\textwidth}  
        {\centering \includegraphics[width=\textwidth]{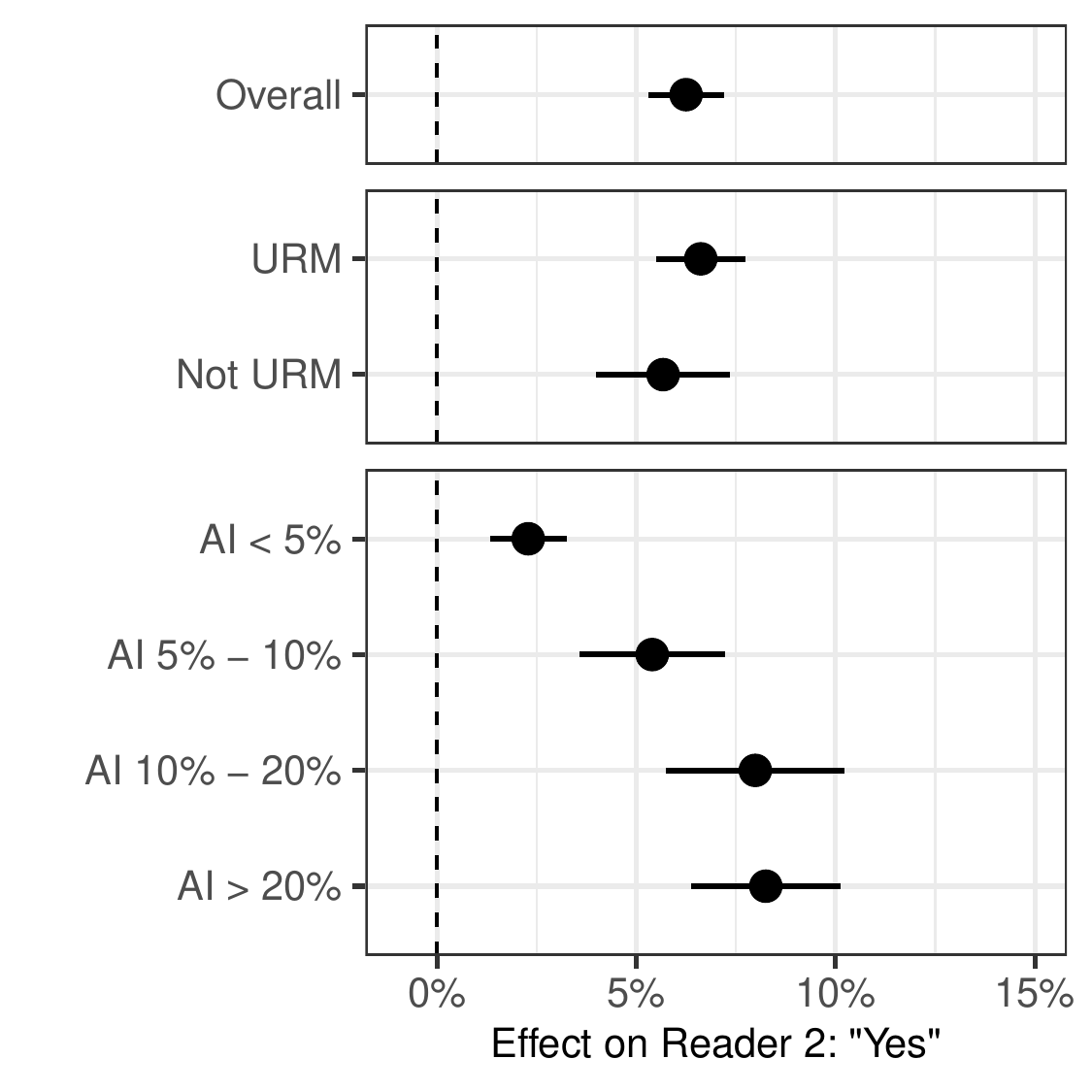} 
}
      \caption{Within-subect design}
      \label{fig:within_estimates_marginal}
      \end{subfigure}
      \caption{Effects on second reader scores overall, by URM status, and by AI, estimated via (a) the partially pooled balancing weights estimator and (b) the within-subject design.} 
      \label{fig:reader2_marginal}
  \end{figure}    

\begin{figure}[tbp]
\centering
\begin{subfigure}[t]{0.45\textwidth}  
  {\centering \includegraphics[width=\textwidth]{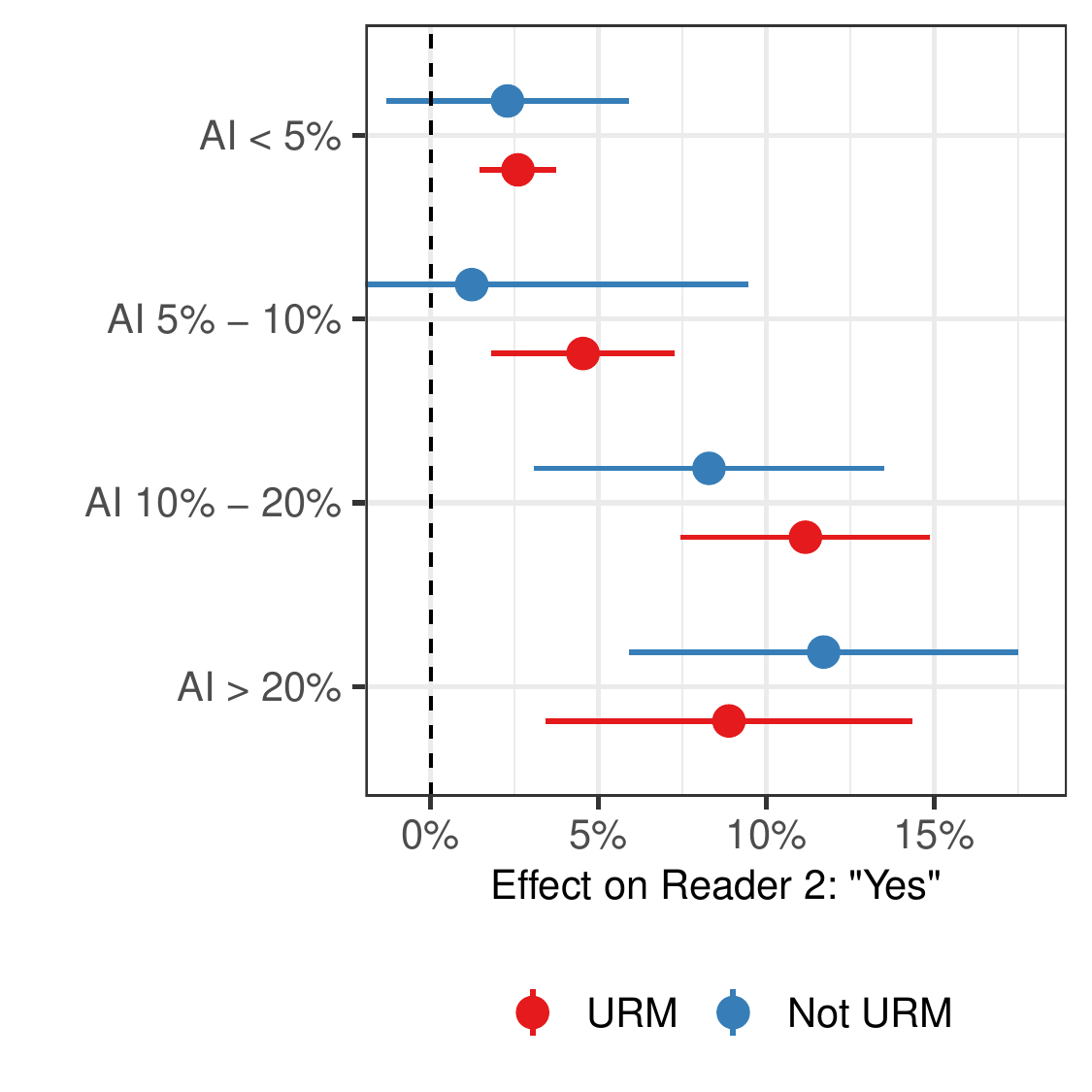} 
      }
\caption{Partially pooled balancing weights} 
\label{fig:reader2_estimates_interact}
\end{subfigure}%
~
\begin{subfigure}[t]{0.45\textwidth}  
  {\centering \includegraphics[width=\textwidth]{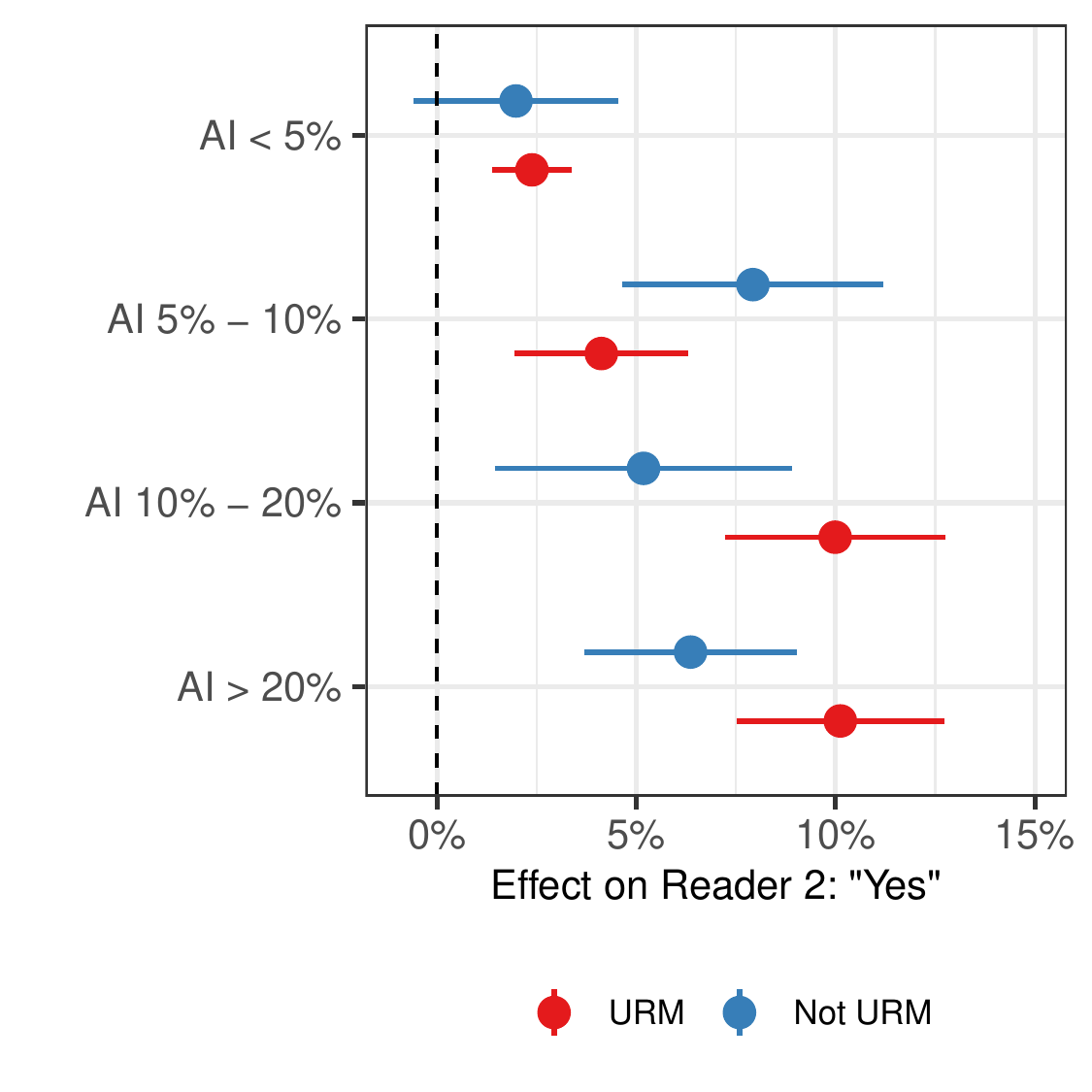} 
  }

\caption{Within-subject design}
\label{fig:within_estimates_interact}
\end{subfigure}
\caption{Effects on second reader scores by URM status interacted with AI, estimated via (a) the partially pooled balancing weights estimator and (b) the within-subject design.} 
\label{fig:reader2_interact}
\end{figure}

\begin{figure}[tbp]
  \centering
      \begin{subfigure}[t]{0.45\textwidth}  
  {\centering \includegraphics[width=\textwidth]{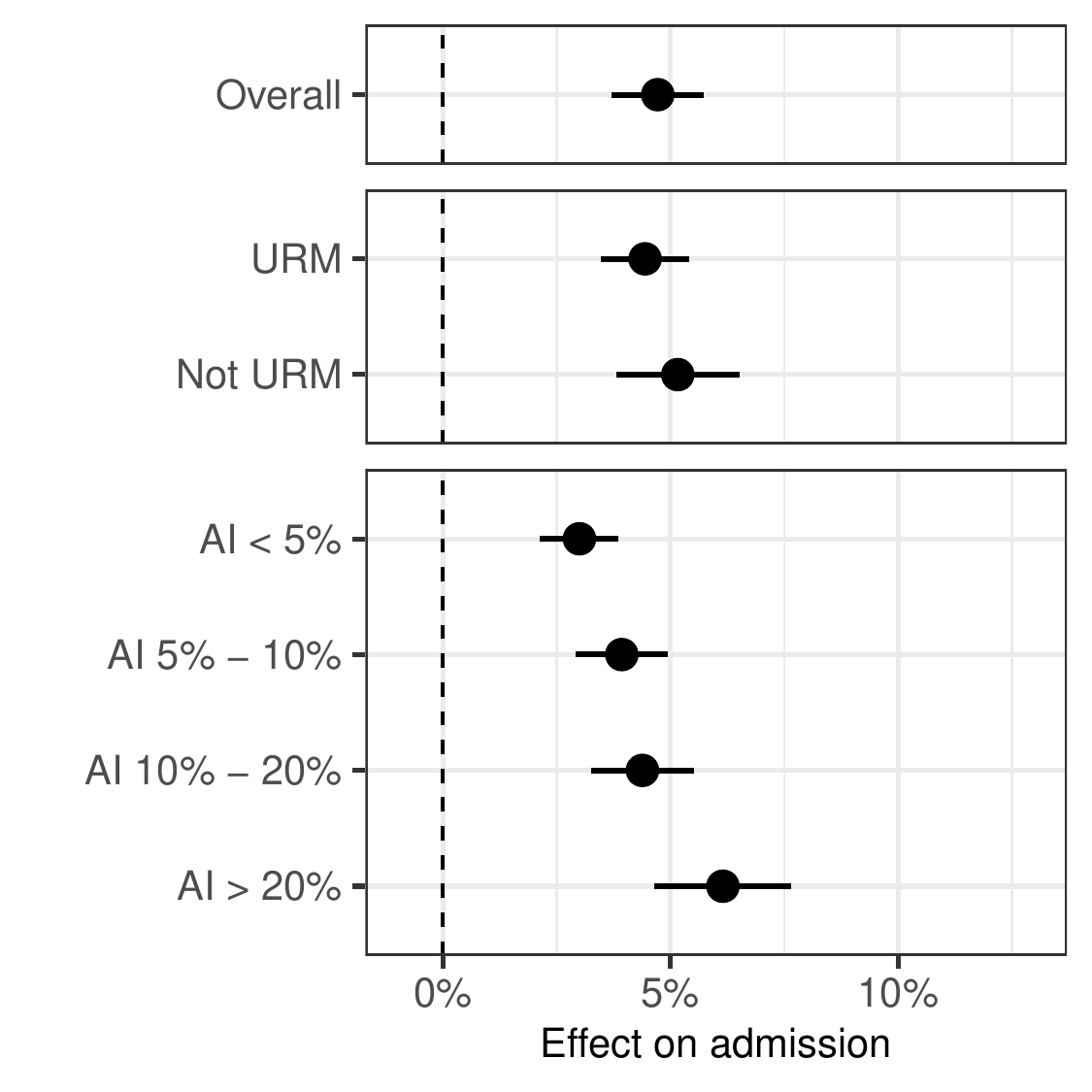} 
  }
  \caption{Overall and by URM status and AI.} 
      \label{fig:bcf_estimates_marginal}
      \end{subfigure}%
      ~
      \begin{subfigure}[t]{0.45\textwidth}  
      {\centering \includegraphics[width=\textwidth]{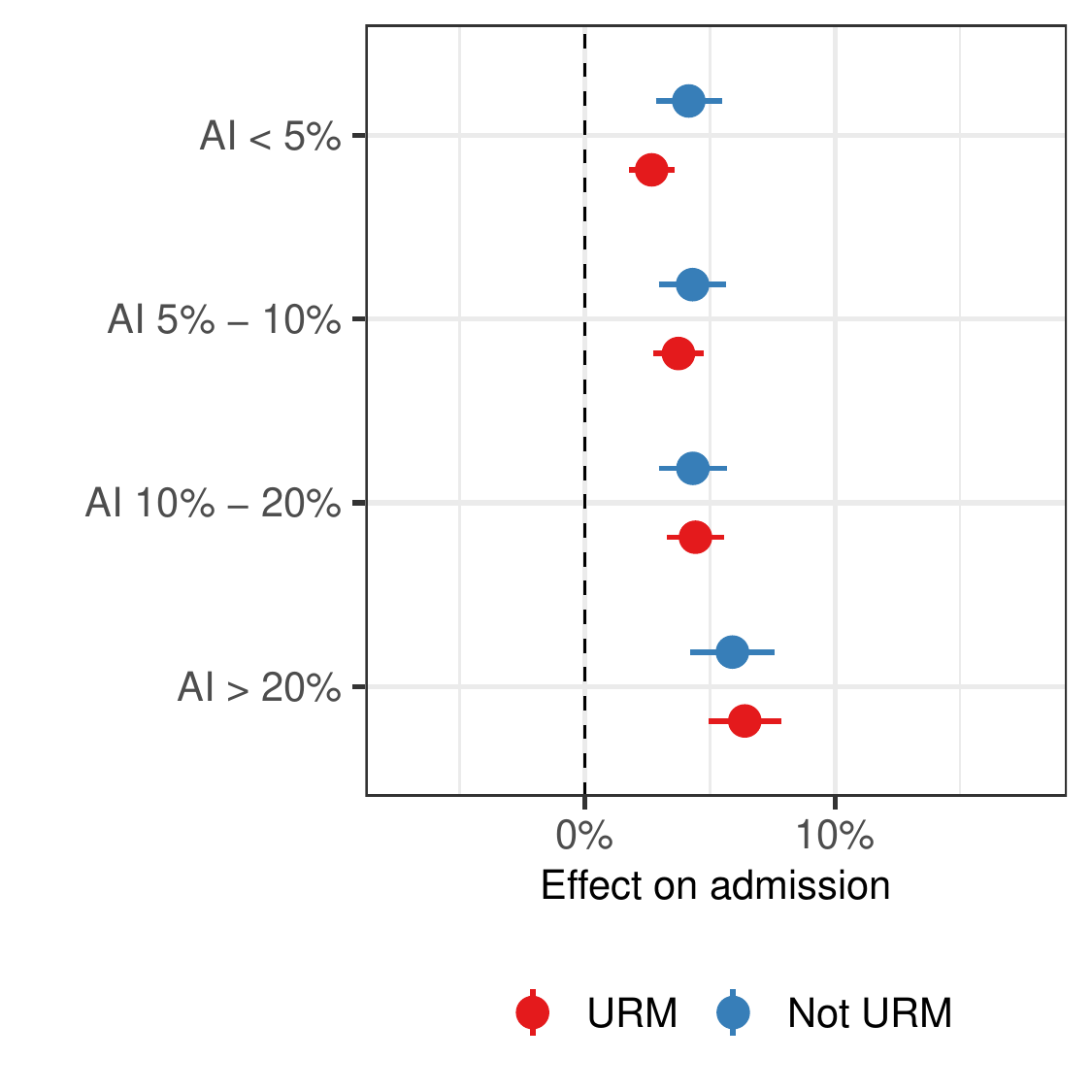} 
      }
      \caption{By URM status interacted with AI.}
      \label{fig:bcf_estimates_interact}
      \end{subfigure}
      \caption{Estimated effect of letters of recommendation on admission rates via Bayesian Causal Forests \citep{Hahn2020}. } 
      \label{fig:bcf_estimates}
  \end{figure}

  \begin{figure}[tbp]
    \centering
    {\centering \includegraphics[width=0.4\textwidth]{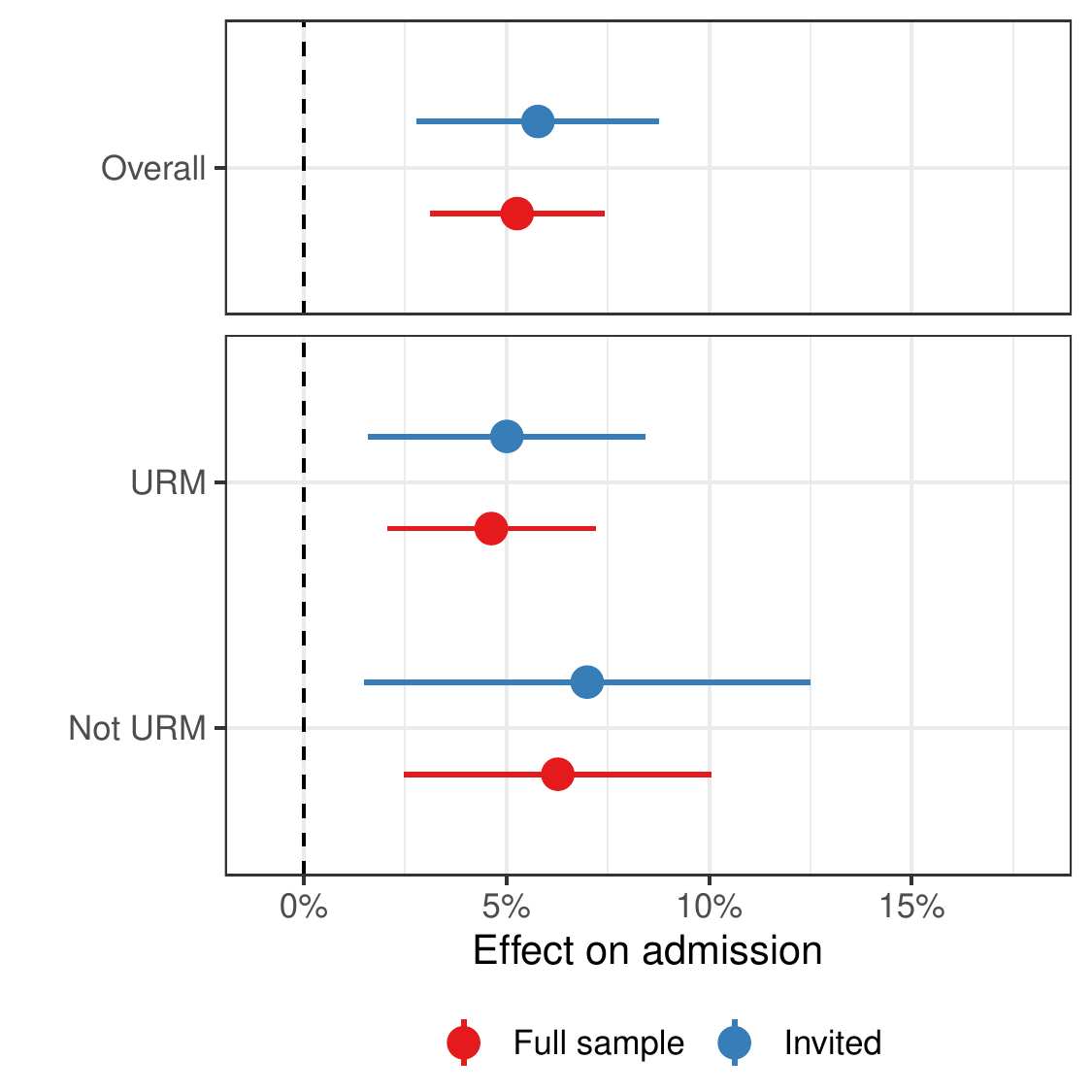} 
    }
        \caption{Estimated effect of letters of recommendation on admission rates via weighting in the full sample and restricting to applicants who were invited to submit an LOR} 
        \label{fig:subset_estimates}
    \end{figure}

\begin{figure}[tbp]
  \centering
  {\centering \includegraphics[width=.6\textwidth]{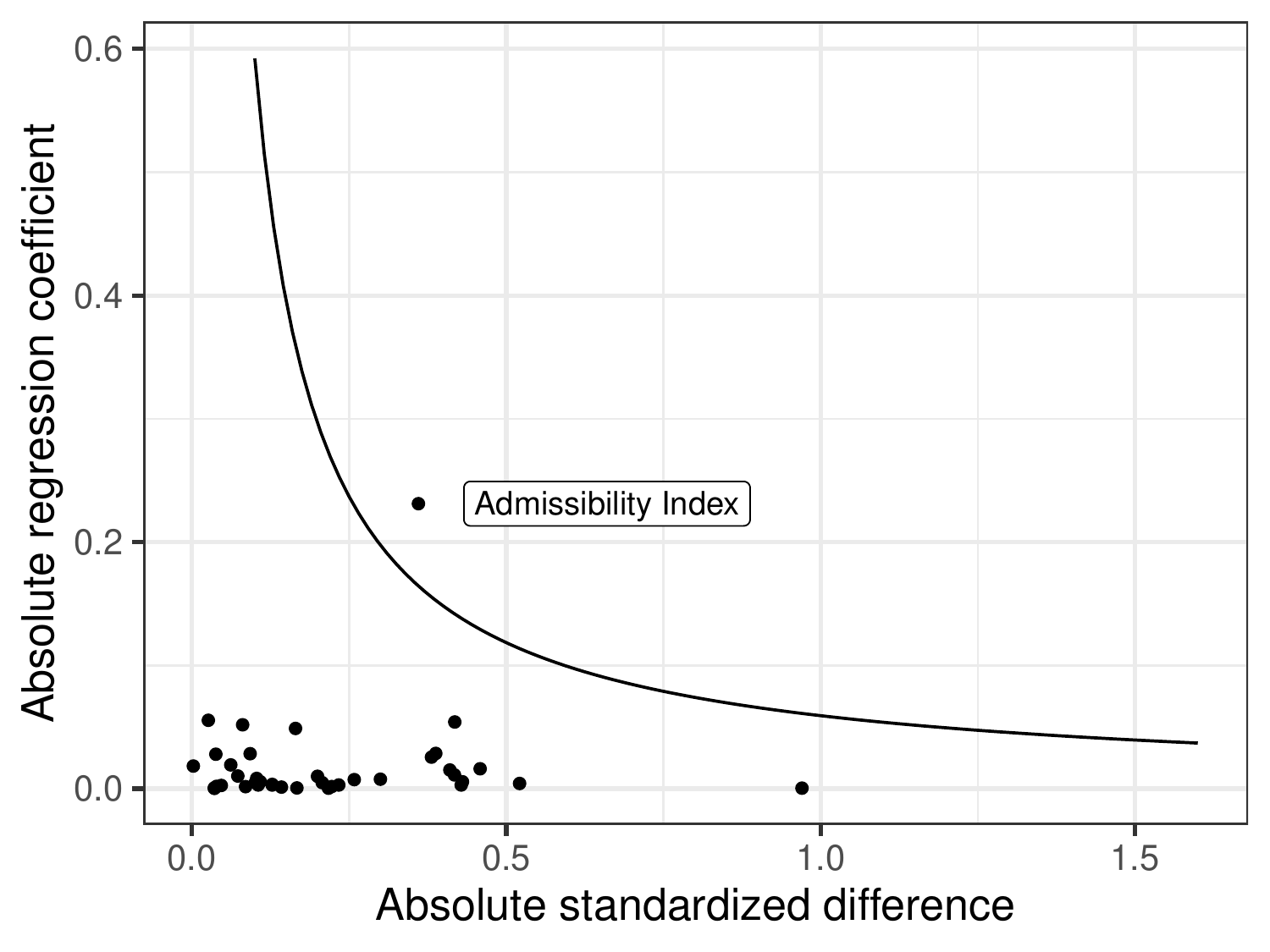}} 
  \caption{Amplification of a sensitivity analysis. The line shows the magnitude of the regression coefficient and the magnitude of the imbalance in an unmeasured standardized covariate required to produce enough bias to remove the effect. Points correspond to the regression coefficients and imbalance before weighting for the 51 components of $\phi(X)$.} 
  \label{fig:sens}
  \end{figure}

%%%
%%% PROOFS
%%%

\clearpage
\section{Proofs}

\begin{proof}[Proof of Proposition \ref{prop:dual}]
  First, we will augment the primal optimization problem in Equation \eqref{eq:primal} with auxiliary covariates $\calE_1,\ldots, \calE_j$ so that $\calE_g = \sum_{G_i = g, W_i = 0}\gamma_i \phi(X_i) - \sum_{G_i = g, W_i = 1} \phi(X_i)$. Then the optimization problem becomes:
\begin{equation}
  \label{eq:primal_aux}
  \begin{aligned}
    \min_{\gamma} \;\;\;\;\;\;\; & \sum_{z=1}^J \frac{1}{2\lambda_g}\left\|\calE_j\right\|_2^2 + \frac{\lambda_g}{2}\sum_{Z_i=z,W_i=0} \gamma_i^2 + \I(\gamma_i \geq 0)\\
    \text{subject to } & \sum_{W_i = 0} \gamma_i \phi(X_i) = \sum_{W_i = 1}\phi(X_i)\\
    & \calE_j = \sum_{G_i = g, W_i = 0} \gamma_i \phi(X_i) - \sum_{G_i = g, W_i = 1} \phi(X_i), \;\;\; z=1,\ldots,J\\
    & \sum_{G_i = g, W_i = 0} \gamma_i = n_{1g},
\end{aligned}
\end{equation}
where $\I(x \geq 0) = \left \{ \begin{array}{cc} 0 & x \geq 0\\ \infty & x < 0 \end{array}\right.$ is the indicator function.
The first constraint induces a Lagrange multiplier $\mu_\beta$, the next $J$ constraints induce Lagrange multipliers $\delta_1,\ldots,\delta_J$, and the sum-to-one constraints induce Lagrange multipliers $\alpha_1,\ldots,\alpha_J$. Then the Lagrangian is
\begin{equation}
  \label{eq:lagrangian}
  \begin{aligned}
    \calL(\gamma, \calE, \mu_\beta, \delta, \alpha) & = \sum_{z=1}^J \left[\frac{1}{2\lambda_g}\|\calE_j\|_2^2 - \calE_j \cdot \delta_j + \sum_{G_i = g, W_i = 0} \frac{1}{2}\gamma_i^2 + \I(\gamma_i \geq 0) - \gamma_i (\alpha + (\mu_\beta + \delta_j) \cdot \phi(X_i)) \right]\\
    & \;\;\; +  \sum_{z=1}^J \sum_{G_i = g, W_i = 1} (1 + (\mu_\beta + \delta_j) \cdot \phi(X_i))
  \end{aligned}
\end{equation}
The dual objective is:
\begin{equation}
  \label{eq:dual_obj}
  \begin{aligned}
    q(\mu_\beta, \delta, \alpha) & = \sum_{z=1}^J \left[\min_{\calE_j} \left\{\frac{1}{2\lambda_g}\|\calE_j\|_2^2 - \calE_j \cdot \delta_j \right\} + \sum_{G_i = g, W_i = 0} \min_{\gamma_i \geq 0}\left\{\frac{1}{2}\gamma_i^2 - \gamma_i (\alpha + (\mu_\beta + \delta_j) \cdot \phi(X_i))\right\} \right]\\
    & \;\;\; +  \sum_{z=1}^J \sum_{G_i = g, W_i = 1} (1 + (\mu_\beta + \delta_j) \cdot \phi(X_i))
  \end{aligned}
\end{equation}
Note that the inner minimization terms are the negative convex conjugates of $\frac{1}{2}\|x\|_2^2$ and ${\frac{1}{2}x^2 + \I(X \geq 0)}$, respectively. Solving these inner optimization problems yields that
\begin{equation}
  \label{eq:dual_obj2}
  \begin{aligned}
    q(\mu_\beta, \delta, \alpha) & = - \sum_{z=1}^J \left[\frac{\lambda_g}{2} \|\delta_j\|_2^2 + \sum_{G_i = g, W_i = 0} \left[\alpha_j + (\mu_\beta + \delta_j) \cdot \phi(X_i)\right]_+^2 \right]\\
    & \;\;\; +  \sum_{z=1}^J \sum_{G_i = g, W_i = 1} (1 + (\mu_\beta + \delta_j) \cdot \phi(X_i))
  \end{aligned}
\end{equation}
Now since there exists a feasible solution to the primal problem \eqref{eq:primal}, from Slater's condition we see that the solution to the primal problem is equivalent to the solution to $\max_{\mu_\beta, \alpha,\delta} q(\mu_\beta, \alpha, \delta)$. Defining $\beta_j \equiv \mu_\beta + \delta_j$ gives the dual problem \eqref{eq:dual}. Finally, note that the solution to the minimization over the weights in Equation \eqref{eq:dual_obj} is $\gamma_i = \left[\alpha_j + \beta_j \cdot \phi(X_i)\right]_+$, which shows how to map from the dual solution to the primal solution.
\end{proof}

\clearpage
\bibliography{citations.bib}
\bibliographystyle{chicago}

% \end{document}

\end{document}